\documentclass{iopart}
\usepackage{iopams} 
\usepackage{mathabx}
\usepackage{graphicx}
\newtheorem{Theorem}{Theorem}[section]
\newtheorem{Proposition}[Theorem]{Proposition}

\def\F{{\mathcal F}}
\def\DD{{\mathbb D}}

\def\KK{{\mathbb K}}
\def\NN{{\mathbb N}}
\def\MM{{\mathbb M}}
\def\P{{\mathbb{P}}}
\def\RR{{\mathbb{R}}}
\def\HH{{\mathbb{H}}}

\begin{document}

\title{Wave Computation on the Poincar\'e dodecahedral space}

\author{Agn\`{e}s BACHELOT-MOTET}

\address{Universit\'e de Bordeaux, Institut de Math\'ematiques, UMR CNRS 5251, F-33405 Talence Cedex}

\ead{agnes.bachelot@math.u-bordeaux1.fr}

\begin{abstract}
We compute the waves propagating on a compact $3$-manifold of constant positive curvature with a non trivial topology: the Poincar\'e dodecahedral space  that is a plausible model of  multi-connected universe.
We transform the Cauchy problem to a mixed problem posed on a fundamental domain determined by the quaternionic calculus. We adopt a variational approach using a space of finite elements that is invariant under the action of the binary icosahedral group. The computation of the transient waves is validated with their spectral analysis by computing a lot of eigenvalues of the Laplace-Beltrami operator.
\end{abstract}
\ams{35Q75, 58J45, 65M60}
\maketitle


\section{Introduction.}
Some fundamental open questions regarding the nature of the universe concern its geometry and topology. They are the subject of many articles (\cite{Corn}, \cite{Lumi-2}, \cite{Lumi-1}, \cite{Senden}, \cite{Wee-1} for example) in which models are compared with observations which are more accurate over the years. About geometry, after nine years of investigation, the data by WMAP (Wilkinson Microwave Anisotropy Probe) mission have provided strong evidence suggesting that the universe is nearly flat, with the ratio of its total matter-energy density to the critical value very close to one \cite{WMAP}, but without fixing the sign of its curvature. The Poincar\'e dodecahedral space (PDS) is a plausible multi-connected universe model with positive spatial curvature. More precisely, PDS is the quotient of the unit 3-sphere $\mathcal{S}^3$ by the group $\mathcal{I}^*$ of covering isometries. We suppose that the local geometry of the Universe is described by a Friedmann-Lema\^{i}tre metric, except possibly during the virialisation (low redshift) epoch, when use of this metric has been claimed to give rise to dark energy as an artefact of assuming a homogeneous metric despite the
strong growth of non-linear inhomogeneities (\cite{Bo}, \cite{Bu}, \cite{Kolb}, \cite{Rouk2}, \cite{Wi-Da}). With this assumption, PDS (denoted by $\mathbf{K}$ in equations) is endowed with the spherical metric $ds_{\mathbf{K}}^2$ induced by $ds_{\mathcal{S}^3}^2$. It is a three dimensional $C^{\infty}$ compact manifold, without boundary. Such a manifold can conveniently be converted to a description as a fundamental domain. It will have the shape of a dodecahedron, with pairs of faces identified. 

Many authors have studied PDS (\cite{Aur}, \cite{C-LR-L-L-R-W}, \cite{G-L-Lu-Uz-W}, \cite{Lumi-2}, \cite{Lumi-Wee},  \cite{Rouk}, \cite{Senden}, \cite{Wee-2}). A stationary approach has been used in all these articles. The exact knowledge of the eigenvalues of the Laplace-Beltrami operator $\Delta_{\mathbf{K}}$ on PDS allows them to predict some results (for example the cosmic microwave background temperature anisotropies), and to compare with the WMAP observations; but we note these theoretical results depend on assumptions about the statistical nature of the power spectrum of density perturbations that may not be fully valid in PDS \cite{Rouk3}.

In this paper we are interested in solving numerically the linear wave equation on PDS
\begin{equation}
\boxvoid_g\Psi:=\partial^2_t\Psi-\Delta_{\mathbf{K}}\Psi=0.
 \label{zeeque}
\end{equation}
This equation plays a fundamental role in General Relativity. We recall that given a lorenzian manifold $({\mathcal{M}},g_{\mu\nu}dx^{\mu}dx^{\nu})$, the gravitational fluctuations are described, at the first order, by the linearized Einstein equations in vacuum
$$
\boxvoid_gh_{\mu \nu}=0,\;\;\boxvoid_g:=\frac{1}{\sqrt{|g|}} \partial_\mu \left(g^{\mu \nu}
\sqrt{|g|}\partial_\nu.\right).
$$
Another situation in which this equation appears, is the construction of the harmonic coordinates $x^{\mu񓩌}$ that are solutions of $\boxvoid_gx^{\mu}=0$. Hence we may consider (\ref{zeeque}) in particular as the equation of the gravitational waves in the PDS universe $(\RR_t\times\mathbf{K}, dt^2-ds^2_{\mathbf K})$ that is the multiconnected analogue of the Einstein universe $(\RR_t\times {\mathcal S}^3, dt^2-ds^2_{{\mathcal S}^3})$ : both these manifolds have the same metric, they differ only by the topology.

To solve equation (\ref{zeeque}), taking advantage of the selfadjointness of the Laplace operator, we could expand any normalizable solution $\Psi\in C^0\left(\RR_t;L^2(\mathbf K)\right)$ on a hilbertian basis of eigenfunctions $\Psi_q$ satisfying $-\Delta_{\mathbf K}\Psi_q=q^2\Psi_q$, $0<q$, $\Vert\Psi_q\Vert_{L^2}=1$ :
$$
\Psi(t,x)=a+bt+\sum_{q\neq 0}\left(c_qe^{iqt}+c'_qe^{-iqt}\right)\Psi_q(x),
$$
where the complex numbers $a,b,c_q,c'_q$ are determined by the initial data. This method is powerful since the eigenvalues are explicit (see \cite{Aur}, \cite{Ike}, \cite{Lach}, \cite{Leh-W-U-G-Lu}) and the drawback of this spectral method, namely the numerical computation of the eigenmodes $\Psi_q$, has been recently overcome in \cite{Aurich2}. Moreover, we can prove that the accuracy of the approximate solution 
$\Psi_{[N]}(t,x):=a-bt-\sum_{0<q<N}\left(c_qe^{iqt}+c'_qe^{-iqt}\right)\Psi_q(x)$ is estimated as
$$
\sup_{t\in\RR,\;x\in\mathbf{K}}\left\vert \Psi(t,x)-\Psi_{[N]}(t,x)\right\vert
\lesssim\left(\sum_{q\geq N}q^4\left(\mid c_q\mid^2+\mid c'_q\mid^2\right)\right)^{\frac{1}{2}}.
$$
Since the sequences $c_q$ and $c'_q$ go to zero as $q\rightarrow\infty$ faster than any inverse power of $q$ when the initial data are $C^{\infty}$, this scheme provides a very precise approximation for the smooth solutions. Nevertheless, in this paper we adopt the finite element method previously developed in \cite{Bach}, where we studied a toy model of hyperbolic universe. This method is less accurate than the spectral approach since the accuracy does not increase with the regularity of the initial data: it is first order. But it has several advantages: there is no heavy precomputing such as the calculation of the eigenmodes, and above all, this algorithm could be extended much more easily than the spectral expansion, to investigate numerically the non-linear dynamics, under the simplifying (standard) assumption that the Friedmann-Lema\^{i}tre metric remains valid despite the non-linearity.

The construction of PDS is detailled in the next section.  In the third part we get a practical description of the particular fundamental domain $\mathcal{F}$ that contains $(1,0,0,0)$ of $\RR^4$, and also we get a numerical description of the projection $\mathcal{F}_v$ of $\mathcal{F}$ in $\RR^3$. We mainly use the quaternionic calculus. Next we consider the scalar wave operator $\partial^2_t-\Delta_{\mathbf{K}}$ on the dodecahedral universe $\RR_t\times\mathbf{K}$. Then, we compute the solutions of the wave equation in the time domain, by using a variational method and a discretization with finite elements.
The domain of calculus is $\mathcal{F}_v$, therefore the initial Cauchy problem on the manifold without boundary $\mathbf{K}$, becomes a mixed problem
on  $\mathcal{F}_v$ with suitable periodic boundary conditions on $\partial\mathcal{F}_v$. These boundary constraints are implemented in the choice of
the basis of finite elements. 
We validate our results by performing a Fourier analysis of the transient waves that allows to find a lot of eigenvalues of the Laplace-Beltrami operator $\Delta_{\mathbf{K}}$ on PDS.

\section{The Poincare Dodecahedral Space.}
To be able to perform computations on PDS we need to describe it accurately. This section recalls the properties of $\mathcal{S}^3$, $\mathcal{I}^*$ and $\mathcal{S}^3 / \mathcal{I}^*$ that we need to know. The 3-sphere is the submanifold of four-dimensional Euclidean space such that $x_0^2+x_1^2+x_2^2+x_3^2=1$. The comoving spatial distance $d$ between any two points $x$ and $y$ on $\mathcal{S}^3$ is given by:
\[ d(x,y)=\arccos [ x^i y_j] \] 
We also use the parametrisation: 
\[ \begin{array}{ll}
x_0=& \cos \chi\\
x_1=& \sin \chi \sin \theta \sin\varphi\\
x_2=& \sin \chi \sin \theta \cos\varphi\\
x_3=& \sin \chi \cos \theta
\end{array}
\]
with $0\leq \chi\leq \pi$, $0\leq \theta\leq \pi$ and $0\leq \varphi\leq 2\pi$. These parametrisation leads to a convenient way to visualize $\mathcal{S}^3$: two balls in $\RR^3$ glued together by their boundary. The $\theta$ and $\varphi$ coordinates are the standard ones for $\mathcal{S}^2$. For the first ball $\chi$ runs from $0$ at the center through $\frac{\pi}{2}$ at the surface. For the second ball $\chi$ runs from $\frac{\pi}{2}$ at the surface through $\pi$ at the center. We use a projection $p$ of $\mathcal{S}^3$ in $\RR^3$:
\[
\begin{array}{lclc}
p:& \mathcal{S}^3& \rightarrow &\RR^3\\
&(x_0,x_1,x_2,x_3)&\mapsto & (x_1,x_2,x_3)\\
\end{array}
\]
 That is to say, to represent a point of coordinates $(x_i)_{\{i=0,1,...,3\}}$ of $\mathcal{S}^3$, we consider only the coordinates $(x_i)_{\{i=1,...,3\}}$ and discard $x_0$. The two points of coordinates $(x_0,x_1,x_2,x_3)$ and $(-x_0,x_1,x_2,x_3)$ have the same coordinates $(x_1,x_2,x_3)$; they are represented by the same point. In the sequel,for any $A\subset \mathcal{S}^3$, we use $A_v$ to denote the visualization in $\RR^3$ of $A$, that is $A_v :=p(A)$.

To define a discrete fixed-point free subgroup $\Gamma \subset SO(4)$ of isometries of $\mathcal{S}^3$, one makes use of the fact that the unit 3-sphere $\mathcal{S}^3 $ is identified with the multiplicative group $\HH_1$ of unit quaternions by 
\[x=(x_0,x_1,x_2,x_3) \Longleftrightarrow q=x_0\mathbf{1}+x_1\mathbf{i}+x_2\mathbf{j}+x_3\mathbf{k}.\] The four basic quaternions $\{\mathbf{1},\mathbf{i},\mathbf{j},\mathbf{k\}}$ of the set $\HH$ of all quaternions, satisfy the multiplication rules $\mathbf{i}^2=\mathbf{j}^2 =\mathbf{k}^2=-1$, and $\mathbf{k}=\mathbf{i}\mathbf{j}=-\mathbf{j}\mathbf{i}$. They commute with every real number. The norm of $q$ is defined by $\|q\|:=x_0^2+x_1^2+x_2^2+x_3^2$. The conjugate $\bar{q}$ of $q=a \mathbf{1}+b\mathbf{i}+c\mathbf{j}+d\mathbf{k}$ is defined by $\bar{q}:=a \mathbf{1}-b\mathbf{i}-c\mathbf{j}-d\mathbf{k}$.

In this paper, we are interested in the subgroup $\mathcal{I}^*$ of $SO(4)$ called the binary icosahedral group (\cite{We-Sei}, \cite{Wolf}, \cite{Sco}, \cite{G-L-Lu-Uz-W}). It is a two sheeted covering of the icosahedral group $\mathcal{I}\subset SO(3)$ consisting of all orientation-preserving symmetries of a regular icosahedron. Indeed, for any $q$ in the multiplicative group of unit quaternions $\HH_1$ we consider $p_q:\, \HH\rightarrow \HH$ defined by $p_q(q'):= q q' q^{-1}$. It fixes the identity quaternion $\mathbf{1}$, so in effect its action is confined to the equatorial 2-sphere spanned by the remaining basis quaternions $(\mathbf{i}, \mathbf{j}, \mathbf{k})$. Thus, by viewing $\RR^3$ as the space of pure imaginary quaternions, subspace of $\HH$ with basis $\{\mathbf{i},\mathbf{j},\mathbf{k}\}$, we get a rotation in $\RR^3$ . So $p_q$ belongs to $SO(3)$. The map $\pi:\,\HH_1\rightarrow SO(3)$ defined by $\pi(q)=p_q$ is a homomorphism of $\HH$ onto $SO(3)$. $\pi$ is a two to one homomorphism as $\pi(q)=\pi(-q)$. By definition $\mathcal{I}^*$ is the pre-image of $\mathcal{I}$ by $\pi$. The order of $\mathcal{I}$ is $60$, so the order of $\mathcal{I}^*$ is $120$. $\mathcal{I}^*$ contains only right-handed Clifford translations $\gamma$, in other words $\gamma$ acts on an arbitrary unit quaternion $q \in \mathcal{S}^3$ by left multiplication and translates all points $q_1, q_2 \in \mathcal{S}^3$ by the same distance $\chi$, i.e. $d(q_1,\gamma q_1)=d(q_2,\gamma q_2)=\chi$. The right-handed Clifford translations act as right-handed cork screw fixed-point free rotations of $\mathcal{S}^3$. \\
Let take $\gamma=a \mathbf{1}+b\mathbf{i}+c\mathbf{j}+d\mathbf{k}$ an element of $\mathcal{I}^*$ (with $\arccos a=\chi$), and $q=x_0\mathbf{1}+x_1\mathbf{i}+x_2\mathbf{j}+x_3\mathbf{k}$ in $\mathcal{S}^3 $. We can express $\gamma $ as a matrix and we have:
\[
\gamma \; q=\left( \begin{array}{rrrr}
a &-b &-c &-d\\
b & a & -d & c \\
c& d & a & -b \\
d & -c & b & a 
\end{array}
\right)
\left( \begin{array}{r}
x_0\\
x_1\\
x_2\\
x_3 
\end{array}
\right)
\]
$\mathcal{I}^*$ is generated by two isometries $s$ and $\gamma$. Denoting by $\sigma=(1+\sqrt{5})/2$ the golden number we have: 
\begin{eqnarray}
\fl \mathcal{I}^*&=\left< s,\gamma\; |\; (s\gamma)^2=s^3=\gamma^5\right>\nonumber\\
\fl&=\left\{\pm \mathbf{1},\pm \mathbf{ i},\pm \mathbf{j},\pm \mathbf{k},\frac12 (\pm \mathbf{1} \pm \mathbf{i} \pm \mathbf{j} \pm \mathbf{k}),\frac12(0 \mathbf{1} \pm \mathbf{i} \pm\frac{1}{\sigma} \mathbf{j} \pm\sigma \mathbf{k})\, (with\; even\; permutations) \right\}.\nonumber
\end{eqnarray}

We see that every element of $\mathcal{I}^*$ is a right-handed Clifford translation with $\chi$ equal to $0$, $\pi$, $\frac{\pi}{2}$, $\frac{\pi}{3}$, $\frac{2\pi}{3}$, $\frac{\pi}{5}$, $\frac{2\pi}{5}$, $\frac{3\pi}{5}$, or $\frac{4\pi}{5}$. We deduce that $\frac{\pi}{5}$ is the smallest non-zero translation distance. For example $\mathcal{I}^*$ is generated by $s=\frac12(\mathbf{1}+\mathbf{i}+\mathbf{ j}+\mathbf{k})$ and $\gamma=\frac{\sigma}{2} \mathbf{1}+\frac{1}{2\sigma}\mathbf{ j} -\frac12 \mathbf{k}$. $s$ is a right-handed Clifford translation with $\chi$ equal to $\frac{\pi}{3}$. Also, $p_s$ is a rotation in $\RR^3$, its angle is $\frac{2\pi}{3}$ and its axis is directed by the vector $\frac{1}{\sqrt{3}}(1,1,1)$. $\gamma$ is a Clifford translation with $\chi$ equal to $\frac{\pi}{5}$. Moreover $p_\gamma$ is a rotation in $\RR^3$, its angle is $\frac{2\pi}{5}$ and its axis is directed by the vector $\frac{1}{\sqrt{3-\sigma}}(0,\frac{1}{\sigma} ,-1 )$.

We now consider $\mathcal{S}^3 / \mathcal{I}^*$ the quotient of the 3-sphere $\mathcal{S}^3 $ under the action of the discrete fixed-point free subgroup $\mathcal{I}^*$ of isometries of $\mathcal{S}^3$, with $\mathcal{I}^*$ acting by left multiplication. This is the Poincar\'e dodecahedral space (PDS) (\cite{We-Sei}, \cite{Wolf}, \cite{Sco}, \cite{G-L-Lu-Uz-W}).
 To perform the computations of the waves on $\mathcal{S}^3 / \mathcal{I}^*$, it is very
useful to represent it by a fundamental domain $\mathcal{F}\subset\mathcal{S}^3$ and an equivalence relation $\sim$
such that 
\begin{equation}
\mathcal{S}^3 / \mathcal{I}^*=\mathcal{F}/ \sim.
\label{SIM}
\end{equation}
$\mathcal{F}$ is such that:
\begin{equation}
\label{SIM2}
\fl  \mathcal\displaystyle{{\mathcal{S}}^3=\bigcup_{t\in \mathcal{I}^*} t(\mathcal{F})},\qquad {\mbox and} \qquad
 \forall t\in \mathcal{I}^*,\,\forall t' \in \mathcal{I}^*,\;\stackrel{\circ}{ t(\mathcal{F})}\cap \stackrel{\circ}{ t'(\mathcal{F})}=\emptyset.
\end{equation}
$\mathcal{F} $ is a regular spherical dodecahedron (dual of a regular icosahedron), and $\sim$ is obtained by identifying any pentagonal face of $\mathcal{F}_v $ with its opposite face, after rotating by $\frac{\pi}{5}$ in the clockwise direction around the outgoing axis orthogonal to this last face. $120$ such spherical dodecahedra tile the $3$-sphere in the pattern of a regular $120$-cell (see figure \ref{DODIM}). 
\section{Fundamental domain $\mathcal{F}$ of PDS, and its visualization $\mathcal{F}_v$.} 
We have choosen the unique fundamental domain that contains $(1,0,0,0)$ to simplify calculations and visualization. In the following it will be denoted $\mathcal{F}$. 
In order to perform computations we need to know the coordinates of all points in $\mathcal{F}$, and the equations of its edges and faces. We also deduce the characteristics of $\mathcal{F}_v:=p(\mathcal{F})$, which is our domain of calculus and visualization.
\begin{Proposition}\label{vertices-F}
The unique fundamental domain that contains $(1,0,0,0)$ is the geodesic convex hull in $\mathcal{S}^3 $ of these $20$ vertices: 
$\frac{1}{2\sqrt{2}} (\sigma ^2,\pm \frac{1}{\sigma ^2},0,\pm 1)$, 
$\frac{1}{2\sqrt{2}} (\sigma ^2,0,\pm 1,\pm \frac{1}{\sigma ^2})$, 
$\frac{1}{2\sqrt{2}} (\sigma ^2,\pm\frac{1}{\sigma},\pm\frac{1}{\sigma},\pm\frac{1}{\sigma})$ and
$\frac{1}{2\sqrt{2}} (\sigma ^2,\pm 1,\pm \frac{1}{\sigma ^2},0)$. 

Each of the $12$ faces $F_i$ of $\mathcal{F}$ is a regular pentagon in $\mathcal{S}^3 $ included in \[\{(x_0,x,y,z)\in \mathcal{S}^3,\quad a_ix+b_iy+c_iz=\frac{x_0}{\sigma^2}\},\]
with $(a_i,b_i,c_i)$ equal to $(\pm \frac{1}{\sigma},\pm 1,0)$, up to an even permutation.

The set of all barycenters in $\RR^4$ of the vertices of $F_i$, denoted by $F_i^b$, is included in a $2$-plane of equations $x_0=\frac{\sigma^2}{2\sqrt{2}}$ and $a_ix+b_iy+c_iz=\frac{x_0}{\sigma^2}$.
\end{Proposition}
{\it Proof}: $\mathcal{F}$ is the geodesic convex hull in $\mathcal{S}^3 $ of its $20$ vertices, so we begin to search the coordinates of the vertices. As the shortest translation distance of the elements of $\mathcal{I}^*$ is $\frac{\pi}{5}$, the shortest translation distance in PDS is $\frac{\pi}{5}$. Two opposite faces must be the image from each other by a Clifford translation with $\chi$ equal to $\frac{\pi}{5}$. $\mathcal{I}^*$ has $12$ such elements denoted by $g_i$ in the following parts:
\[
\fl \frac12 (\sigma \mathbf{1}+0\,\mathbf{ i}\pm\frac{1}{\sigma} \mathbf{j }\pm \mathbf{k})\; (with\; even\; permutations\; of\; the\; three\; last\; coordinates). 
\]
So the isometries of $SO(3)$ denoted by $p_{g_i}$ leave invariant $p(\mathcal{F}):=\mathcal{F}_v$. Furthermore $(0,0,0)$ belongs to $\mathcal{F}_v$ because $(1,0,0,0)$ belongs to $\mathcal{F}$. So $OB_i:=\frac{1}{\sqrt{3-\sigma}}(0,\pm\frac{1}{\sigma},\pm 1)$, with even permutations, are the orthogonal axes to each pair of opposite faces of $\mathcal{F}_v$. $B_i$ are the vertices of an icosahedron that is the dual of a regular dodecahedron whose vertices are the barycenters of three equidistant vertices of the icosahedron. We get a pentagonal face having $OB_i$ as symetry axis by finding the five vertices that are at the same minimal distance from $B_i$. Among these five points, two vertices are adjacent if their distance is the smallest one. We deduce that, in $\RR^4$, the three last coordinates of the vertices of $\mathcal{F}$ could be: 
\[
 (  \pm \frac16 \sigma,\pm \frac16 \sigma,\pm \frac16 \sigma),\quad 
\mbox{or}\quad(  0,\pm \frac16 \sigma^2,\pm \frac16)\; (with\; even\; permutations).
\]
So the coordinates of the $20$ vertices $C_i$ belonging to $\mathcal{S}^3$ and being the vertices of a regular dodecahedron that has the same symetry axes are of the form:
\[
\begin{array}{ll}
&    \sqrt{1-3\left(\lambda \frac{\sigma}{6}\right)^2}\mathbf{1}\;+ \lambda (\pm \frac16 \sigma \mathbf{i}\pm \frac16 \sigma \mathbf{j}\pm \frac16 \sigma \mathbf{k}),\\
or\; &\sqrt{1-\lambda^2\frac{\sigma^2}{12} }\mathbf{1}\;+\lambda( 0\,\mathbf{ i}\pm \frac16 \sigma^2 \mathbf{j} \pm \frac16 \mathbf{k})\; (with\; even\; permutations),
\end{array}
\]
with $\lambda\in \RR$ such that two opposite faces can be the image from each other by a Clifford translation with $\chi$ equal to $\frac{\pi}{5}$. Consider a face $F$ of $\mathcal{F}$ such that $\frac12 \left(-\frac{1}{\sigma} i-j\right)$ is a symetry axis of $\mathcal{F}_v$ orthogonal to the induced face of $\mathcal{F}_v$. Its adjacent vertices satisfy:
\[
\begin{array}{ll}
\fl C_1= \sqrt{1-3\left(\lambda \frac{\sigma}{6}\right)^2}\mathbf{1}+ \lambda (\frac16 \left(-\sigma \mathbf{ i}-\sigma \mathbf{ j}+\sigma \mathbf{ k} \right)),&C_2= \sqrt{1-3\left(\lambda \frac{\sigma}{6}\right)^2}\mathbf{1}+ \lambda (\frac16 \left(-\sigma^2 \mathbf{i}-\mathbf{j }\right) ),\\
\fl C_3= \sqrt{1-3\left(\lambda \frac{\sigma}{6}\right)^2}\mathbf{1}+ \lambda (\frac16 \left(-\sigma \mathbf{i}-\sigma \mathbf{j}-\sigma \mathbf{k} \right) ), &C_4= \sqrt{1-3\left(\lambda \frac{\sigma}{6}\right)^2}\mathbf{1}+ \lambda (\frac16 \left( -\sigma^2\mathbf{ j}- \mathbf{k} \right)), \\
\fl C_5= \sqrt{1-3\left(\lambda \frac{\sigma}{6}\right)^2}\mathbf{1}+ \lambda (\frac16 \left(-\sigma^2 \mathbf{j}+\mathbf{k} \right) ).&
\end{array}
\]
And the opposite face having the same symetry axis has the following adjacent vertices:
\[
\begin{array}{ll}
\fl C_6= \sqrt{1-3\left(\lambda \frac{\sigma}{6}\right)^2}\mathbf{1}+ \lambda (\frac16 \left(\sigma \mathbf{i}+\sigma \mathbf{j}-\sigma \mathbf{k} \right)), & C_7= \sqrt{1-3\left(\lambda \frac{\sigma}{6}\right)^2}\mathbf{1}+ \lambda (\frac16 \left(\sigma^2\mathbf{ i}+\mathbf{j} \right)), \\
\fl C_8= \sqrt{1-3\left(\lambda \frac{\sigma}{6}\right)^2}\mathbf{1}+ \lambda (\frac16 \left(\sigma \mathbf{i}+\sigma \mathbf{j}+\sigma \mathbf{k} \right)),  &C_9= \sqrt{1-3\left(\lambda \frac{\sigma}{6}\right)^2}\mathbf{1}+ \lambda (\frac16 \left(\sigma^2 \mathbf{j}+\mathbf{ k }\right)), \\
\fl C_{10}= \sqrt{1-3\left(\lambda \frac{\sigma}{6}\right)^2}\mathbf{1}+ \lambda (\frac16 \left(\sigma^2 \mathbf{j}-\mathbf{k} \right)). &
\end{array}
\] 
We are searching $\lambda$ such that $C_1$ has  $C_8$ as image by the Clifford translation $\frac12 (\sigma \mathbf{1}+\frac{1}{\sigma} \mathbf{i} +\mathbf{ j})$. So the spherical distance between $C_1$ and $C_8$ is equal to $\frac{\pi}{5}$. We get $\lambda^2=\frac92 (5- 3\sigma)=\frac92\frac{1}{\sigma^4}$. Due to the symetry of $\mathcal{F}_v$, the sign of $\lambda$ is indifferent. So we have found only one $\lambda >0$ which is suitable for this face. Next we verify that it is also suitable for the others faces. We can then deduce the $20$ vertices of $\mathcal{F}$. 

Now we construct the faces $F_i$ of $\mathcal{F}$. Each of them has five edges which are the shortest geodesic $\mathcal{G}$ joining two adjacent vertices $S_i$ and $S_j$. The geodesics can be described as follows \cite{Sco}. A path $l$ on $\mathcal{S}^3$ is a geodesic if and only if there is a 2-dimensional plane $\Pi$ in $\RR^4$ passing through the origin such that $l\subset \Pi \cap \mathcal{S}^3$. All the geodesics are circles.
\begin{eqnarray}
\fl q\in  \mathcal{G}&\Longleftrightarrow \exists \alpha \geq 0,\; \exists \beta \geq 0 \quad q=\alpha OS_i+\beta OS_j \quad &and\quad \|q\|=1 \label{geod}\\
&\Longleftrightarrow \exists \alpha \geq 0,\; \exists \beta \geq 0 \quad q=\alpha OS_i+\beta OS_j \quad &and \quad 1=\alpha^2+\beta^2+ \sigma \alpha \beta. \nonumber
\end{eqnarray}
Then a face is the set of shortest geodesics joining two points of the edges. One may also say that a face is the set of the projection on $\mathcal{S}^3$ of all barycenters of its five vertices in $\RR^4$. If we note $F_i^b$ the set of all barycenters in $\RR^4$ of the vertices $S_i^1,...,S_i^5$ of $F_i$, we have:
\begin{eqnarray} 
\label{Fi}
F_i=\left\{ \frac{1}{ \sqrt{{x'_0}^2+x'^2+y'^2+z'^2} } (x'_0,x',y',z'),\; (x'_0,x',y',z')\in F_i^b \right\}, 
\end{eqnarray}
and 
\begin{eqnarray*}  
\fl (x'_0,x',y',z')\in F_i^b \Leftrightarrow \exists (\lambda_1,..,\lambda_5)\in [0,1]^5, \sum_{j=1,..,5}\lambda_j OS_i^j=\frac{1}{\sum_{j=1,..,5}\lambda_j}(x'_0,x',y',z')
\end{eqnarray*}
As $F_i^b$ is included in a $2$-plane of equations $x'_0=\frac{\sigma^2}{2\sqrt{2}}$ and $ax'+by'+cz'=\frac{x'_0}{\sigma^2}$ with even permutations of $(a,b,c)=(\pm \frac{1}{\sigma},\pm 1,0)$, we deduce from \eref{Fi} :
\[
\forall (x_0,x,y,z)\in F_i,\quad ax+by+cz=\frac{x_0}{\sigma^2}.
\]
(See Appendix A for a detailed description of the faces). The proof is complete.\\

\begin{Proposition}
The set of $t(S_i)$ for all $t$ in $\mathcal{I}^*$, and all $S_i$ of $\mathcal{F} $ has $600$ vertices given by:
\begin{enumerate}
\item a set of $24$ vertices given by $\frac{1}{2\sqrt{2}} (\pm 2,\pm 2,0,0)$ and all its permutations,
\item a set of $64$ vertices given by $\frac{1}{2\sqrt{2}} (\pm \sqrt{5},\pm 1,\pm 1,\pm 1)$ and all its permutations,
\item a set of $64$ vertices given by $\frac{1}{2\sqrt{2}} (\pm \sigma,\pm \sigma,\pm \sigma,\pm \frac{1}{\sigma ^2})$ and all its permutations,
\item a set of $64$ vertices given by $\frac{1}{2\sqrt{2}} (\pm \sigma^2,\pm \frac{1}{\sigma},\pm \frac{1}{\sigma},\pm \frac{1}{\sigma})$ and all its permutations,
\item a set of $96$ vertices given by $\frac{1}{2\sqrt{2}} (\sigma ^2,\pm \frac{1}{\sigma ^2},0,\pm 1)$ and all its even permutations,
\item a set of $96$ vertices given by $\frac{1}{2\sqrt{2}} (\pm \sqrt{5},\pm \frac{1}{\sigma} ,0,\pm \sigma)$ and all its even permutations,
\item a set of $192$ vertices given by $\frac{1}{2\sqrt{2}} (\pm 2,\pm1, \pm \frac{1}{\sigma},\pm \sigma)$ and all its even permutations.
\end{enumerate}
They are the vertices of $120$ regular dodecahedra which tesselate $\mathcal{S}^3$. 
\end{Proposition}
This result is obtained by an explicit calculus of $t(S_i)$ for all $t$ in $\mathcal{I}^*$ and all vertices $S_i$ of $\mathcal{F} $. Anyone of these regular dodecahedra is a fundamental domain. Following figure \ref{120Cell} shows the three last coordinates of these $600$ vertices, viewed from a face of the centered dodecahedron. Straight lines between two vertices symbolize edges of pentagonal faces.
\begin{figure}[!ht]
\includegraphics[scale=0.4]{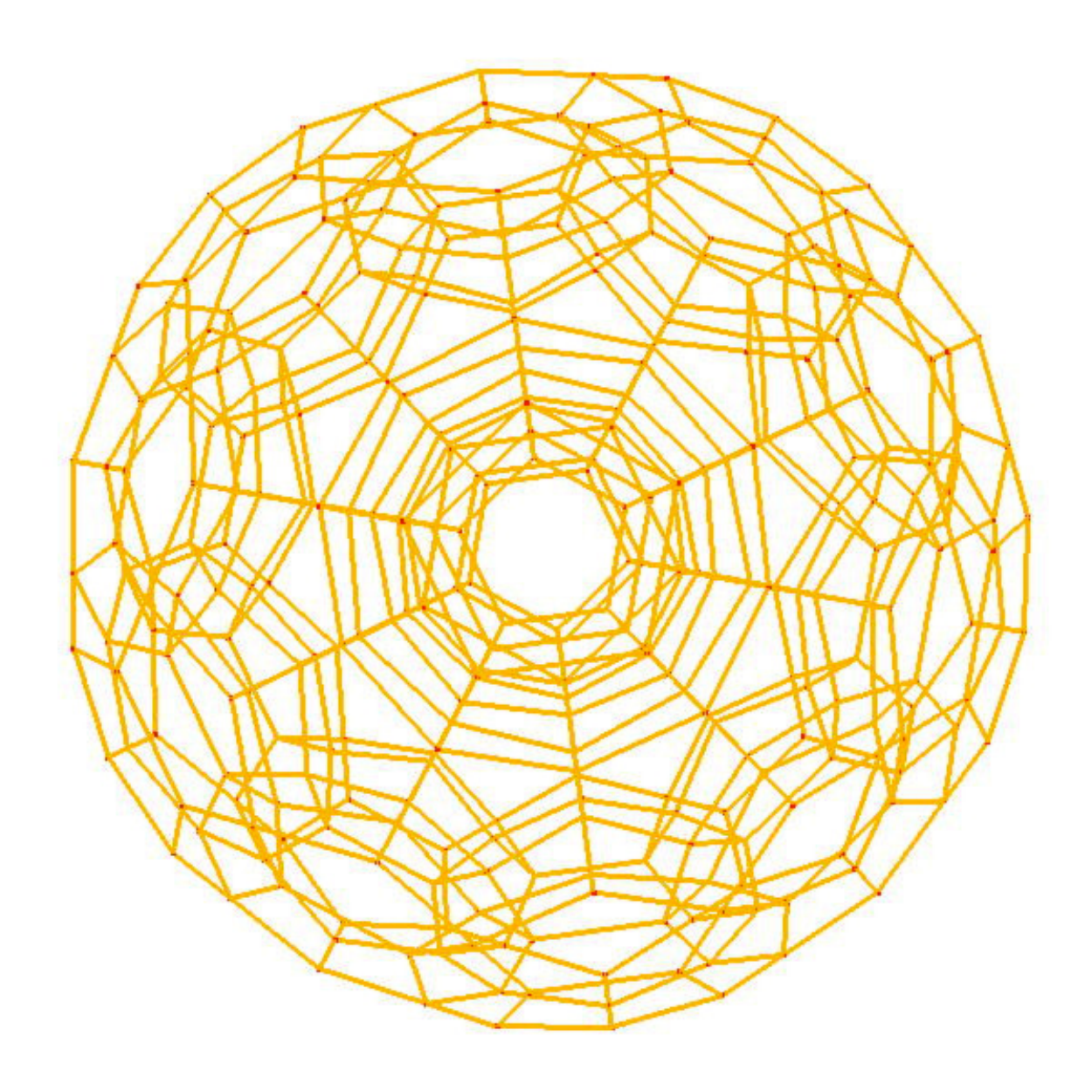}
\caption{$120$ Cell}
\label{120Cell}
\end{figure}
We note that these coordinates look like those of Coxeter \cite{Cox}, up to an odd permutation of the three last coordinates. They are adapted to $\mathcal{I}^*$, unlike those of \cite{Cox}. 
From proposition \ref{vertices-F} we deduce the following:

\begin{Proposition}
$p(\mathcal{F})$, denoted $\mathcal{F}_v$, is a centered regular dodecahedron in $\RR^3$ such that
\[
(x,y,z)\in \mathcal{F}_v \Longleftrightarrow (+\sqrt{1-x^2-y^2-z^2},x,y,z) \in \mathcal{F}.
\]
It is included in the first ball of the visualization of $\mathcal{S}^3$. Its $20$ vertices are
$\frac{1}{2\sqrt{2}} (\pm \frac{1}{\sigma ^2},0,\pm 1)$, 
$\frac{1}{2\sqrt{2}} (0,\pm 1,\pm \frac{1}{\sigma ^2})$, 
$\frac{1}{2\sqrt{2}} (\pm\frac{1}{\sigma},\pm\frac{1}{\sigma},\pm\frac{1}{\sigma})$ and
$\frac{1}{2\sqrt{2}} (\pm 1,\pm \frac{1}{\sigma ^2},0)$. 

Each face $F_{i,v}$ of $\mathcal{F}_v$ is a regular pentagon included in an ellipsoid 
\[
F_{i,v} \subset \left\{ (x,y,z)\in \RR^3,\; \sigma^4\,(ax+by+cz)^2=1-x^2-y^2-z^2 \right\}.
\]
with $(a_i,b_i,c_i)$ equal to $(\pm \frac{1}{\sigma},\pm 1,0)$, up to an even permutation. 

The set of all barycenters in $\RR^3$ of the vertices of $F_{i,v}$, denoted by $F_{i,v}^b$, is included in a $2$-plane of equation $a_ix+b_iy+c_iz=\frac{1}{2\sqrt{2}}$.
\end{Proposition}

See Appendix A for a detailed description of $\mathcal{F}_v$. The following figure is a diagramm, and not a visualization, of $\mathcal{F}$ (or $\mathcal{F}_v$) because their faces are not in a plane of $\RR^4$ (or $\RR^3$).
\begin{figure}[!ht]
\includegraphics[scale=0.5]{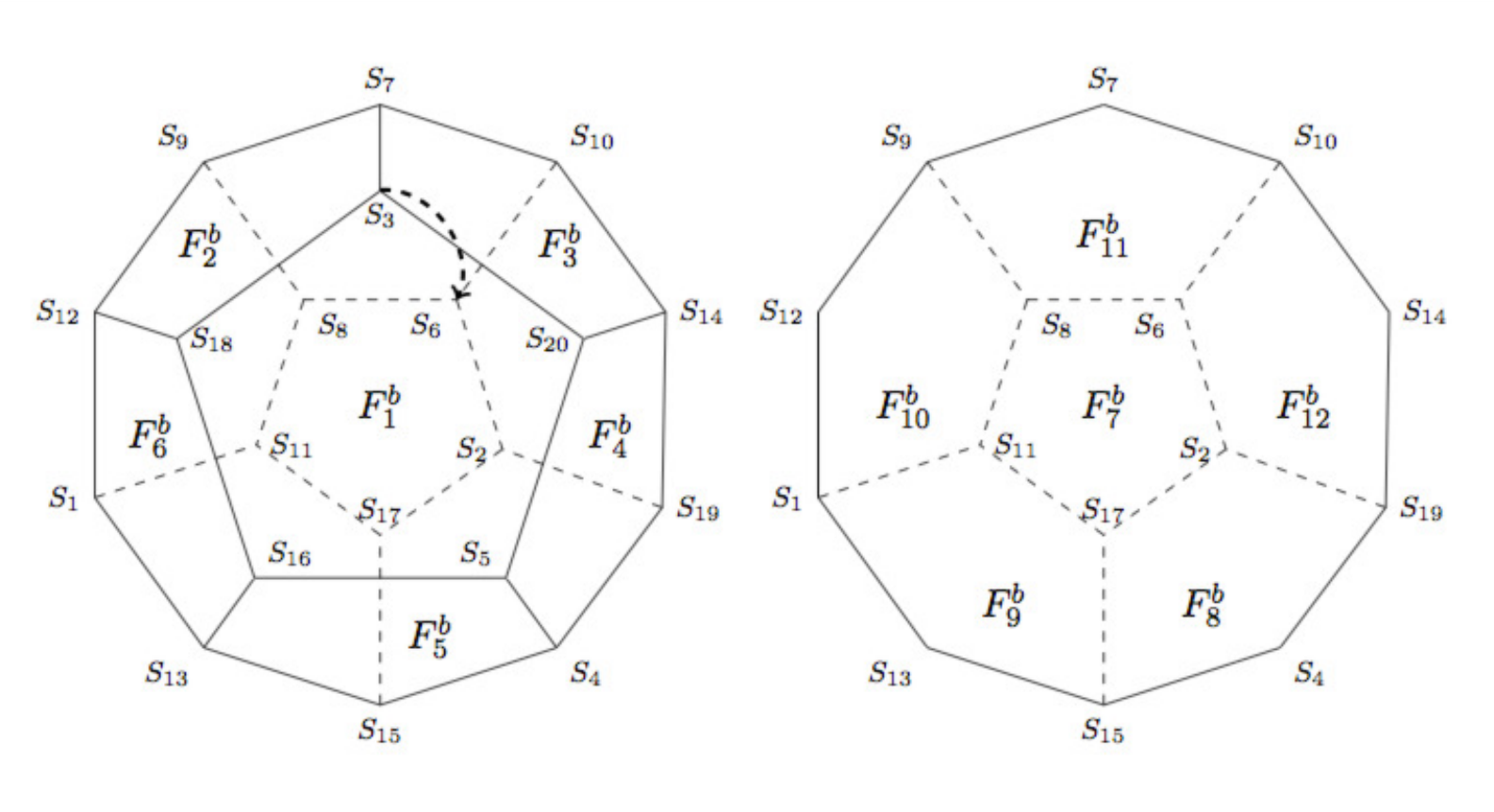}
\caption{Faces $F_{i,b}$ viewed from $F_{1,b}$. The dashed lines are hidden.}
\label{DODSurf}
\end{figure}

\section{Equivalence relations on $\mathcal{F}$ and $\mathcal{F}_v$}
In order to have relation \ref{SIM}, the equivalence relation must identify any pentagonal face of $\mathcal{F}$ with its opposite face, after rotating by $\frac{\pi}{5}$ in the clockwise direction around the outgoing axis orthogonal to this last face. First specify our notations. We consider the Clifford translations $g_i$ that have been used for the construction of $\mathcal{F}$, with: 
\begin{eqnarray}
\fl g_1:=\frac12 \sigma \mathbf{1}+\frac12 \frac{1}{\sigma} \mathbf{i}+\frac12 \mathbf{j},\quad &\; g_2:=\frac12 \sigma \mathbf{1}+\frac12\mathbf{ i} -\frac12 \frac{1}{\sigma} \mathbf{k},\quad &\; g_3:=\frac12 \sigma \mathbf{1}+\frac12 \frac{1}{\sigma}\mathbf{ j}-\frac12 \mathbf{k},\nonumber\\
\fl g_4:=\frac12 \sigma \mathbf{1}-\frac12 \frac{1}{\sigma}\mathbf{ i}+\frac12 \mathbf{j},\quad &\; g_5:=\frac12 \sigma \mathbf{1}+\frac12 \frac{1}{\sigma}\mathbf{ j}+\frac12 \mathbf{k},\quad &\; g_6:=\frac12 \sigma \mathbf{1}+\frac12 \mathbf{i}+\frac12 \frac{1}{\sigma} \mathbf{k}. \nonumber
\end{eqnarray}
They are such that: $\quad \forall i \in \{1,...,6\},\; g_i(F_i)=F_{i+6}$. The inverses of these six first translations are the six other translations:
\begin{eqnarray}
\fl g_7:=(g_1)^{-1}=\frac12 \sigma \mathbf{1}-\frac12 \frac{1}{\sigma} \mathbf{i}-\frac12 \mathbf{j},&\; g_8:=(g_2)^{-1}=\frac12 \sigma \mathbf{1}-\frac12\mathbf{ i} +\frac12 \frac{1}{\sigma} \mathbf{k},\\
\fl g_9:=(g_3)^{-1}=\frac12 \sigma \mathbf{1}-\frac12 \frac{1}{\sigma}\mathbf{ j}+\frac12 \mathbf{k},&\; g_{10}:=(g_4)^{-1}=\frac12 \sigma \mathbf{1}+\frac12 \frac{1}{\sigma}\mathbf{ i}-\frac12 \mathbf{j}, \nonumber\\
\fl g_{11}:=(g_5)^{-1}=\frac12 \sigma \mathbf{1}-\frac12 \frac{1}{\sigma}\mathbf{ j}-\frac12 \mathbf{k},&\; g_{12}:=(g_6)^{-1}=\frac12 \sigma \mathbf{1}-\frac12 \mathbf{i}-\frac12 \frac{1}{\sigma} \mathbf{k}. \nonumber
\end{eqnarray}
They all have a translation distance $\chi$ equal to $\frac{\pi}{5}$. We have for the vertices (see figure \ref{DODSurf} for notations):\\
$g_1(S_{3}) =S_{6},\quad g_1( S_{18})=S_{8},\quad g_1(S_{16})=S_{11},\quad g_1(S_{5} )=S_{17},\quad g_1(S_{20})=S_{2}$,\\ 
$g_2(S_{18}) =S_{15},\quad g_2( S_{12})=S_{17},\quad g_2(S_{9})=S_{2},\quad g_2(S_{7} )=S_{19},\quad g_2(S_{3})=S_{4}$, \\
$g_3(S_{3}) =S_{1},\quad g_3( S_{7})=S_{11},\quad g_3(S_{10})=S_{17},\quad g_3(S_{14} )=S_{15},\quad g_3(S_{20})=S_{13}$,\\ 
$g_4(S_{20}) =S_{9},\quad g_4( S_{14})=S_{8},\quad g_4(S_{19})=S_{11},\quad g_4(S_{4} )=S_{1},\quad g_4(S_{5})=S_{12}$, \\
$g_5(S_{5}) =S_{10},\quad g_5( S_{4})=S_{6},\quad g_5(S_{15})=S_{8},\quad g_5(S_{13} )=S_{9},\quad g_5(S_{16})=S_{7}$, \\
$g_6(S_{16}) =S_{19},\quad g_6( S_{13})=S_{2},\quad g_6(S_{1})=S_{6},\quad g_6(S_{12} )=S_{10},\quad g_6(S_{18})=S_{14}$.\\

We define the relation $\sim$ by specifying the equivalence classes $\dot{q}$ of any $q\in\mathcal{F}$:
\[\quad \forall q\in \mathcal{F},\quad \dot{q}:=\mathcal{I}^* (\{q\})\cap\mathcal{F}.\]
$\mathcal{F}$ has been constructed such that $\frac{\pi}{5}$ is the translation distance beetwen two opposite faces. Otherwise $\frac{\pi}{5}$ is the smallest translation distance of the elements of $\mathcal{I}^*  $. So:
\[
q \in \mathcal{F}\Rightarrow \;\dot{q}=\left\{\frac{}{}g_i(q),\; i\in \{1,...,12\} \right\}\cap {\mathcal{F}}.
\]
It follows that:\begin{itemize}
\item If $q$ belongs to $\stackrel{\circ}{\mathcal{F}}$, then $\dot{q}$ has only one element,
\item If $q$ is a vertex of $\mathcal{F}$, then $\dot{q}$ has four elements,
\item If $q$ belongs to an edge of a face, without beeing a vertex, then $\dot{q}$ has three elements,
\item If $q$ belongs to a face and does not belong to an edge, then $\dot{q}$ has two elements.
\end{itemize}
So we have:
\begin{Proposition}
We define the equivalence classes on $\mathcal{F}$ by:
\[
\fl \begin{array}{l}
q \in \stackrel{\circ}{\mathcal{F}}\quad \Rightarrow \dot{q}=\{q\},\\
\exists (i,j,k), \; q\in F_i\cap F_j \cap F_k \quad  \Rightarrow \dot{q}=\{q,g_i(q),g_{j}(q),g_{k}(q)\},\\
(\exists (i,j),\; q\in F_i\cap F_j )\quad and \quad (\forall k \neq i,\; j,\;\; q\notin F_k) \quad  \Rightarrow \dot{q}=\{q,g_i(q),g_{j}(q)\},\\
(\exists i,\; q\in F_i)\quad and \quad (\forall j\neq i,\; q\notin F_j) \quad \Rightarrow \dot{q}=\{q,g_i(q)\}.
\end{array}
\]
Here the integers $i,j,k$ belong to $\{1,...,12\}$.
\end{Proposition}
The equivalence relation $\sim$ on  $\mathcal{F}_v$ is easily deduced from this one on $\mathcal{F}$. Let us denote $g_{i,v}$ the application $\RR^3\rightarrow \RR^3$ induced by $g_i$ on $\mathcal{F}_v$. Hence for all $i$ in $\{1,...,6\}$, $g_{i,v}(F_{i,v})=F_{i+6,v}$ and $g_{i,v}^{-1}=g_{i+6,v}$. We have:

\begin{Proposition}
We define the equivalence classes on $\mathcal{F}_v$ by:
\[
\fl \begin{array}{l}
X \in \stackrel{\circ}{\mathcal{F}_v}\quad \Rightarrow \dot{X}=\{X\},\\
\exists (i,j,k) \; X\in F_{i,v}\cap F_{j,v} \cap F_{k,v} \quad  \Rightarrow \dot{X}=\{X,g_{i,v}(X),g_{j,v}(X),g_{k,v}(X)\},\\
(\exists (i,j),\; X\in F_{i,v}\cap F_{j,v}) \quad and \quad (\forall k \neq i,\; j,\;\; X\notin F_{k,v}) \quad \Rightarrow \dot{X}=\{X,g_{i,v}(X),g_{j,v}(X)\},\\
(\exists i,\; X\in F_{i,v})\quad and \quad (\forall j\neq i,\; X\notin F_{j,v})\quad \Rightarrow \dot{X}=\{X,g_{i,v}(X)\}.
\end{array}
\]
Here the integers $i,j,k$ belong to $\{1,...,12\}$.
\end{Proposition}
The geometrical meaning of  this equivalence relation is clear : we identify any pentagonal face of $\mathcal{F}_v $ with its opposite face, after rotating by $\frac{\pi}{5}$ in the clockwise direction around the outgoing axis orthogonal to this last face (see Figure \ref{DODSurf}).

\section{The Wave Propagation on the Dodecahedral Space.}
We consider the lorentzian manifold
$\RR_t\times \mathcal{S}^3 / \mathcal{I}^*$ endowed with the metric 
\begin{equation*}
g_{\mu\nu}dx^{\mu}dx^{\nu}=dt^2-ds_{\mathbf{K}}^2,
  \label{}
\end{equation*}
and we study the scalar covariant wave equation associated to this metric :
\begin{equation}
\partial_t^2\Psi-\Delta_{\mathbf{K}}\Psi=0.\;\;
  \label{eq}
\end{equation}
Here $\Delta_{\mathbf{K}}$ is the Laplace Beltrami operator on $\mathbf{K}$, that is defined by
\[
\Delta_{\mathbf{K}}:=\frac{1}{\sqrt{|g|}} \partial_\mu g^{\mu \nu}
\sqrt{|g|}\partial_\nu,\;\;g^{-1}=(g^{\mu \nu})\;,\qquad |g|=|\det
g_{\mu \nu}|.
\]
Since $\mathbf{K}$ is a
smooth compact manifold without boundary, $-\Delta_{\mathbf{K}}$ endowed with
its natural domain $\{u\in L^2(\mathbf{K});\;\;\Delta_{\mathbf{K}}u\in
    L^2(\mathbf{K})\}$ is a densely defined, positive, self-adjoint operator on $L^2(\mathbf{K})$ and the global Cauchy problem is
well posed by the spectral functional calculus :
\[
\Psi(t)=\cos\left(t\sqrt{-\Delta_{\mathbf K}}\right)\Psi(0)+\frac{\sin\left(t\sqrt{-\Delta_{\mathbf K}}\right)}{\sqrt{-\Delta_{\mathbf K}}}\partial_t\Psi(0).
\]
We shall use the functional framework of the finite energy spaces. Given
$m\in\NN$, we introduce the Sobolev space
\begin{equation*}
  H^m(\mathbf{K}):=\left\{u\in
    L^2(\mathbf{K}),\;\nabla_{\mathbf{K}}^{\alpha}u\in
    L^2(\mathbf{K}),\;\mid\alpha\mid\leq m\right\}.
\end{equation*}
where $\nabla_{\mathbf{K}}$  are the covariant derivatives. We can also interpret
this space as the set of the distributions $u\in
H^m(\mathcal{S}^3)$ such that $u\circ g=u$ for any
$g\in\mathcal{I}^*$. Then the standard spectral theory assures that for all
$\Psi_0\in H^1(\mathbf{K})$, $\Psi_1\in L^2(\mathbf{K})$, there exists
a unique $\Psi\in C^0\left(\RR^+_t;H^1(\mathbf{K})\right)\cap
C^1\left(\RR^+_t;L^2(\mathbf{K})\right)$
solution of (\ref{eq}) satisfying
\begin{equation}
\Psi(t=0)=\Psi_0,\;\;\partial_t\Psi(t=0)=\Psi_1,
  \label{ci}
\end{equation}
and we have
\begin{equation*}
\int_{\mathbf{K}}\mid\partial_t\Psi(t)\mid^2+\mid\nabla_{\mathbf{K}}\Psi(t)\mid^2d\mu_K=Cst.
  \label{}
\end{equation*}
Also we have a result of regularity : when $\Psi_0\in H^2(\mathbf{K})$, $\Psi_1\in H^1(\mathbf{K})$, then $\Psi\in C^0\left(\RR_t^+;H^2(\mathbf{K})\right)\cap
C^1\left(\RR_t^+;H^1(\mathbf{K})\right)\cap
C^2\left(\RR_t^+;L^2(\mathbf{K})\right)$.\\

Obviously $\Psi$ is entirely determined on $\RR_t\times\mathbf{K}$ by its restriction to $\RR_t\times\mathcal{F}$. To perform the numerical computation of this solution,  we take the domain of visualization $\mathcal{F}_v\subset \RR^3$  of the fundamental polygon $\mathcal{F}\subset\mathcal{S}^3$ as the domain of calculus. Therefore we introduce the map $f$ that is one-to-one from  $\mathcal{F}_v\subset \RR^3$ onto  $\mathcal{F}\subset \mathcal{S}^3$ defined by
\begin{eqnarray}
\fl f(x,y,z)=\left( \sqrt{1-x^2-y^2-z^2}, x,y,z \right)=\left(f_1(x,y,z),f_2(x,y,z),f_3(x,y,z),f_4(x,y,z)\right),\nonumber
\end{eqnarray}
and we put
\begin{equation}
\psi(t,x,y,z):=\Psi(t,f(x,y,z)).
 \label{psipsi}
\end{equation}
We introduce the usual Sobolev space $H^m$ for the euclidean metric of $\RR^3$:
$$
H^m(\mathcal{F}_v)=\{u\in L^2(\mathcal{F}_v),\;\;\forall
\alpha\in\NN^2,\;\;\mid\alpha\mid\leq
m,\;\;\partial_{x,y,z}^{\alpha}u\in L^2(\mathcal{F}_v)\}.
$$
\begin{Proposition}
$\Psi\in C^0\left(\RR^+_t;H^1(\mathbf{K})\right)\cap
C^1\left(\RR^+_t;L^2(\mathbf{K})\right)$ is solution of (\ref{eq}) iff $\psi$ belongs to $C^0\left(\RR^+_t;H^1(\mathcal{F}_v)\right)\cap
C^1\left(\RR^+_t;L^2(\mathcal{F}_v)\right)$ and satisfies the equation
\begin{equation}
\partial_{tt}\psi-\Delta_{\mathcal{F}_v}\psi=0,\;\;(t,x,y,z)\in\RR^+\times\mathcal{F}_v,
  \label{eqxy}
\end{equation}
where
\begin{eqnarray}
\Delta_{\mathcal{F}_v}&=&(1-x^2-y^2-z^2) \,\partial_{11}^2+(1-y^2)\, \partial_{22}^2+(1-z^2)\, \partial_{33}^2\nonumber\\
& \,&-2 xy\, \partial_{12}^2 -2 xz \,\partial_{13}^2 -2 yz \,\partial_{23}^2-3 x\,\partial_1-3y\,\partial_2-3z\,\partial_3,\nonumber
\end{eqnarray}
and the boundary conditions
\begin{equation}
\forall (t,X,X')\in\RR\times\partial\mathcal{F}_v\times\partial\mathcal{F}_v,\;\;X\sim X'\Rightarrow \psi(t,X)=\psi(t,X').
  \label{cl}
\end{equation}
Moreover $\Psi\in C^0\left(\RR_t^+;H^2(\mathbf{K})\right)\cap
C^1\left(\RR_t^+;H^1(\mathbf{K})\right)\cap
C^2\left(\RR_t^+;L^2(\mathbf{K})\right)$ iff $\psi\in C^0\left(\RR_t^+;H^2(\mathcal{F}_v)\right)\cap
C^1\left(\RR_t^+;H^1(\mathcal{F}_v)\right)\cap
C^2\left(\RR_t^+;L^2(\mathcal{F}_v)\right)$.
\end{Proposition}

{\it Proof.}
We denote $g$ the metric induced on $\mathcal{F}_v$ by the metric of $\mathcal{S}^3$ and the map $f$.
We get that the coefficients of $g_{ij}$ are:
$$g_{ii} := -\sum_{j=1}^4\, f_j \left( x,y,z \right) {\partial_{ii} ^{2}} f_j \left( x,y,z \right),$$
$$g_{ik} := -\sum_{j=1}^4\, f_j \left( x,y,z \right) {\partial_{ik} ^{2}} f_j \left( x,y,z \right).$$
So
 \[
\fl g_{ij}=\left(
\begin{array}{ccc}
\frac {1-{y}^{2}-{z}^{2}}{1-{x}^{2}-{y}^{2}-{z}^{2}}& \frac {xy}{1-{x}^{2}-{y}^{2}-{z}^{2}} & \frac {xz}{1-{x}^{2}-{y}^{2}-{z}^{2}}\\
\frac {xy}{1-{x}^{2}-{y}^{2}-{z}^{2}} & \frac {1-{x}^{2}-{z}^{2}}{1-{x}^{2}-{y}^{2}-{z}^{2}} & \frac {yz}{1-{x}^{2}-{y}^{2}-{z}^{2}} \\
\frac {xz}{1-{x}^{2}-{y}^{2}-{z}^{2}}&\frac {yz}{1-{x}^{2}-{y}^{2}-{z}^{2}} & \frac {1-{x}^{2}-{y}^{2}}{1-{x}^{2}-{y}^{2}-{z}^{2}}
\end{array}
\right),\;
g^{ij }=\left(
\begin{array}{ccc}
1-x^2&-xy&-xz\\
-xy & 1-y^2 & -yz\\
-xz & -yz & 1-z^2
\end{array}
\right),
\]
and 
$$
\det g=\frac{1}{1-{x}^{2}-{y}^{2}-{z}^{2}}.
$$
As $\Delta_{\mathcal{F}_v}:=\frac{1}{\sqrt{|g|}} \partial_\mu g^{\mu \nu}\sqrt{|g|}\partial_\nu $ we obtain the expression of $\Delta_{\mathcal{F}_v}$.

Given $u\in H^1(\mathcal{F}_v)$, the trace of $u$ on $\mathcal{F}_v$ is
well defined since the domain  $\mathcal{F}_v$ is Lipschitz and $C^{\infty}$ piecewise. $\psi$ satisfies the boundary conditions (\ref{cl}) iff its pull-back $\Psi$ on $\mathcal{F}$ can be extended in a solution defined on the whole Poincar\'e dodecahedron $\mathbf K$. The extension to the smooth solutions is straightforward. The proof is completed.\\

To handle the boundary when applying the finite element method, it is
very convenient to take into account the boundary condition (\ref{cl})
by a suitable choice of the functional space.
We introduce
the spaces $W^m(\mathcal{F}_v)$ that correspond to the spaces $H^m(\mathbf{K})$ :
\begin{eqnarray*}
\fl W^0(\mathcal{F}_v):=L^2(\mathcal{F}_v,(1-{x}^{2}-{y}^{2}-{z}^{2})^{-\frac{1}{2}}dxdydz), \\
\fl 1\leq m,\; W^m(\mathcal{F}_v):=\left\{u\in H^m(\mathcal{F}_v),\;\; \forall
  (X,X')\in \partial\mathcal{F}_v^2,\; X\sim X'\Rightarrow u(X)=u(X')\right\},
  \label{}
\end{eqnarray*}
endowed with the norm
\begin{equation*}
\|u\|_{W^m}^2:=\sum_{\mid\alpha\mid\leq m}\|\partial^{\alpha}u\|^2_{W^0(\mathcal{F}_v)}.
  \label{}
\end{equation*}
In particular, we have
\begin{equation*}
W^1(\mathcal{F}_v)=\left\{u\in H^1(\mathcal{F}_v),\;\;X\sim X'\Rightarrow
  u(X)=u(X')\right\},
  \label{}
\end{equation*}
and 
$$
\psi\in C^k\left(\RR_t, W^m(\mathcal{F}_v)\right)\Longleftrightarrow \Psi\in C^k\left(\RR_t; H^m(\mathbf{K})\right).
$$

The numerical method to solve the Cauchy problem will be based on its variational formulation.
\begin{Theorem}
Given $\psi_0\in W^2(\mathcal{F}_v)$, $\psi_1\in W^1(\mathcal{F}_v)$, there exists a unique $\psi\in C^0\left(\RR_t^+;W^2(\mathcal{F}_v)\right)\cap
C^1\left(\RR_t^+;W^1(\mathcal{F}_v)\right)\cap
C^2\left(\RR_t^+;W^0(\mathcal{F}_v)\right)$
solution of the equation (\ref{eqxy}), and satisfying
\begin{equation}
\psi(0,.)=\psi_0(.),\;\;\partial_t\psi(0,.)=\psi_1(.).
\label{condinit}
\end{equation}
$\psi$ is the unique function in $C^0\left(\RR_t^+;W^2(\mathcal{F}_v)\right)\cap
C^1\left(\RR_t^+;W^1(\mathcal{F}_v)\right)\cap
C^2\left(\RR_t^+;W^0(\mathcal{F}_v)\right)$ satisfying (\ref{condinit}) and such that for any $\phi\in W^1(\mathcal{F}_v)$, we have :
\begin{eqnarray}
\fl 0=\frac{d^2}{dt^2}\int_{\mathcal{F}_v}(1-x^2-y^2-z^2)^{-\frac12}\psi(t,x,y,z)\phi(x,y,z)\,dx\,dy\,dz  \label{pbvaria}\\
+\int_{\F_v} (1-x^2-y^2-z^2)^{-\frac12}\nabla \psi(t,x,y,z) \cdot \nabla \phi(x,y,z)\,dx\,dy\,dz\nonumber\\
-\int_{\mathcal{F}_v}(1-x^2-y^2-z^2)^{-\frac12}\left[(x,y,z)\cdot \nabla \psi(t,x,y,z)\right]\nonumber\\
\qquad \left[(x,y,z) \cdot \nabla \phi(x,y,z)\right]\,dx\,dy\,dz.\nonumber
\end{eqnarray}
 \label{}
\end{Theorem}

{\it Proof:} The existence and the uniqueness of the solution of the Cauchy problem are given by the previous proposition since the boundary conditions are imposed by our choice of space $W^1(\mathcal{F}_v)$. Now the mixed
problem can be expressed as a variational problem. $\psi$ is solution
iff for all $\phi\in W^1(\mathcal{F}_v)$, we have :
\begin{eqnarray}
\fl 0=\left<\partial_t^2\psi-\Delta_{\mathcal{F}_v}\psi;\phi\right>_{W^0(\mathcal{F}_v)}\nonumber
\end{eqnarray} 
\begin{eqnarray}
\fl=\frac{d^2}{dt^2} \int_{\F_v}\frac{1}{\sqrt{1-x^2-y^2-z^2}}\psi(t,x,y,z)\phi(x,y,z)\,dx\,dy\,dz \nonumber\\
-\int_{\F_v}\frac{1}{\sqrt{1-x^2-y^2-z^2}}\left(\Delta_{\mathcal{F}_v}\psi\right)(t,x,y,z)\,\phi(x,y,z)\,dx\,dy\,dz.
  \label{pv}
\end{eqnarray} 
To get the formulation (\ref{pbvaria}), the key point consists in expressing in a symetric manner the last integral, without using an integral on the boundary. To invoke the Green formula, we denote $\nu(x,y,z)$ the unit outgoing normal at $(x,y,z)$ belonging to face $ F_{i,v}$. We know explicit form of $g_{i,v}$, the application $\RR^3\rightarrow \RR^3$ induced by $g_i$ on $\mathcal{F}_v$. Thanks to relations written in Appendix A, we have:
\[
\fl g_{1,v}(x,y,z)=\frac12\left(\begin{array}{rrr}
\frac{1}{\sigma}&-\sigma &1\\
-\sigma&-1&-\frac{1}{\sigma}\\
-1&\frac{1}{\sigma}&\sigma
\end{array} \right) 
\left(\begin{array}{l}
x\\
y\\
z
\end{array} \right),
g_{2,v}(x,y,z)=\frac12\left(\begin{array}{rrr}
-1&\frac{1}{\sigma}&\sigma \\
-\frac{1}{\sigma}&\sigma&-1\\
-\sigma&1&\frac{1}{\sigma}
\end{array} \right) 
\left(\begin{array}{l}
x\\
y\\
z
\end{array} \right),
\]
\[
\fl g_{3,v}(x,y,z)=\frac12\left(\begin{array}{rrr}
\sigma&1&\frac{1}{\sigma}\\
-1&\frac{1}{\sigma}&\sigma\\
-\frac{1}{\sigma}&\sigma&-1
\end{array} \right) 
\left(\begin{array}{l}
x\\
y\\
z
\end{array} \right),
g_{4,v}(x,y,z)=\frac12\left(\begin{array}{rrr}
\frac{1}{\sigma}&\sigma&1 \\
\sigma&-1&\frac{1}{\sigma}\\
-1&-\frac{1}{\sigma}&\sigma
\end{array} \right) 
\left(\begin{array}{l}
x\\
y\\
z
\end{array} \right),
\]
\[
\fl g_{5,v}(x,y,z)=\frac12\left(\begin{array}{rrr}
\sigma&-1&\frac{1}{\sigma}\\
1&\frac{1}{\sigma}&-\sigma\\
-\frac{1}{\sigma}&-\sigma&-1
\end{array} \right) 
\left(\begin{array}{l}
x\\
y\\
z
\end{array} \right),
g_{6,v}(x,y,z)=\frac12\left(\begin{array}{rrr}
-1&-\frac{1}{\sigma}&-\sigma \\
\frac{1}{\sigma}&\sigma&-1\\
-\sigma&1&\frac{1}{\sigma}
\end{array} \right) 
\left(\begin{array}{l}
x\\
y\\
z
\end{array} \right).
\]
Moreover we know $\nu(x,y,z)$ thanks to the equation of the ellipsoids (Appendix A).
We verify that:
$$
g_{i,v}(\nu(x,y,z))=-\nu(g_{i,v}(x,y,z)),
$$
Since $(u\circ g_{i,v})_{\vert F_{i,v}}=u _{\vert F_{i,v}}$, we have for
$u\in W^2(\mathcal{F}_v)$
$$
\partial_{\nu(X)}u(x,y,z)=g_{i,v}[\nu(x,y,z)].\nabla u(g_{i,v}(x,y,z))=-\partial_{\nu(g_{i,v}(x,y,z))}u(g_{i,v}(x,y,z)).
$$
We deduce that for $u\in W^2(\mathcal{F}_v)$, $v\in W^1(\mathcal{F}_v)$,
we have
\begin{eqnarray}
\fl \int_{ {F}_{i,v}}v(x,y,z)\partial_{\nu(x,y,z)}u \,d\sigma(x,y,z)&=-\int_{g_{i,v}({F}_{i,v})}v(x,y,z)\partial_{\nu(x,y,z)}u\, d\sigma(x,y,z)\nonumber\\
&=-\int_{{F}_{i+6,v}}v(x,y,z)\partial_{\nu(x,y,z)}u\, d\sigma(x,y,z),
\end{eqnarray}
and therefore
$$
\int_{\partial\mathcal{F}_v}v(x,y,z)\partial_{\nu(x,y,z)}u\, d\sigma(x,y,z)=0,
$$
where $d\sigma$ denotes the volume measure on $(\partial\mathcal{F}_v, g)$.
We conclude that :
\[
\fl \int_{\F_v}\left(\Delta_{\mathcal{F}_v}\psi\right)(t,x,y,z)\phi(x,y,z)\,\sqrt{\det g} \, dx\,dy\,dz=-\int_{\F_v} g({\mathrm{grad}}\, \psi,{\mathrm{grad}} \,\phi)\, \sqrt{\det g} \; dx\,dy\,dz.
\]
To simplify the writing we shall note in the following $X$ instead of $(x,y,z)$ and $|X|^2$ instead of $x^2+y^2+z^2$ for $(x,y,z)\in \mathcal{F}_v$; and $dX$ will designate $dx\,dy\,dz$. Therefore we compute :
\begin{eqnarray}
\fl \int_{\F_v}(1-|X|^2)^{-\frac12}\Delta_{\mathcal{F}_v}\psi(t,X)\phi(X)\,dX=\nonumber\\
-\int_{\F_v}      (1-|X|^2)^{-\frac12}      \left[      (1-x^2) \,\partial_{1}\psi(t,X) \partial_{1}\phi(t,X)        \right. \nonumber\\
\qquad \left.      +(1-y^2)\, \partial_{2}\psi(t,X) \partial_{2}\phi(t,X)        +(1-z^2)\, \partial_{3}\psi(t,X) \partial_{3}\phi(t,X)        \right]\,dX \nonumber\\
+\int_{\F_v}      (1-|X|^2)^{-\frac12}      \left[     xy  \left(\partial_{1}\psi(t,X) \partial_{2}\phi(t,X)+\partial_{2}\psi(t,X) \partial_{1}\phi(t,X)\right)      \right. \nonumber\\
\qquad    \left.     +xz   \left(\partial_{1}\psi(t,X) \partial_{3}\phi(t,X) +\partial_{3}\psi(t,X) \partial_{1}\phi(t,X)\right)        \right. \nonumber\\
\qquad  \left.       +yz   \left(\partial_{2}\psi(t,X) \partial_{3}\phi(t,X)+\partial_{3}\psi(t,X) \partial_{2}\phi(t,X)\right)       \right]\,dX. \nonumber
\end{eqnarray}
And finally we calculate :
\begin{eqnarray}
\fl \int_{\F_v}(1-|X|^2)^{-\frac12}\Delta_{\mathcal{F}_v}\psi(t,X)\phi(X)\,dX=\nonumber\\
-\int_{\F_v} (1-|X|^2)^{-\frac12}\nabla \psi(t,X) \cdot \nabla \phi(t,X)\, dX \nonumber\\
+\int_{\F_v} (1-|X|^2)^{-\frac12}\left(X\cdot \nabla \psi(t,X)\right)\left(X \cdot \nabla \phi(t,X)\right)\, dX. \nonumber
\end{eqnarray}
The proof of the theorem is complete.\\

We solve this variational problem by the usual way. We take a family $V_h$, $0<h\leq h_0$, of finite dimensional vector subspaces of
$W^1(\mathcal{F}_v)$. We assume that
\begin{equation*}
\overline{\cup_{0<h\leq h_0}V_h}=W^1(\mathcal{F}_v).
  \label{}
\end{equation*}
We choose sequences $\psi_{0,h},\;\psi_{1,h}\in V_h$ such that
$$
\psi_{0,h}\rightarrow\psi_0\;\;in\;\;W^1(\mathcal{F}_v),\;\psi_{1,h}\rightarrow\psi_1\;\;in\;\;L^2(\mathcal{F}_v).
$$
We consider the solution $\psi_h\in C^{\infty}(\RR_t;V_h)$ of
\begin{eqnarray}
\fl \forall\phi_h\in V_h,\;\;
\frac{d^2}{dt^2} \int_{\F_v}(1-|X|^2)^{-\frac12}\psi_h(t,X)\phi_h(X)dX\nonumber\\
+\int_{\F_v} (1-|X|^2)^{-\frac12}\nabla \psi_h(t,X) \cdot \nabla \phi_h(t,X)\, dX\nonumber\\
-\int_{\mathcal{F}_v}(1-|X|^2)^{-\frac12}\left(X\cdot \nabla \psi_h(t,X)\right)\left(X \cdot \nabla \phi_h(t,X)\right)\, dX=0,\nonumber
\end{eqnarray}
satisfying $\psi_h(0,.)=\psi_{0,h}(.)$, $\partial_t\psi_h(0,.)=\psi_{1,h}(.)$.
Thanks to the conservation of the energy,
\begin{equation*}
 \int_{\mathcal{F}_v}\left(\mid\partial_{t}\psi(t,x,y,z)\mid^2+\mid\nabla_{\mathcal{F}_v}\psi(t,x,y,z)\mid^2 \right) \,\sqrt{\det g} \; dx\,dy\,dz\,=\,Cst,
  \label{}
\end{equation*}
 this scheme is stable :
$$
\forall T>0,\;\;\sup_{0<h\leq h_0}\sup_{0\leq t\leq T}\left\|\psi_h(t)\right\|_{W^1}+\left\|\frac{d}{dt}\psi_h(t)\right\|_{L^2}<\infty.
$$
Moreover, when $\psi\in C^2\left(\RR^+_t;W^1(\mathcal{F}_v)\right)$, it
is also converging :
$$
\forall T>0,\;\;\sup_{0\leq t\leq
  T}\left\|\psi_h(t)-\psi(t)\right\|_{W^1}+\left\|\frac{d}{dt}\psi_h(t)-\frac{d}{dt}\psi(t)\right\|_{L^2}\rightarrow 0,\;\;h\rightarrow 0.
$$
If we take a basis $\left(e_j^h\right)_{1\leq j\leq N_h}$ of $V_h$, we
expand $\psi_h$ on this basis :
$$
\psi_h(t)=\sum_{j=1}^{N_h}\psi_j^h(t)e_j^h,
$$
and we introduce 
\[
\fl U(t):= \;^t\!\left(\psi_1^h,\psi_2^h,\cdots,\psi_{N_h}^h\right)
\]
\[
\fl \MM=\left(M_{ij}\right)_{1\leq i,j\leq N_h},\;\;\DD=\left(D_{ij}\right)_{1\leq i,j\leq N_h},\;\;\KK=\left(K_{ij}\right)_{1\leq i,j\leq N_h}
\]
\[
\fl M_{ij}:=\int_{\F_v}\frac{1}{\sqrt{1-|X|^2}}e_i^h(X)e_j^h(X) \, dX,
\]
\[
\fl K_{ij}:=\int_{\F_v}\frac{1}{\sqrt{1-|X|^2}}\left(\partial_xe_i^h(X)\partial_xe_j^h(X)+ \partial_ye_i^h(X)\partial_ye_j^h(X) +\partial_ze_i^h(X)\partial_ye_j^h(X)\right) \, dX.
\]
\begin{eqnarray}
\fl D_{ij}:= -\int_{\F_v}\frac{1}{\sqrt{1-|X|^2}}&\left(x\partial_xe_i^h(X)+y\partial_ye_i^h(X)+z\partial_ze_i^h(X)\right) \nonumber\\
&\qquad \left(x\partial_xe_j^h(X)+y\partial_ye_j^h(X)+z\partial_ze_j^h(X)\right)\, dX. \nonumber
\end{eqnarray}
Then the variational formulation is equivalent to 
\begin{equation}
\MM X''+\left(\KK+\DD\right) X=0.
  \label{SumLaplacien}
\end{equation}
This differential system is solved very simply by iteration by solving
\begin{equation*}
\MM(U^{n+1}-2U^n+U^{n-1})+(\Delta T)^2\left(\KK+\DD\right) U^n=0.
  \label{}
\end{equation*}
We know that this scheme is stable, and so convergent by a consequence of the Lax theorem \cite{Lax}, when
\begin{equation*}
\sup_{U\neq 0}\frac{<\left(\KK+\DD\right) U,U>}{<\MM U,U>}< \frac{4}{\Delta T^2}.
  \label{}
\end{equation*}
Therefore if there exists $K>0$ such that
\begin{equation*}
\fl \forall h\in]0,h_0],\;\;\forall \phi_h\in V_h,\;\;
\left\|\frac{\nabla_{x,y,z}\phi_h}{(1-| X|^2)^{\frac14}}\right\|_{L^2(\mathcal{F}_v)}\leq
\frac{K}{h}\left\|\frac{\phi_h}{(1-| X|^2)^{\frac14}}\right\|_{L^2(\mathcal{F}_v)},
  \label{}
\end{equation*}
the CFL condition
\begin{equation}
K\Delta T<\sqrt{2} h,
  \label{CFL}
\end{equation}
is sufficient to assure the stability and the convergence of our scheme.
\section{Numerical calculations}
\subsection{Mesh}
Our goal is to built a mesh of $\partial \mathcal{F}$ such that it has a fixed size in advance, all the edges are splitting in the same way, and the meshes of two opposite faces are the image from each other by the Clifford translation $g_i$ fitting for $\sim$. Of course we also want that all points of the mesh of $F_{i,v} \subset \partial \mathcal{F}_v$ are on the suitable ellipsoid, and that all points of the edges of $F_{i,v}$ are on the right geodesic.\\
First of all we construct the boundary $\partial \mathcal{F}_v $. Each $F_i^b$ is included in a 2-plane of $\RR^4$ and can be easily meshed with a convenient metric. We choose to built a mesh of $F_1^b$, with $F_1:= (S_3,S_{18},S_{16},S_5,S_{20})$ (see \Fref{DODSurf}). We denote $F_{1,v}^b:=p(F_1^b)$ the visualization of $F_1^b$. As we want the mesh vertices on each edge of $F_1$ are equidistant, we choose {\it a priori} the desired spherical distance between two consecutive mesh vertices. This determines the number of mesh vertices on an edge. Then, by a suitable application, we map $F_1^b$ in the 2-plane $z=0$ of $\RR^3$, which allows us to use a 2-D mesh generator able to respect a given metric matrix $M$. Let us recall the inclusions
\begin{eqnarray}
 F_1^b \subset \left\{(x_0,x,y,z) \in \RR^4 ,\quad  x_0=\frac{\sigma^2}{2\sqrt{2}}\;, \; -\frac{1}{\sigma} x-y=\frac{x_0}{\sigma^2}\right\},\nonumber\\
F_{1,v}^b \subset \left\{(x,y,z) \in \RR^3 ,\quad  -\frac{1}{\sigma} x-y=\frac{1}{2\sqrt{2}}\right\}.\nonumber
\end{eqnarray}
We do a translation $t$ of $F_{1,v}^b$ by the vector $\overrightarrow{O M_{5,20}} =(0,-\frac{1}{\sqrt{2}},0)$ where $M_{5,20}$ denotes the middle of $S_5 S_{20}$, followed by a rotation $r$ in $\RR^3$, with an angle $-\frac{\pi}{2}$ and axis $\vec{u}=\overrightarrow{S_{18}M_{5,20}}=(\frac{1}{\sqrt{3-\sigma}},-\frac{1}{\sigma \sqrt{3-\sigma}},0)$. So, thanks to the quaternionic calculus:
\begin{eqnarray}
\fl r \left(x \mathbf{i}+y \mathbf{j}+z \mathbf{k}\right)=\left[\frac{1}{\sqrt{2}}\mathbf{1}-\frac{1}{\sqrt{2}}\left( \frac{1}{\sqrt{3-\sigma}}\mathbf{i}-\frac{\sqrt{2}}{2}\frac{1}{\sigma\sqrt{3-\sigma}}\mathbf{j}\right)\right]\left[x \mathbf{i}+y \mathbf{j}+z \mathbf{k}\right]\nonumber\\
\qquad \qquad \qquad \qquad \left[\frac{1}{\sqrt{2}}\mathbf{1}+\frac{1}{\sqrt{2}}\left( \frac{1}{\sqrt{3-\sigma}}\mathbf{i}-\frac{\sqrt{2}}{2}\frac{1}{\sigma\sqrt{3-\sigma}}\mathbf{j}\right)\right],\nonumber
\end{eqnarray}
and,
\[
r(x,y,z)=\left(
\begin{array}{l}
\frac{1}{3-\sigma} x-\frac{1}{\sigma (3-\sigma)}y+\frac{1}{\sigma \sqrt{3-\sigma}} z\\
-\frac{1}{\sigma (3-\sigma)} x+\frac{1}{\sigma^2 (3-\sigma)} y+\frac{1}{\sqrt{3-\sigma}} z\\
-\frac{1}{\sigma\sqrt{3-\sigma}}x - \frac{1}{\sqrt{3-\sigma}}y
\end{array}
\right).
\]
If $(x,y,z)$ belongs to $F_{1,v}^b$ we get 
\[\fl (r\circ t )\,(x,y,z)=\left(x',y',-\frac{1}{\sigma\sqrt{3-\sigma}}x - \frac{1}{\sqrt{3-\sigma}}\left(y +\frac{1}{2\sqrt{2}}\right)\right)=(x',y',0).
\] 
Therefore $r\circ t \,(F_{1,v}^b)$ is in the 2-plane $z=0$ of $\RR^3$. Of course we want a metric on $\RR^2$ endowed by the metric of $\mathcal{S}^3$.
We denote by $f$ the application that maps $r\circ t \,(F_{1,v}^b)$ to $F_1$, that is $f:$
\[ 
\fl \begin{array}{lclclcl}
r\circ t \,(F_{1,v}^b)&\rightarrow &\RR^3&\rightarrow  &F_{1,v}^b&\rightarrow &F_1^b\subset \RR^4\\
 (x,y) & \mapsto &(x,y,0)&\mapsto&(x',y',z') :=(r\circ t)^{-1}(x,y,0)&\mapsto &\left(\frac{\sigma^2}{2\sqrt{2}},x',y',z'\right)
\end{array}
\]
followed by: 
\[ 
\begin{array}{clcl}
&F_1^b\subset \RR^4&\rightarrow &F_1 \subset \mathcal{S}^3\\
&\left(\frac{\sigma^2}{2\sqrt{2}},x',y',z'\right)& \mapsto & \frac{1}{\|\left(\frac{\sigma^2}{2\sqrt{2}},x',y',z'\right)\|}\left(\frac{\sigma^2}{2\sqrt{2}},x',y',z'\right).
\end{array}
\]
We deduce that the coefficients of $M$ are given by:
\begin{equation}
\fl m_{ii} := -\sum_{j=1,4}\, f_j \left( x,y \right) {\partial_{ii} ^{2}} f_j \left( x,y \right),\quad m_{ik} := -\sum_{j=1,4}\, f_j \left( x,y\right) {\partial_{ik} ^{2}} f_j \left( x,y \right).
\label{metricR2}
\end{equation}
See Appendix B for the values $m_{ij}$.\\

We used the software FreeFem$++$ \cite{FreeFem} for the generation of the 2-D mesh of $r\circ t \,(F_{1,v}^b)$. Then, applying $f$, we get a mesh of $F_1$ with the wanted size. Moreover, as we know the equation of the 2-plane containing $F_{1,v}^b$, we are able to control and to correct the coordinates of the vertices of the mesh of $F_{1,v}^b$ with a great precision. We can also control that the mesh vertices on the edges of $F_1$ are on the convenient geodesic (\ref{geod}), distant from each other with the wanted distance; we correct their coordinates if necessary.

Then, by the use of rotations in $\RR^3$ we get a mesh of the five adjacent faces $F_{2,v}^b$, ....., $F_{6,v}^b$ (see Appendix C). After normalizing we have got a mesh of $F_1$, ..., $F_6$. At last, by the use of $g_1$, ...., $g_6$ we deduce a mesh of $F_7$, ..., $F_{12}$. Once more we control and correct the mesh vertices on edges. As we also know the equation of the ellipsoid that contains each face, we also control all the mesh vertices. Meshes of each $F_{i,v}$ are obtained from those of $F_i$ by discarding the first coordinate of the mesh vertices.

To get the mesh of $ \stackrel{\circ}{\mathcal{F}}$ we used the software Tetgen \cite{Tetgen} in several ways. We have created a mesh with an inner sphere or an inner dodecahedron, in order to impose two different average sizes of tetrahedra depending on the location. We also used Tetgen with a $.mtr$ file that gives us the wanted size near each point. In all cases we employed an option so that there is no point introduced into $\partial \mathcal{F}_v$. We can see the accuracy of our mesh by calculating the sum of the volume of tetrahedra which compose $\mathcal{F}_v$. This sum must be closed to $\frac{2\pi^2}{120}$ that is $\simeq 0.1644934066$. 

\begin{table}[!ht]
\caption{\label{Meshes2}Distances beetween two mesh vertices of $\partial \mathcal{F}_v$.}
\begin{tabular}{@{}lclll}
\br
Name of & Number of mesh vertices between & minimal distance & &maximal distance \\
the mesh&two vertices of an edge of $F_i$&&&\\
\mr
F089&89&$2.06\;10^{-3}$&&$6.11\;10^{-3}$\\
F101&101&$1.82\;10^{-3}$&&$4.90\;10^{-3}$\\
\br
\end{tabular}
\end{table}
\begin{table}[!ht]
\caption{\label{Meshes} Characteristics of these two meshes.}
\begin{tabular}{clllll}
\br
Mesh&\mbox{number} & \mbox{number} & number of &volume of & \mbox{relative error}\\
 &\mbox{of vertices} & \mbox{of nodes} & \mbox{tetrahedra} &all tetrahedra & \\
\mr
F089&$182162$&$730309$&$4642744$&$0.1644927119$&$4.22\;10^{-6}$\\
F101&$233858$&$1600118$&$10198710$&$0.1644928648$&$3.29\;10^{-6}$\\
\br
\end{tabular}
\end{table}

\begin{figure}[!ht]
\centerline{\includegraphics[scale=0.2]{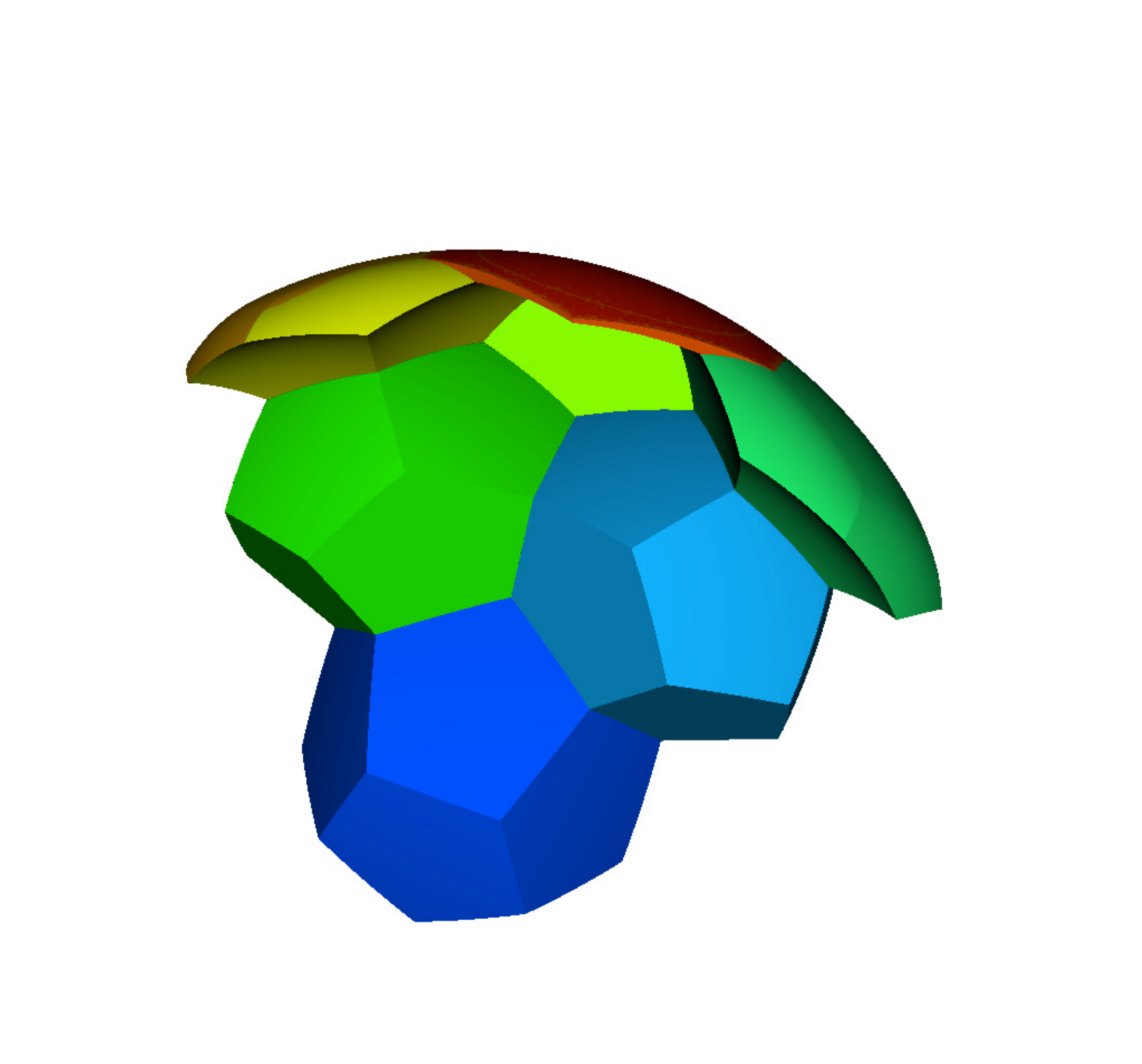}}
\includegraphics[scale=0.35]{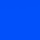}$\mathcal{F}_v$, \includegraphics[scale=0.35]{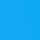}$ g_1 (\mathcal{F}_v)$, \includegraphics[scale=0.35]{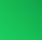}$ g_1 \, g_1 (\mathcal{F}_v)$, \includegraphics[scale=0.35]{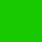}$g_6(\mathcal{F}_v)$, \includegraphics[scale=0.3]{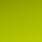}$ g_6 \, g_6 (\mathcal{F}_v)$, \includegraphics[scale=0.35]{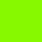}$ g_5 \, g_6(\mathcal{F}_v)$, \includegraphics[scale=0.3]{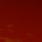}$ g_2\, g_1\, g_1 (\mathcal{F}_v)$
\caption{\label{DODIM}The Fundamental Domain (dark blue) and six images of it.}
\end{figure}

We can also look at the tilling of $\mathcal{S}^3$. In figure \ref{DODIM} we display six images of $\mathcal{F}_v$, as well as $\mathcal{F}$ in the first ball of visualization of $\mathcal{S}^3$. All points of theses images have a positive first coordinate in $\RR^4$, except those of $g_2 \, g_1 \, g_1 (\mathcal{F})$. To fully represent this last set, we should also draw the second ball of visualization of $\mathcal{S}^3$, in which its representation is the same as in the first ball.

\subsection{$V_h$ space}
We construct the finite element spaces $V_h$ of $\P_1$ type. We take into account the boundary condition (\ref{cl}) in the
definition of the finite elements, so that $V_h\subset W^1(\mathcal{F}_v)$.
We note $\mathcal{T}_h$ all tetraedra of a mesh, $\mathcal{ F}_{v,h}$ the set $\displaystyle{\cup_{K\in \mathcal{T}_h} K}$, and $\P_1(K)$ the set of first degree polynomial functions on $K$. Then we
introduce:
$$V_h:=\left\{ v: \mathcal{ F}_{v,h} \rightarrow \RR,\, v \in
\mathcal{C}^0(\mathcal{ F}_{v,h}), \forall K \in \mathcal{T}_h,\,
v_{|K}\in \P_1(K),\, M\sim M'\Rightarrow v(M)=v(M')\right\}.$$
The equivalent points  on $\partial \mathcal{F}_v $ are known thanks our construction of meshes.
The number of nodes $N_h$ is the sum of $\quad \frac{30}{3} \times {n_\mathrm{ve}}\,+\,\frac{12}{2}\times n_{\mathrm{vf}}\,+n_{\mathrm{vi}}\,+\frac{20}{4}$, with $n_{\mathrm{ve}}$ the number of mesh vertices on an edge of a face that are not a vertex of $\mathcal{F}_v $, $n_{\mathrm{vf}}$ the number of mesh vertices on a face that are not on an edge, $n_{\mathrm{vi}}$ the number of mesh vertices in $ \stackrel{\circ}{\mathcal{F}_v}$.

If $j$ is the number of a node and if $M_i$ denotes a vertex of the mesh, we construct a basis $\left(e_j^h\right)_{1\leq j\leq N_h}$ of $V_h$ by:\\
\begin{enumerate}
\item If $j$ is associated to a node that does not belong to $\partial \mathcal{F}_v$ :
$e_j^h(M_i)=\delta _{ij}$. \\
There are $n_{\mathrm{vi}}$ functions of this kind.
\item If $j$ is associated to a node that is a vertex of  $\mathcal{F}_v$: \\
$e_j^h(M_i)=\left\{\begin{array}{ll}
1 & if\ M_i\sim M_j,\\
0 & otherwise.
\end{array}
\right.$\\
There are five functions of this kind.
\item If $j$ is associated to a node that belongs to a face of $\partial \mathcal{F}_v$ and not to an edge:\\
$e_j^h(M_i)=\left\{\begin{array}{ll}
1 & if\ M_i=P_i,\\
0 & otherwise.
\end{array}
\right.$\\
There are $6\times n_{\mathrm{vf}}$ functions of this kind.
\item If $j$ is associated to a node that belongs to an edge of a face of $\partial \mathcal{F}_v$ and is not a vertex of $\mathcal{F}_v$:
$e_j^h(M_i)=\left\{\begin{array}{ll}
1 & if\ M_i=P_i,\\
0 & otherwise.
\end{array}
\right.$\\
There are $10\times n_{\mathrm{ve}}$ functions of this last kind.
\end{enumerate}

\subsection{Matrix form of the problem}
$\KK(i,j)$, $\DD(i,j)$ and $\MM(i,j)$ are found with a numerical integration using CUBPACK \cite{CUB}. These matrices are sparse and symetric. So we choose a Morse storage of their lower part, and all of the calculations will be performed with this storage.
To solve the linear problem we use a preconditioned conjugate gradient method. The preconditioner is an incomplete Choleski factorisation, and the starting point is the solution obtained with a diagonal preconditioner.

In order to control $\KK+\DD$ we verify that $\Delta_{\mathcal{F}_v} X=0$ for an $X$ with all its coordinates equal to $1$.  Consequently, we compute the sums of the elements of each line of $\KK+\DD$; all these sums must be equal to $0$. For F089 mesh we obtain sums less than $5\;10^{-17}$. This maximal sum is reached for very few mesh vertices of the interior of $\mathcal{F}_v$. This result is slightly lower for meshes with more mesh vertices on the edges of $F_i$. Higher order finite elements would improve this result.

\subsection{Initial data}
We choose different initial data, all with $\partial_t\psi(0,.)=\psi_1(.)=0$ in order to simplify. Some of them have a more or less small support near given points, others are chosen randomly. For  the wave depicted in figure \ref{InitCentr}, we have taken
\begin{equation*}
\fl \psi_0(X)=100e^{\frac{d(X,X_0)}{d(X,X_0) -r_0}},\;\;\mbox{for}\;d(X,X_0)< r_0,\;\mbox{and}\;\psi_0(X)=0,\;\;\mbox{for}\;d(X,X_0)\geq r_0.
  \label{data}
\end{equation*}
with $X_0=(0,0,0)$ and $r_0=0.3$. In the following we refer to this initial data as the name $Init_c$. And for the one depicted in figure \ref{InitExc}, we have taken a similar function with a smaller support, and especially a support not centered at the origin. In the following we refer to this another initial data as the name $Init_{exc}$.
Both these initial data have a kink at $X_0$: they are not in ${\cal{C}}^{\infty}(\mathcal{F}_v)$, and by computing their derivatives in the sense of the distributions, we can check that they belong only to $W^2(\mathcal{F}_v)\setminus W^3(\mathcal{F}_v)$. Hence this singularity is  rather weak and we may approximate these functions in $V_h$. The main interest of the kink is its ability of exciting a large amount of eigenmodes, and it provides an efficient tool to compute a lot of eigenvalues with an excellent agreement.

We also have considered a third initial data in ${\cal{C}}^{\infty}(\mathcal{F}_v)$, denoted $Init_{exc, \infty}$ in the sequel, and of the form:
\begin{equation*}
\fl \psi_0(X)=100e^{\frac{d^2(X,X_0)}{d^2(X,X_0) -r_0^2}},\;\;\mbox{for}\;d(X,X_0)< r_0,\;\mbox{and}\;\psi_0(X)=0,\;\;\mbox{for}\;d(X,X_0)\geq r_0.
  \label{data2}
\end{equation*}
Since our scheme is based on the approximation by finite element of order one, we cannot expect a best accuracy but this function will be very convenient if we want use a higher finite element method to improve the accuracy of our computation.

We note that the support of all these initial data is far from $\partial \mathcal{F}_v$, therefore they respect obviously the constraint of the equivalent points. It is interesting to consider initial data involving several equivalent points. On $\mathcal{S}^3$, any function $\Phi_0$ defines an initial data $\Psi_0$ that is invariant under the action of  $\mathcal{I}^*$ by the formula
$$
\Psi_0(x_0,x,y,z)=\sum_{g\in \mathcal{I}^*}\Phi_0(g(x_0,x,y,z)),
$$
hence by (\ref{psipsi}), we can introduce on the unit ball of $\RR^3$ (that is how we visualize $\mathcal{S}^3$) :
$$
\psi_0(X):=\sum_{g\in \mathcal{I}^*}\Phi_0(g(f(X))),\;\;\varphi_0(X):=\Phi_0(f(X)).
$$
After transcribing the property \eref{SIM2} in $\RR^3$, and denoting $g_{v}$ the application $\RR^3\rightarrow \RR^3$ induced by $g\in \mathcal{I}^*$ on $\mathcal{F}_v$, we can see that when the support of $\Phi_0$ is small enough, such an initial data satisfies:
\begin{equation*}
\forall X\in \mathcal{F}_v,\; \exists ! g_v \qquad g_v(X)\in \mbox{supp } \varphi_0
\end{equation*}
Therefore we can define $\psi_0$ on $\mathcal{F}_v$ by
$
\psi_0(X)= \varphi_0(g_v (X)).
$

\begin{figure}[h]
\begin{center}
\includegraphics[scale=0.1]{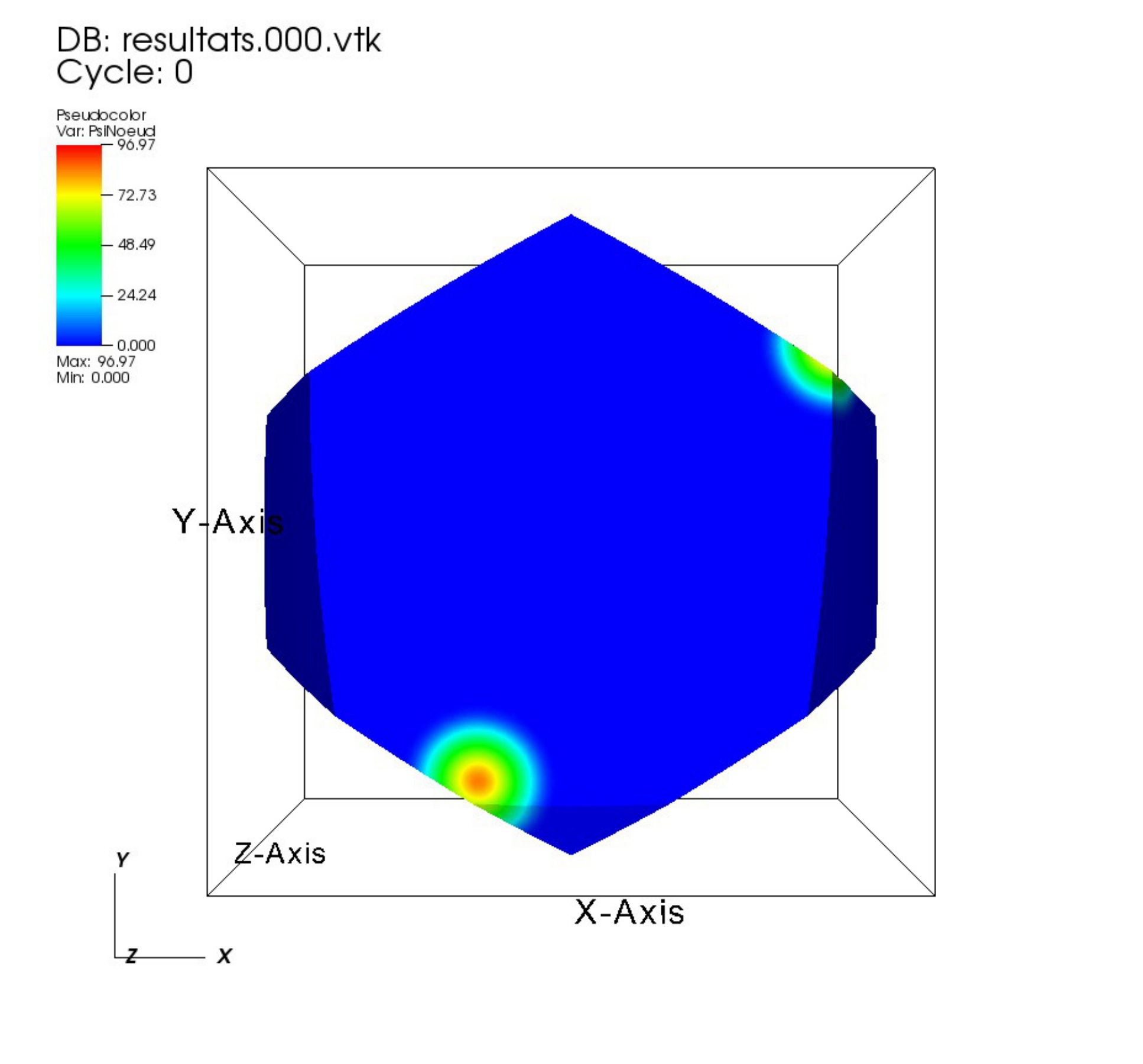}
\caption{\label{InitSum} $Init_{sum}$ on a cutting plane.}
\end{center}
\end{figure}

For example we have made calculations with an initial data denoted by $Init_{sum}$, and defined by:
\[
\psi_0(X)=\left\{
\begin{array}{lll}
\varphi_0(X), &\mbox{if} & X \in \mbox{supp }\varphi_0\cap \mathcal{F}_v\\
\varphi_0(g_{3,v}(X)) &\mbox{if} &  g_{3,v}(X) \in \mbox{supp }\varphi_0 \cap g_{3,v} (\mathcal{F}_v)\\
\varphi_0(g_{1,v}(X)) &\mbox{if} & g_{1,v} (X) \in \mbox{supp }\varphi_0\cap g_{1,v} (\mathcal{F}_v)\\
\end{array}
\right.
\]
with \begin{equation*}
\fl \varphi_0(X)=100e^{\frac{d(X,X_0)}{d(X,X_0) -r_0}},\;\;\mbox{for}\;d(X,X_0)< r_0,\;\mbox{and}\;\varphi_0(X)=0,\;\;\mbox{for}\;d(X,X_0)\geq r_0.\\
\end{equation*}
where $X_0=(-0.1,-0.27,0.16)$ and $r_0=0.1$. $X_0$ belongs to $\mathcal{F}_v$ and is closed to the middle of the edge $\wideparen{S_{20}S_{3}}$ of $\mathcal{F}_v$ (see Appendix A). There are three regions of $\mathcal{F}_v$ where $\psi_0$ is non equal to $0$. On a cutting plane passing through the zone which contains $X_0$ and one of the two others, we can see that the second one is the complement of the first one (\Fref{InitSum}, \Fref{InitSum2}).

\subsection{Time resolution}
\begin{figure}[h]
\begin{center}
\includegraphics[scale=0.2]{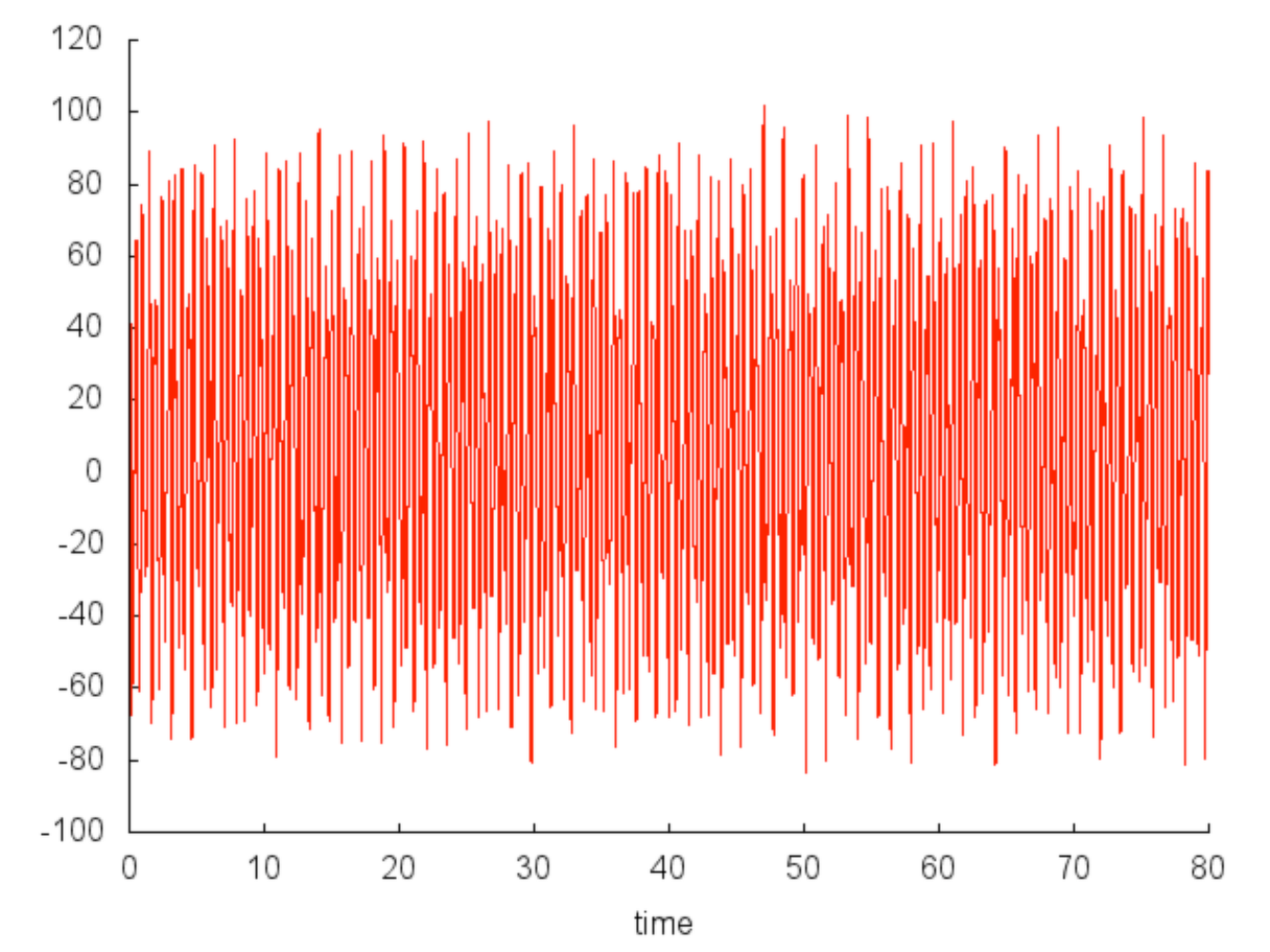}
\caption{\label{Gnuplot}The solution $\psi(t,X_0)$ from $t=0$ to $t=80$.}
\end{center}
\end{figure}

In order to see the stability of our method, we compute $E_d(t)$, the discrete energy associated to our numerical scheme at the time $t$:
$$E_d( t):=\left< \MM \;\frac{X^n-X^{n-1}}{\Delta t}\; , \;
  \frac{X^n-X^{n-1}}{\Delta t}\right>\;+\; \left< (\KK +\DD)\; X^{n-1}\; , \; X^n\right>.$$
It is well known that our scheme is conservative, hence $E_d$ must be invariant as a fonction of time: this is the case in our calculations. For example, with the
previous initial data and the use of the mesh F089 (see \Tref{Meshes}), we obtain:
\[
\begin{array}{|l|l|l|}
\hline
\mbox{Initial data} &  E_d(0) & E_d(80) \\
\hline
Init_c& 9328.34990491949 & 9328.34990491958\\
Init_{exc}& 6248.87928527422 & 6248.87928527418\\
Init_{sum}&10890.5629570901&10890.5629570901\\
\hline
\end{array}
\]
The solution varies extremely fast as can be seen on \Fref{Gnuplot}. This was expected since eigenvalues are large.

\begin{figure}[!h]
\begin{center}
\includegraphics[scale=0.085]{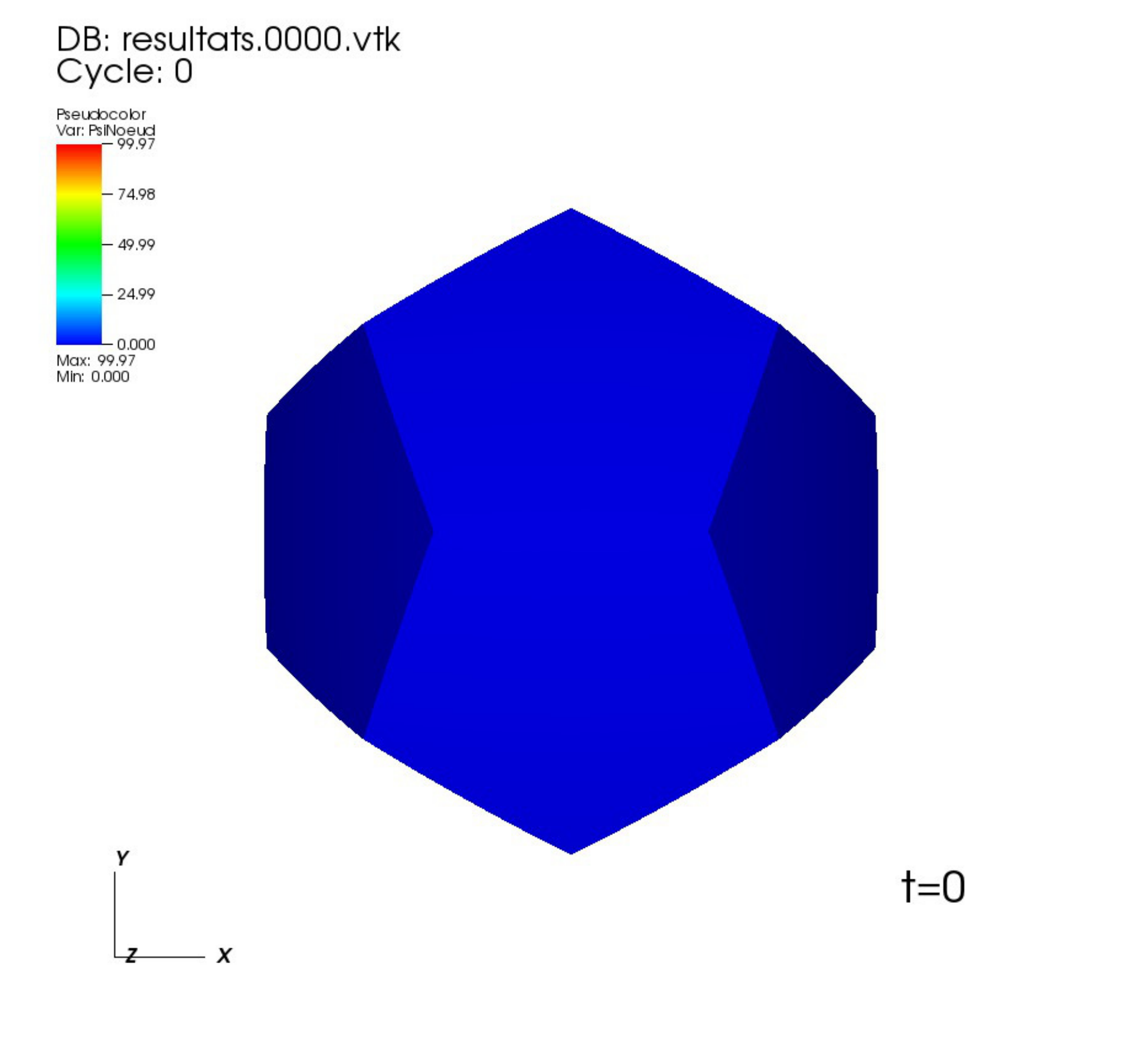}
\includegraphics[scale=0.085]{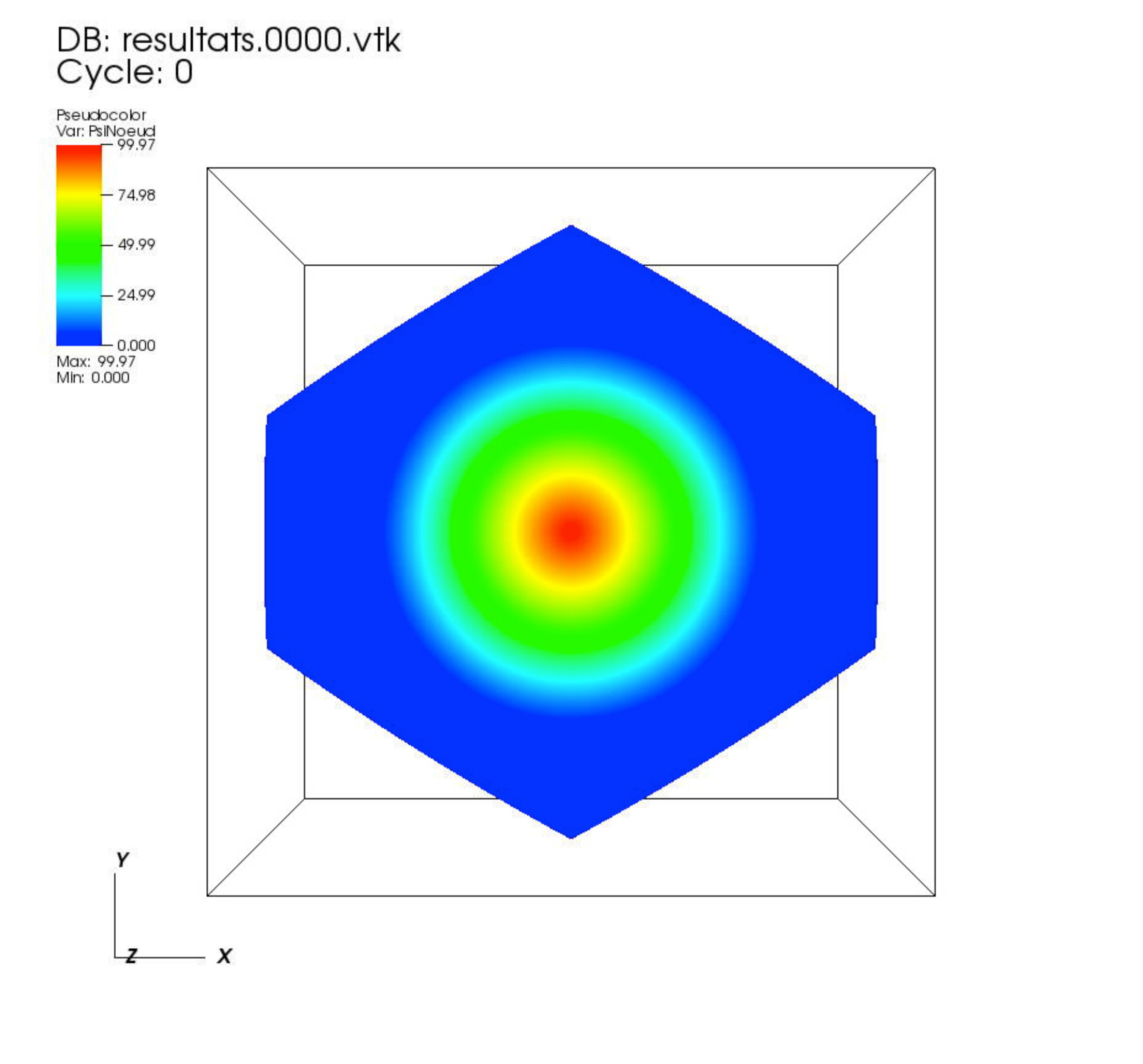}
\includegraphics[scale=0.085]{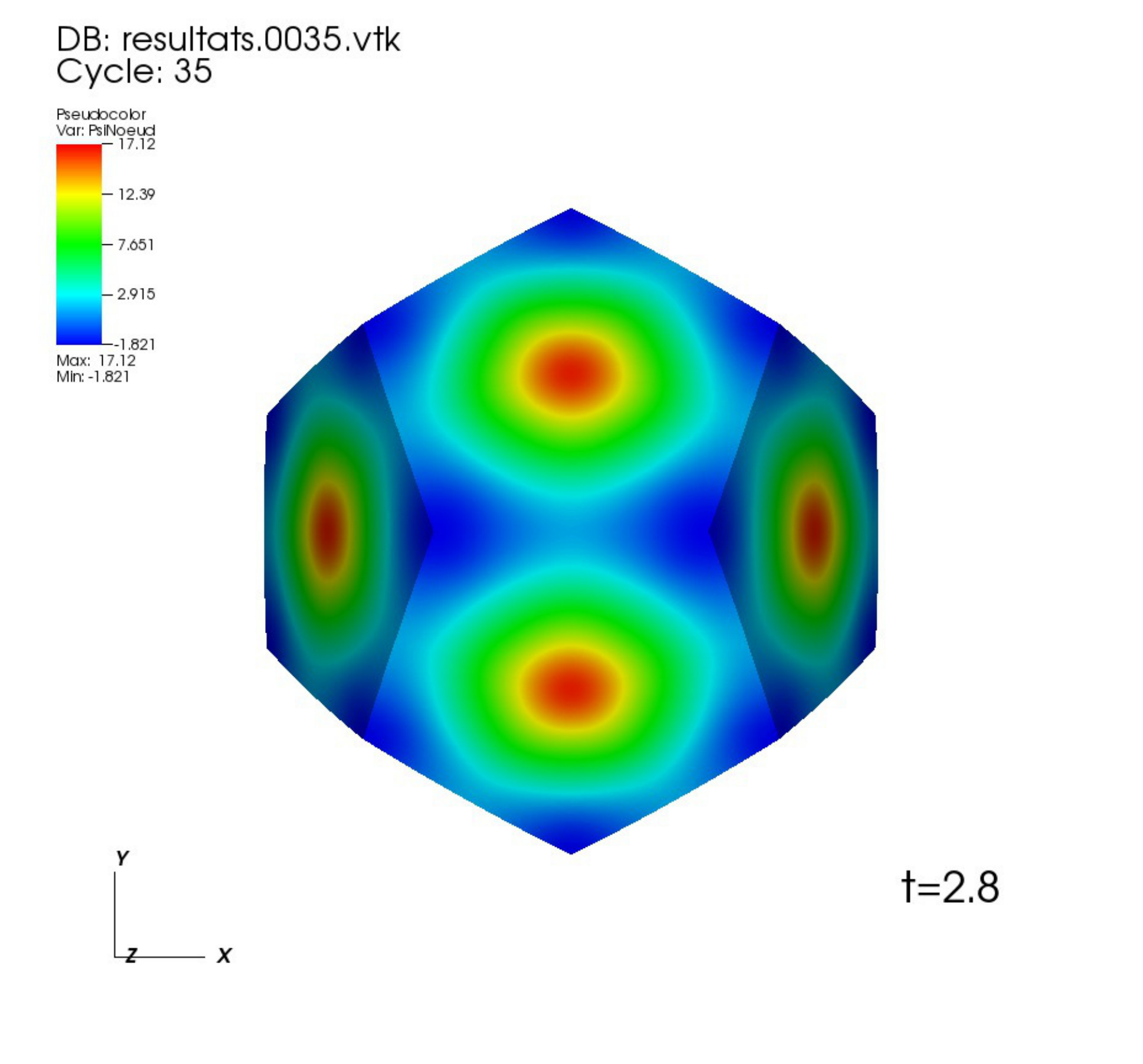}
\includegraphics[scale=0.085]{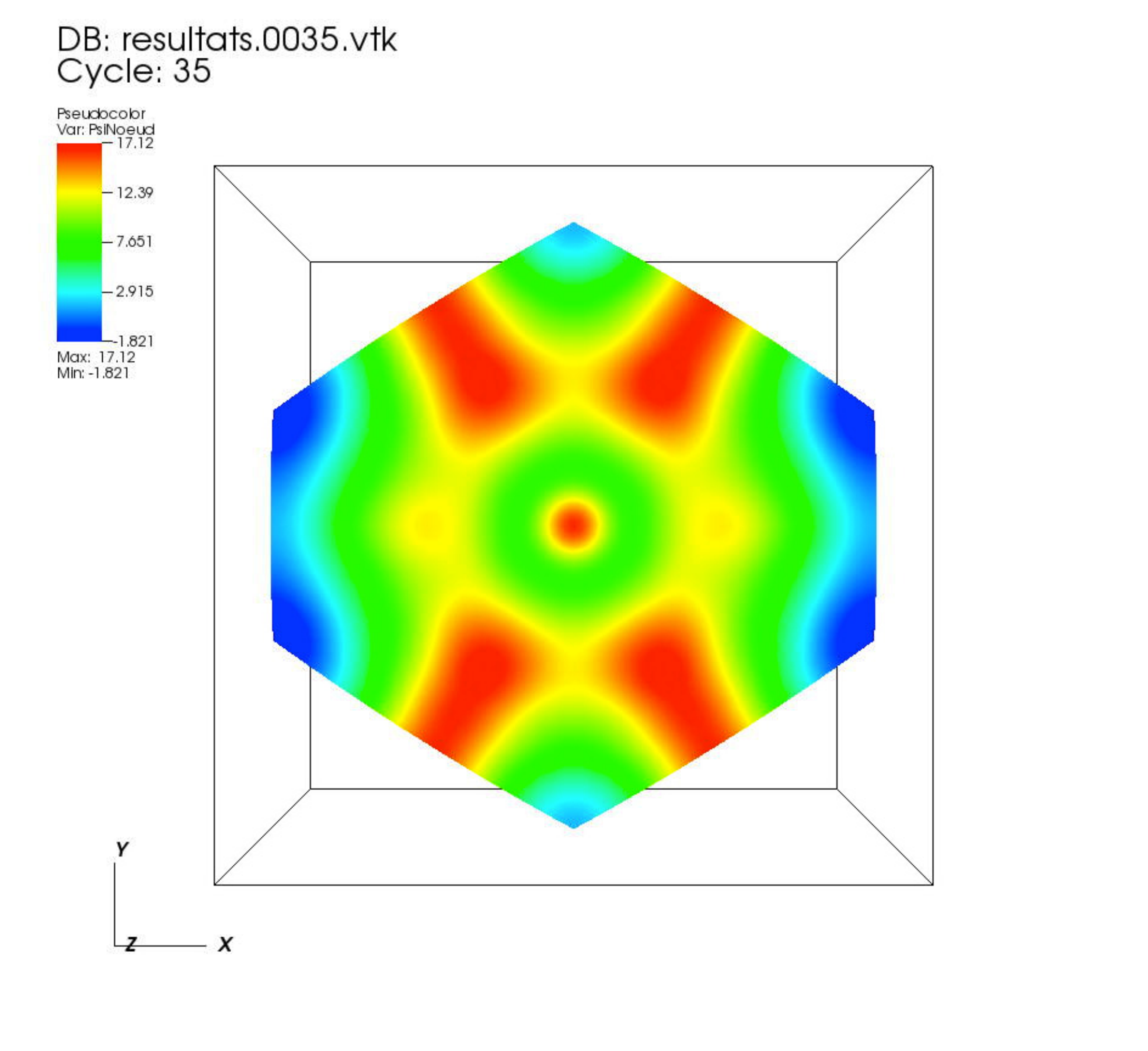}
\includegraphics[scale=0.085]{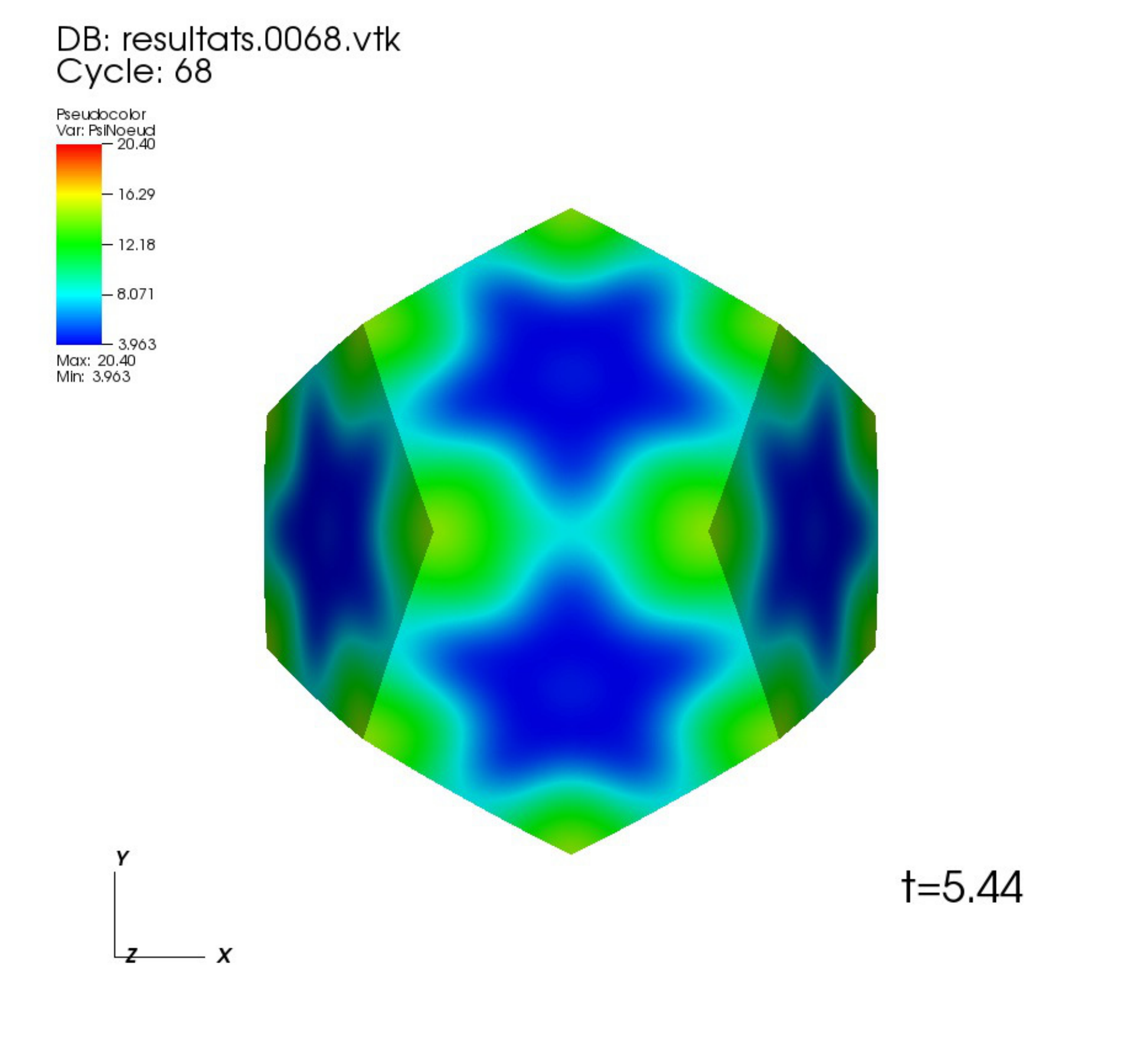}
\includegraphics[scale=0.085]{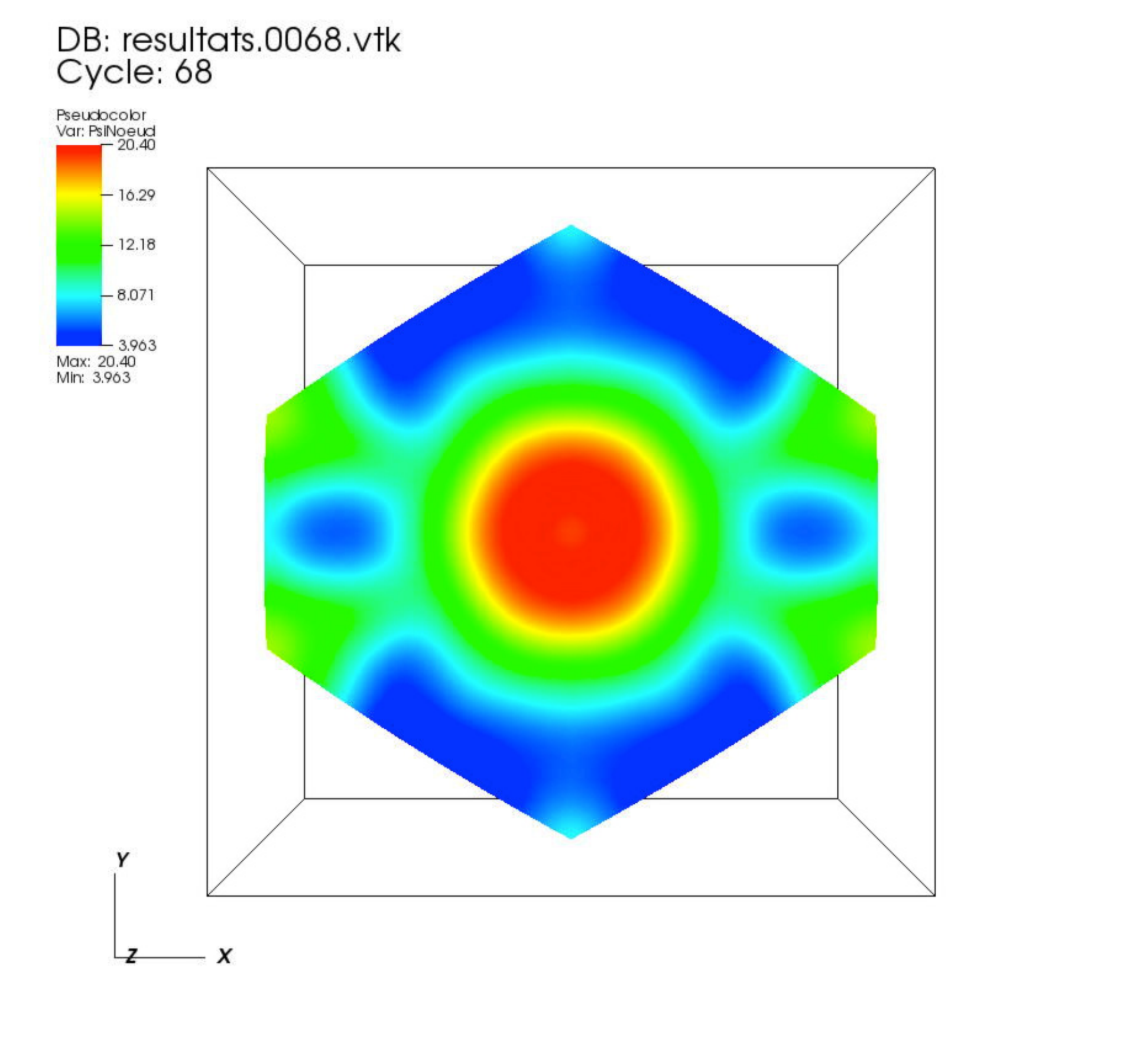}
\includegraphics[scale=0.085]{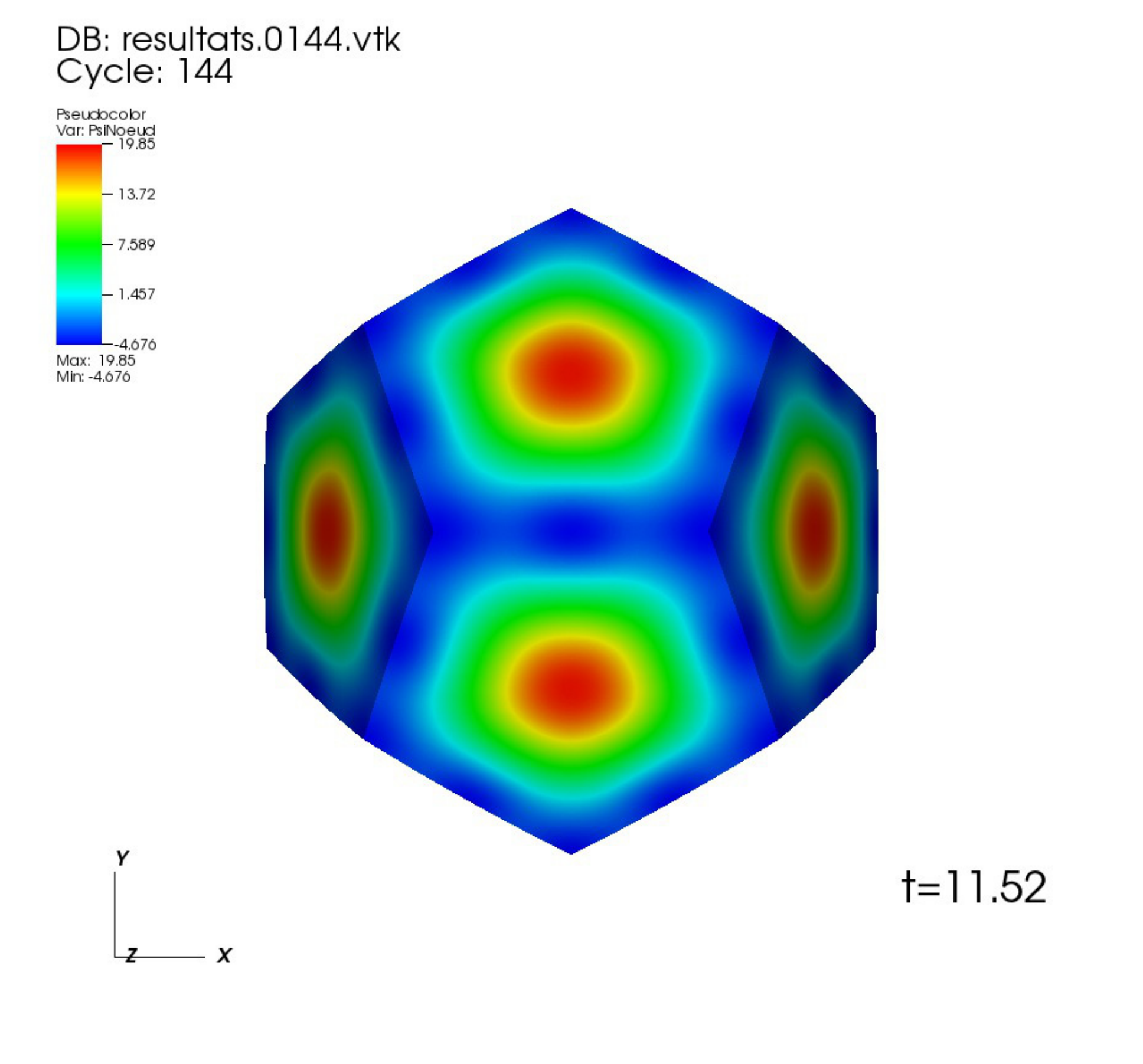}
\includegraphics[scale=0.085]{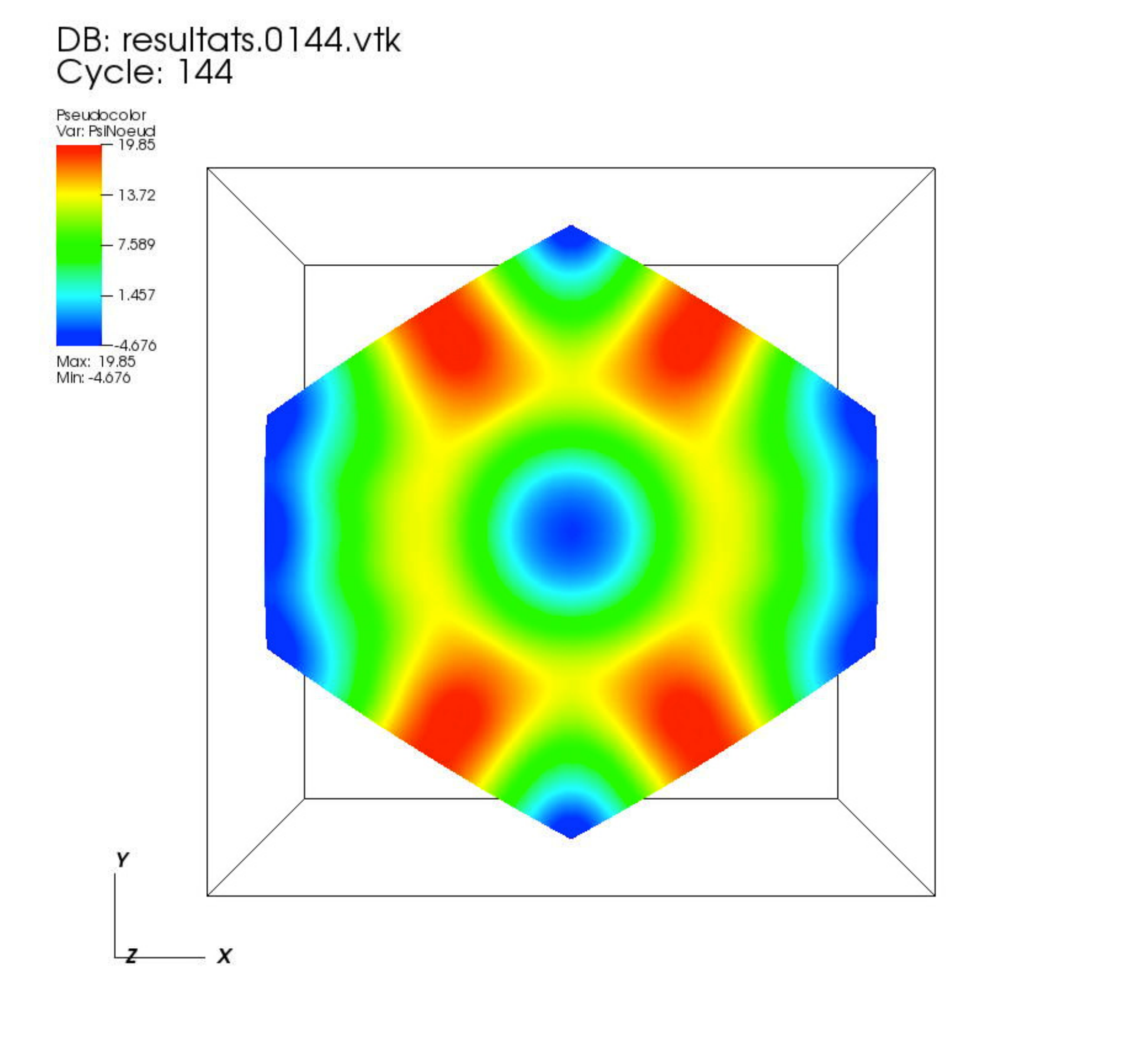}
\includegraphics[scale=0.085]{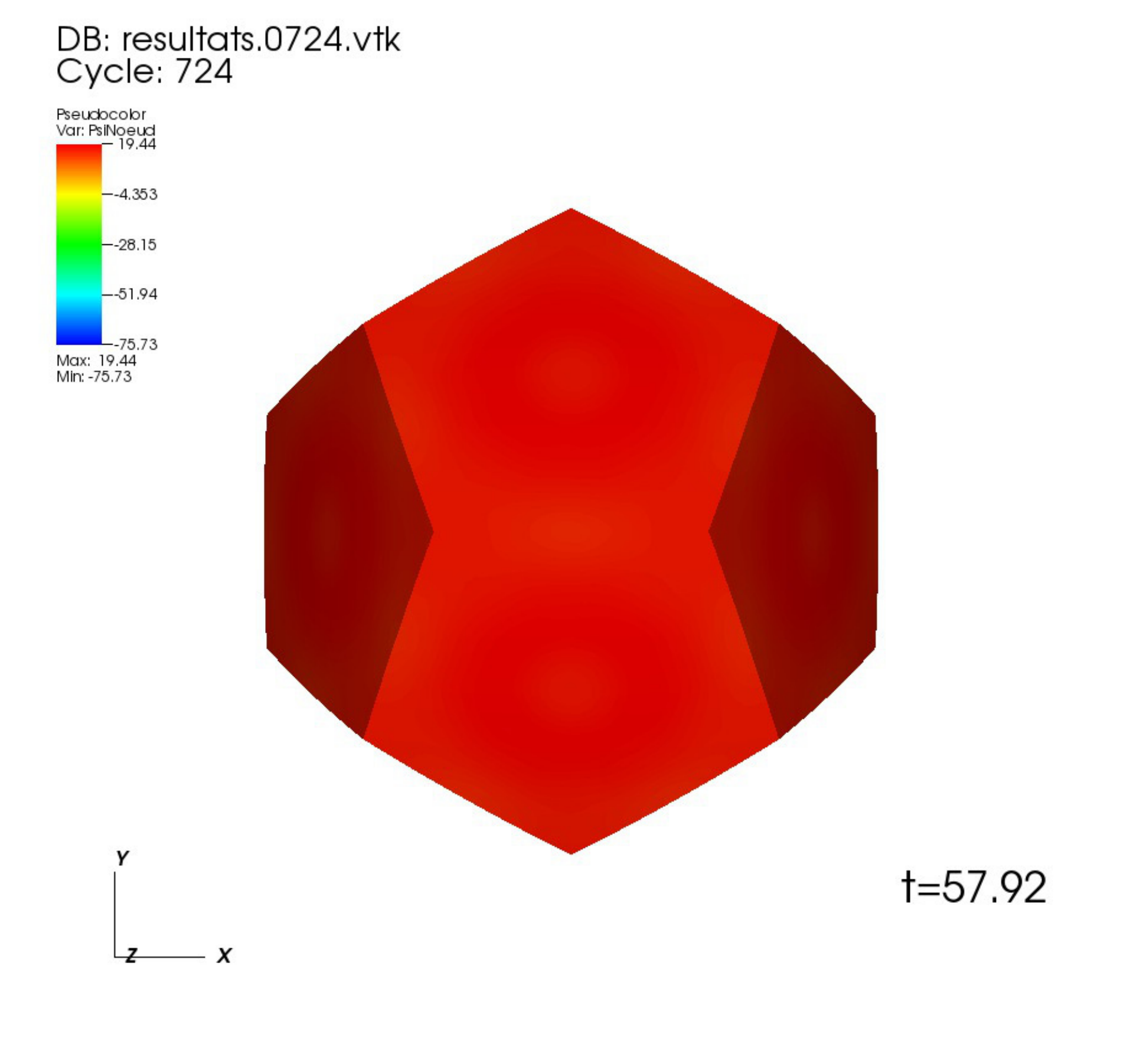}
\includegraphics[scale=0.085]{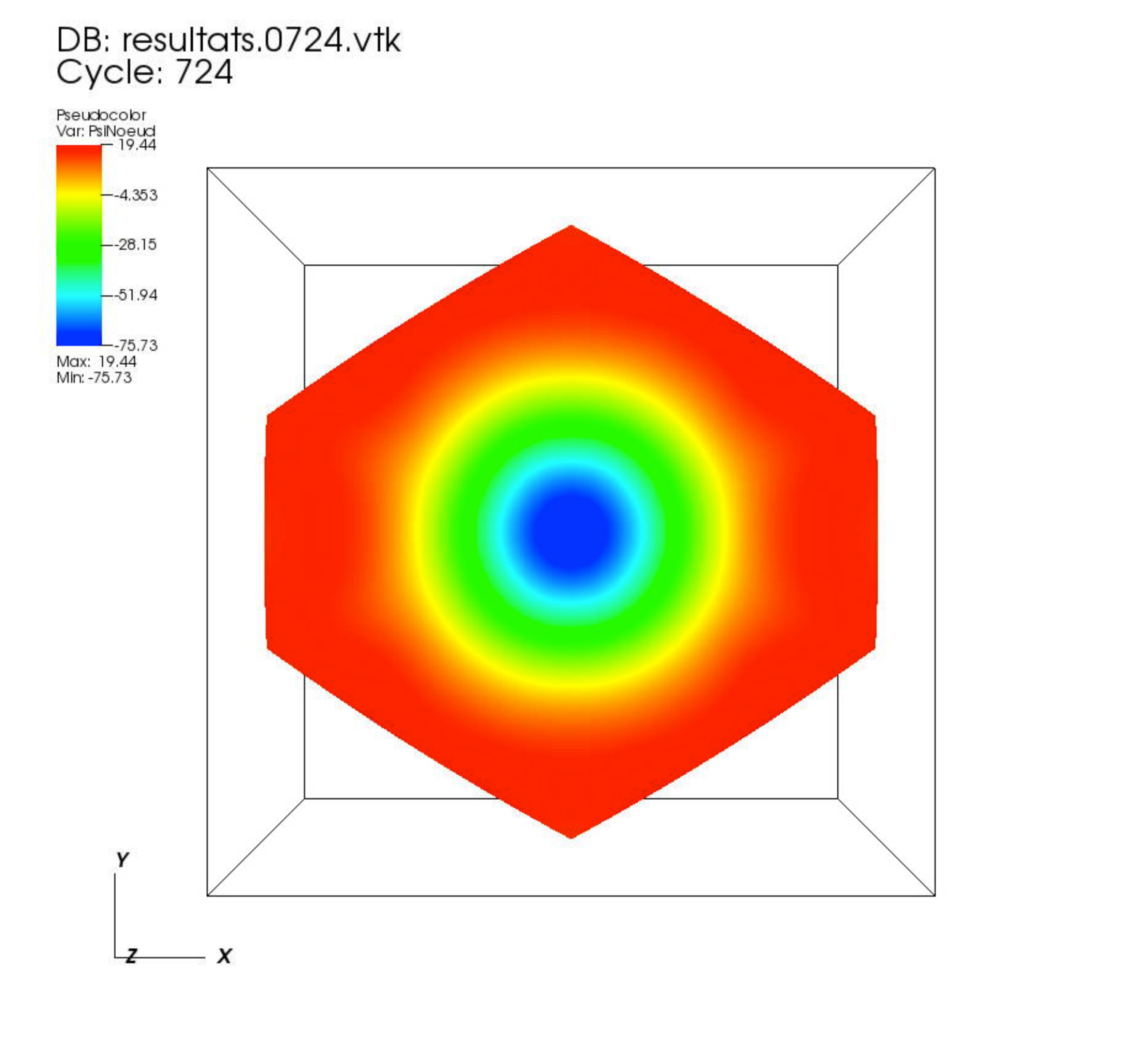}
\includegraphics[scale=0.085]{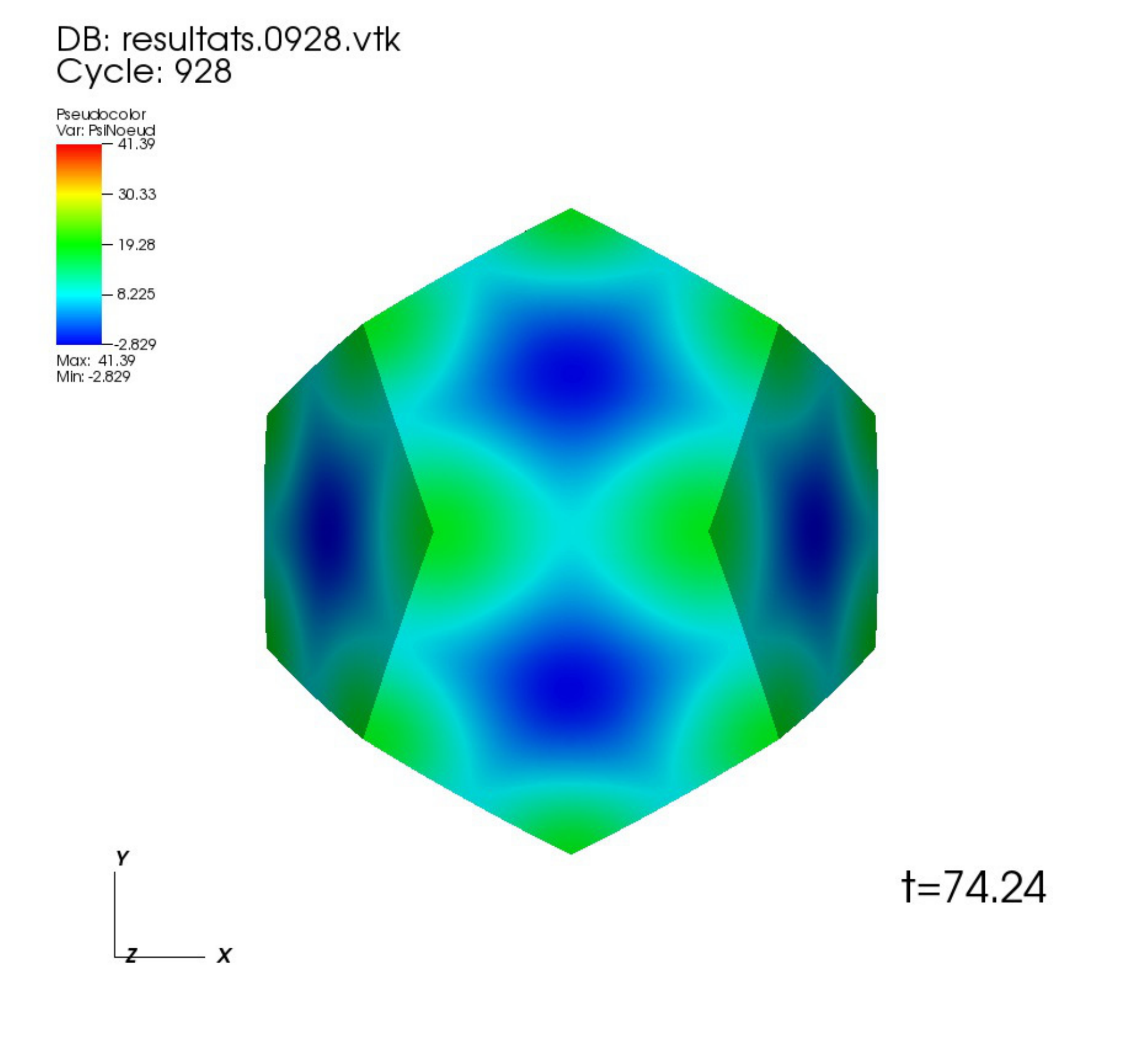}
\includegraphics[scale=0.085]{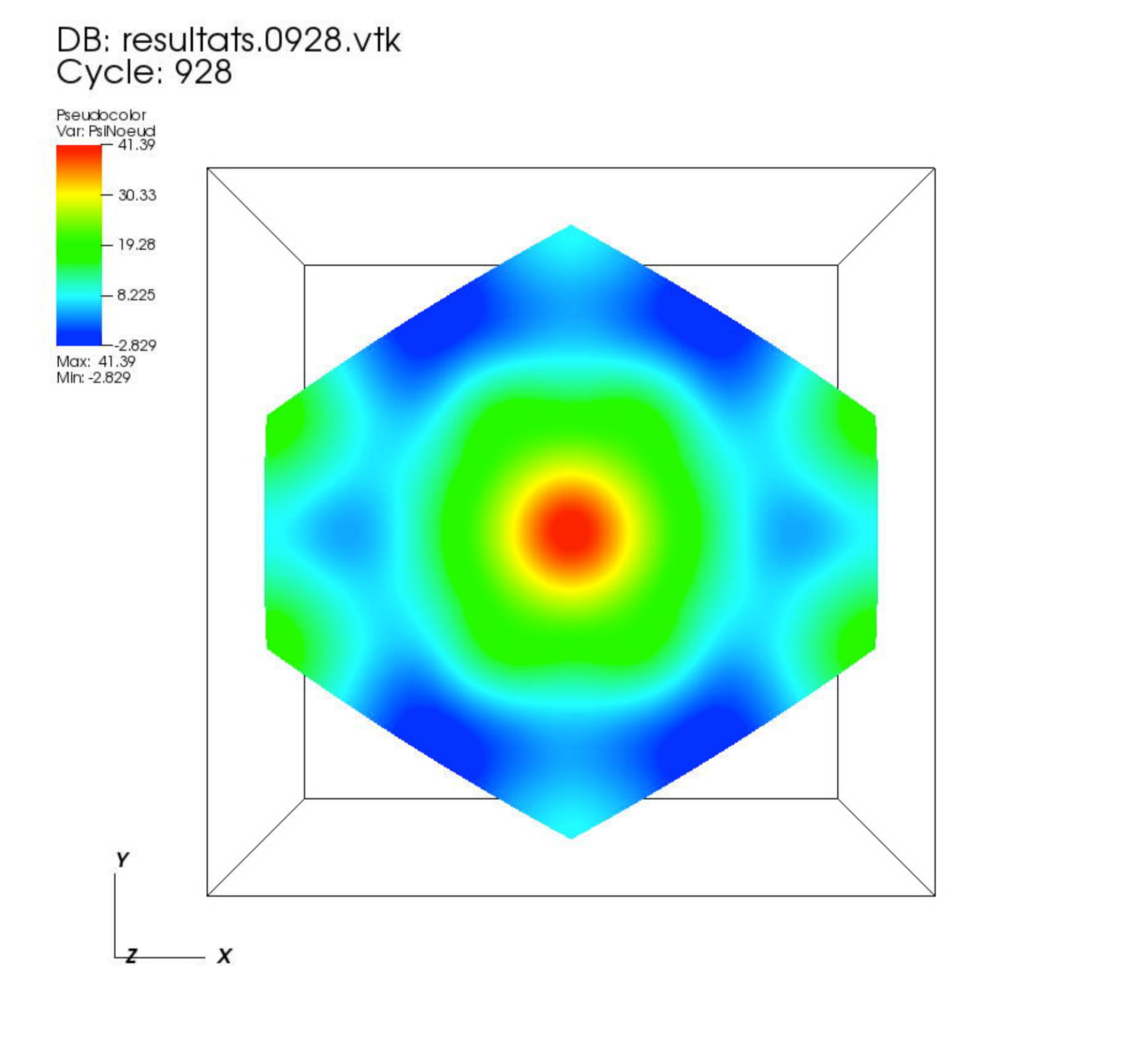}
\caption{\label{InitCentr}The solution $\psi(t,X)$ with the initial data $Init_c$ at $t=0$, $t=2.8$, $t=5.44$, $t=11.52$, $t=57.92$, $t=74.24$.}
\end{center}
\end{figure}

\begin{figure}[!h]
\begin{center}
\includegraphics[scale=0.085]{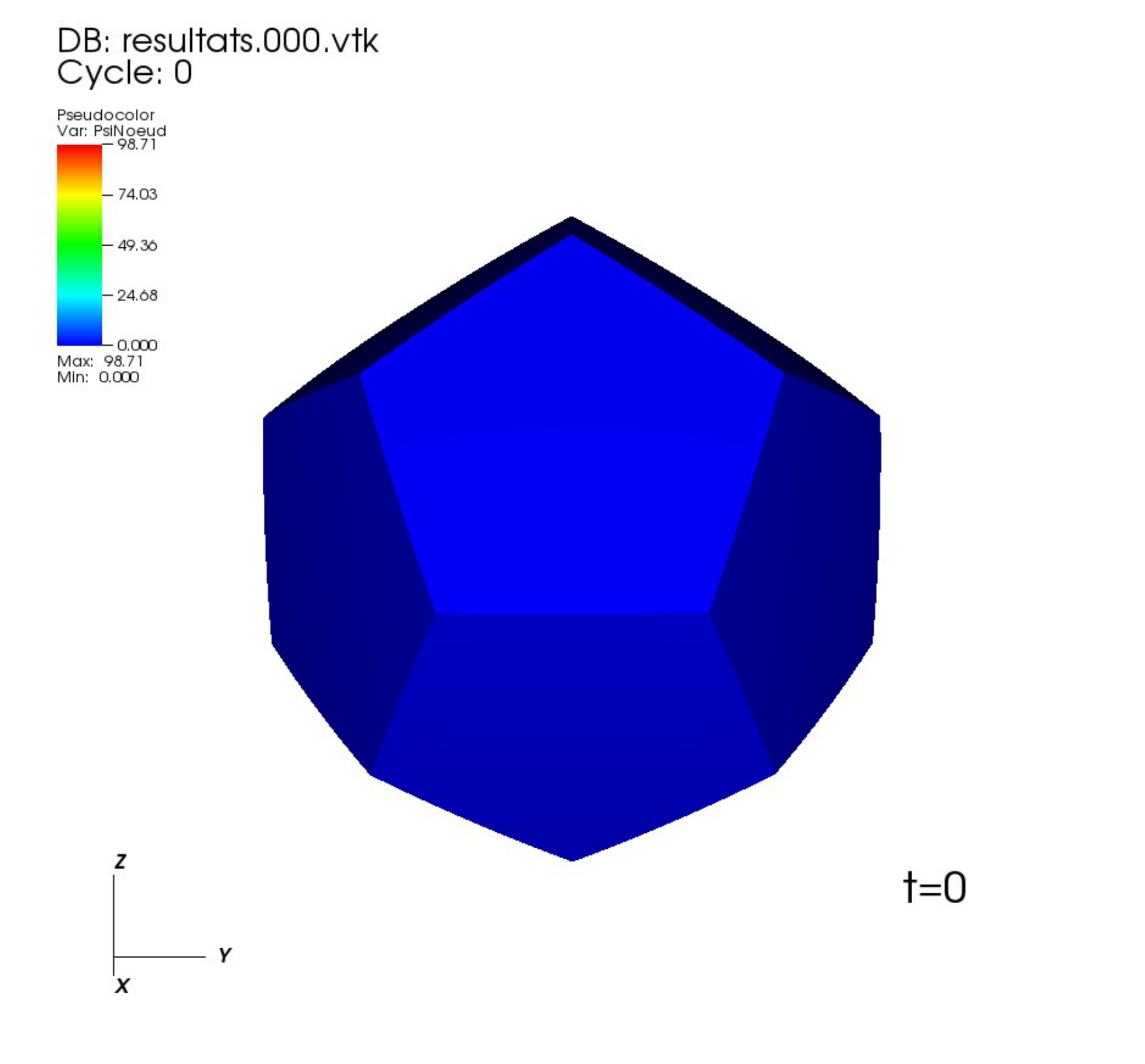}
\includegraphics[scale=0.085]{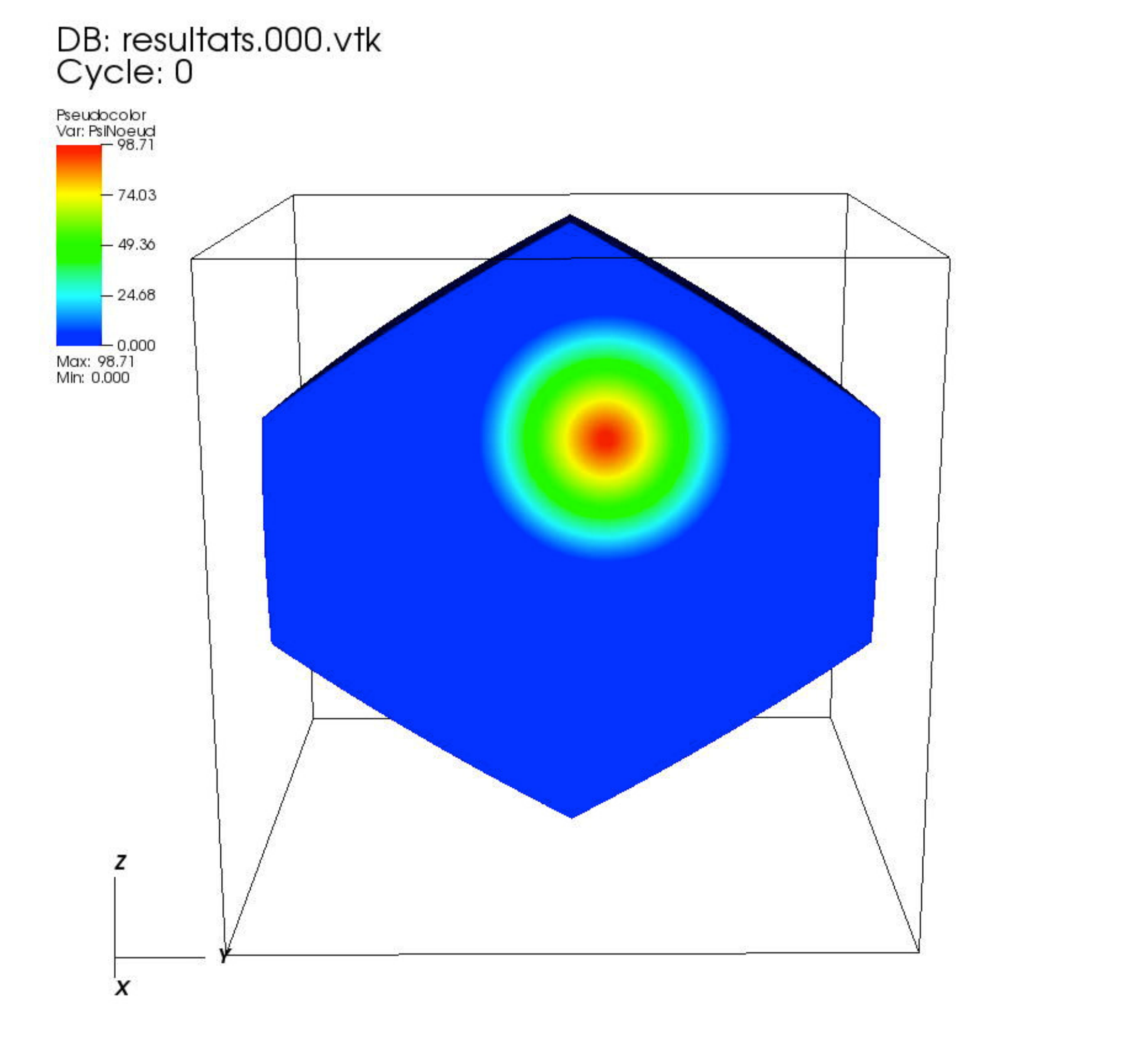}
\includegraphics[scale=0.085]{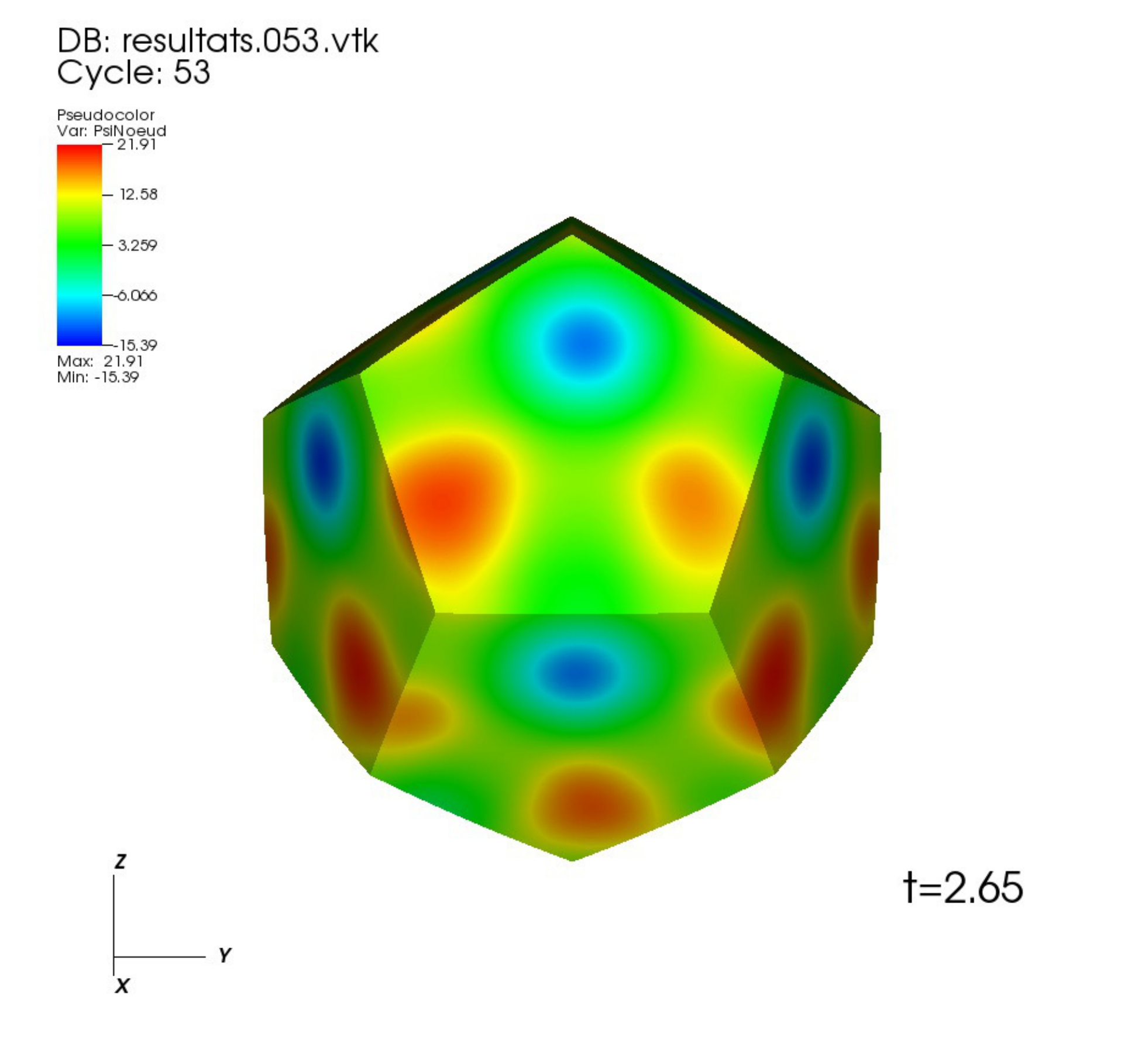}
\includegraphics[scale=0.085]{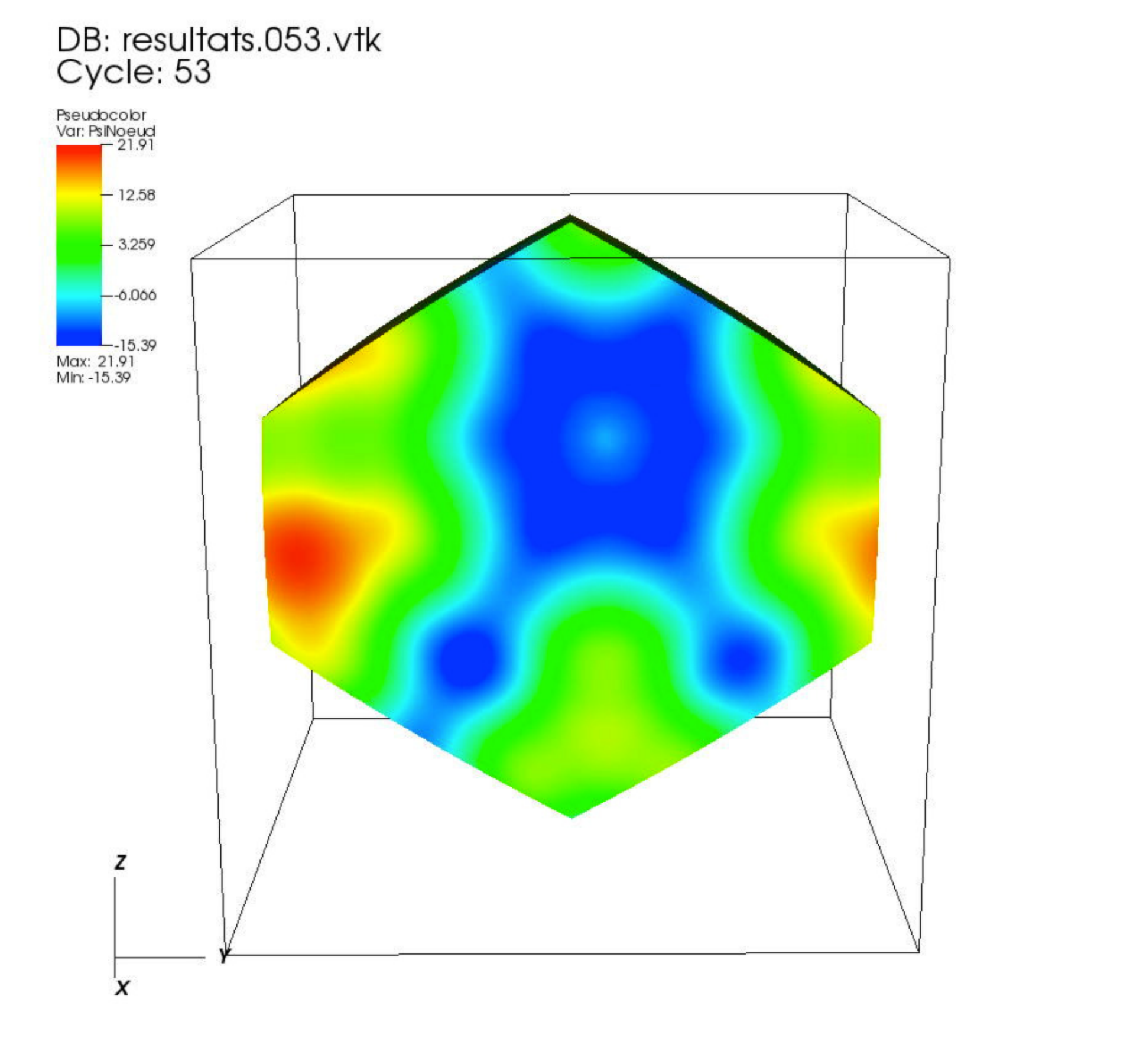}
\includegraphics[scale=0.085]{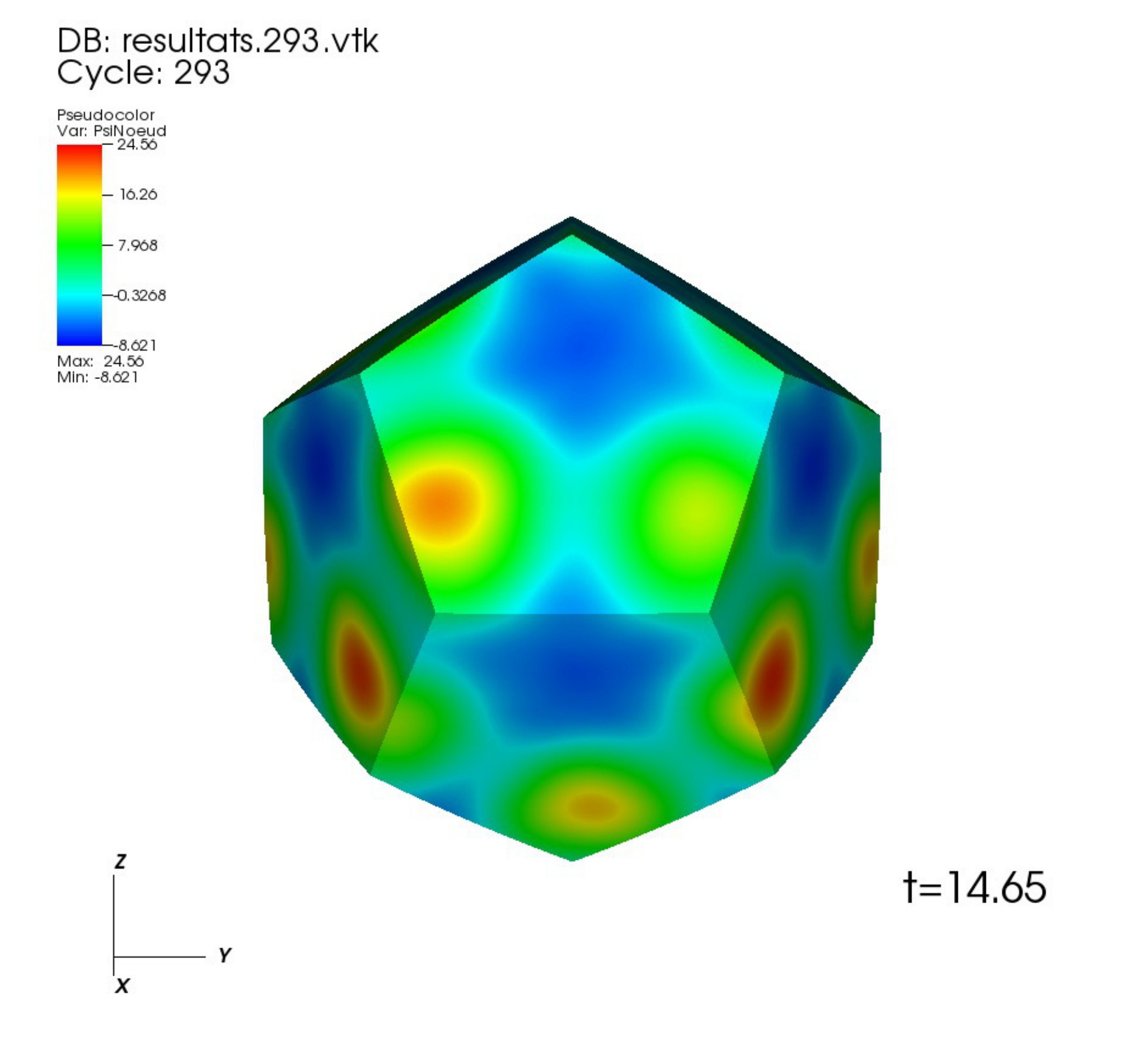}
\includegraphics[scale=0.085]{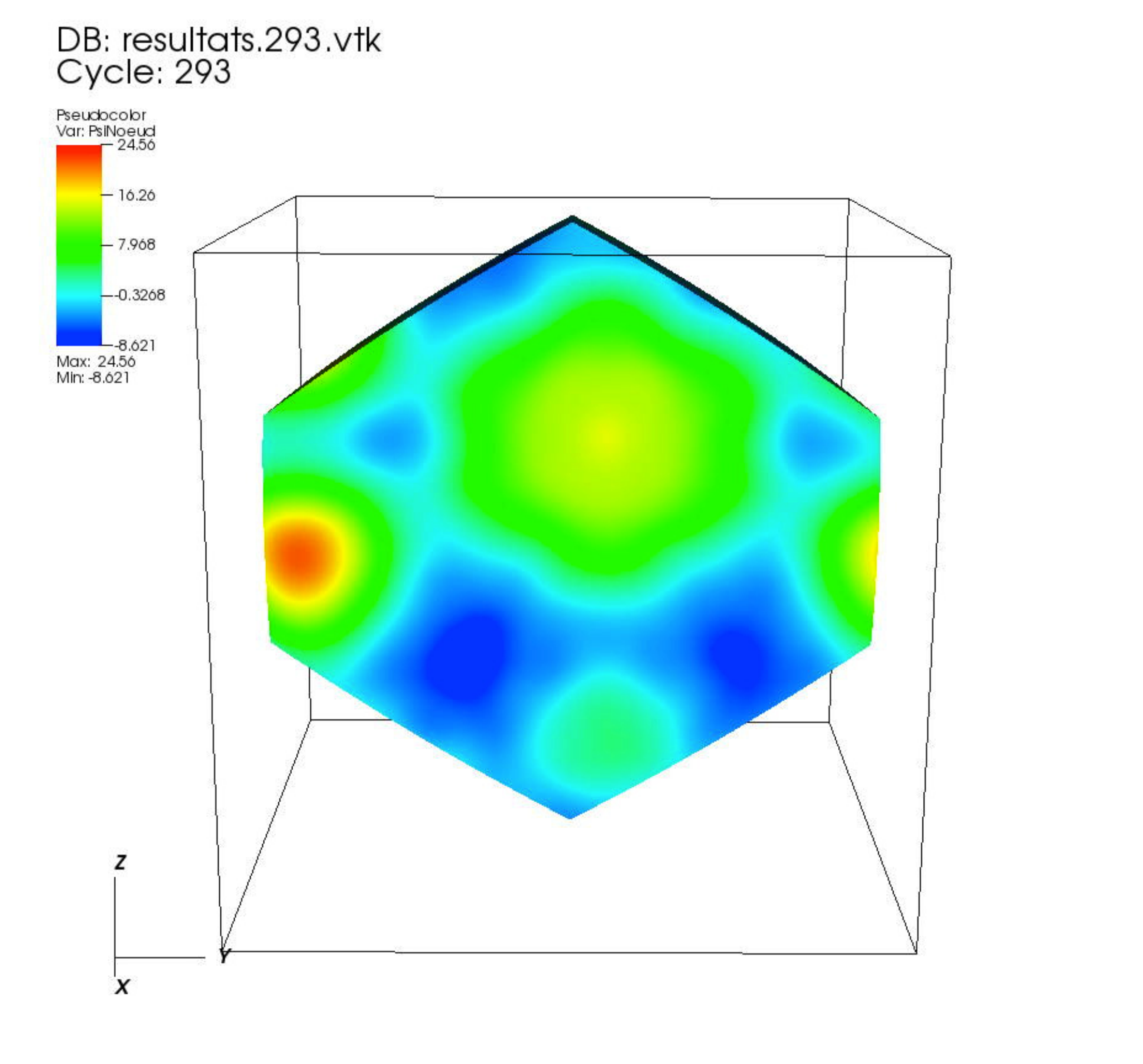}
\includegraphics[scale=0.085]{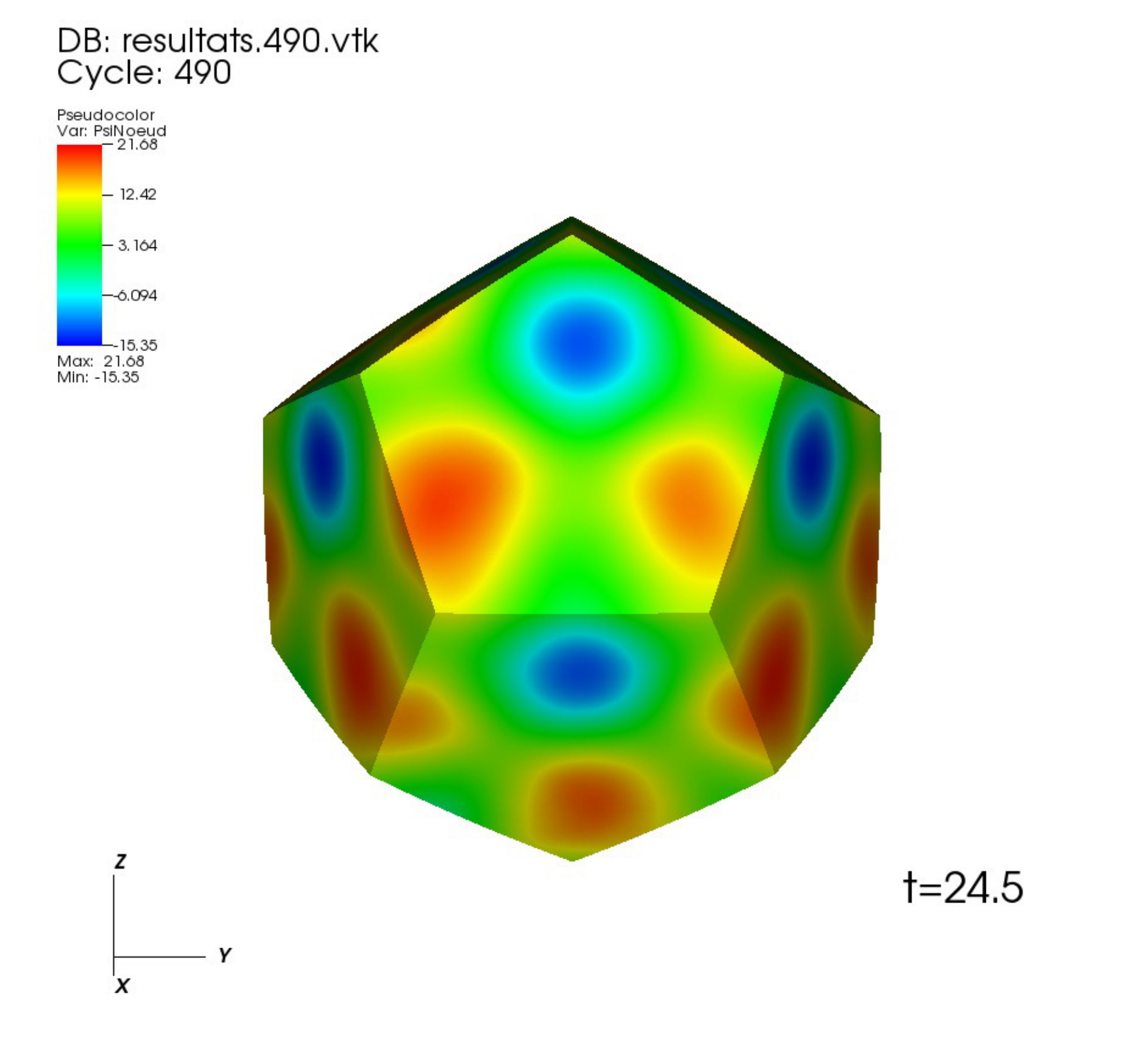}
\includegraphics[scale=0.085]{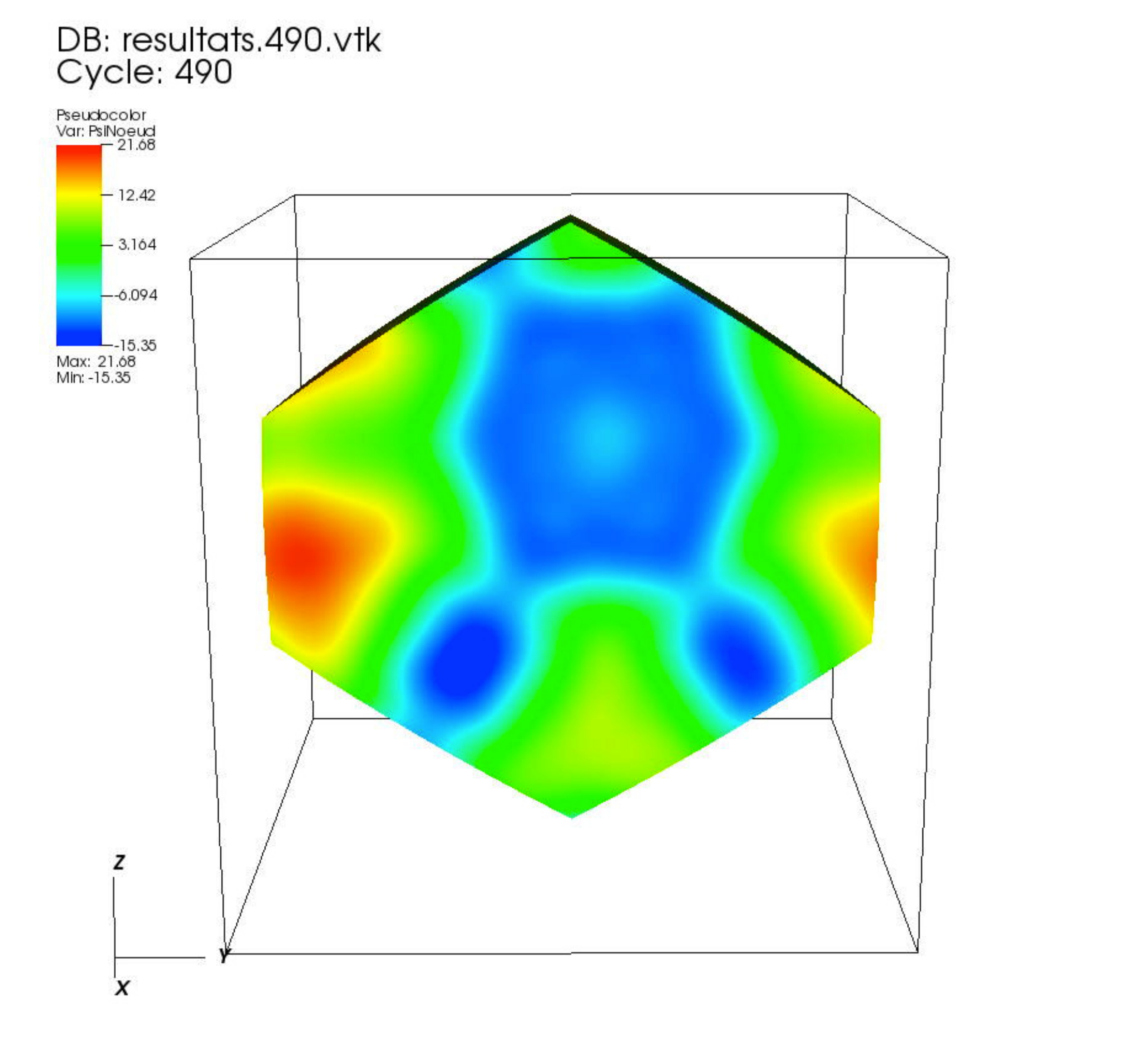}
\includegraphics[scale=0.085]{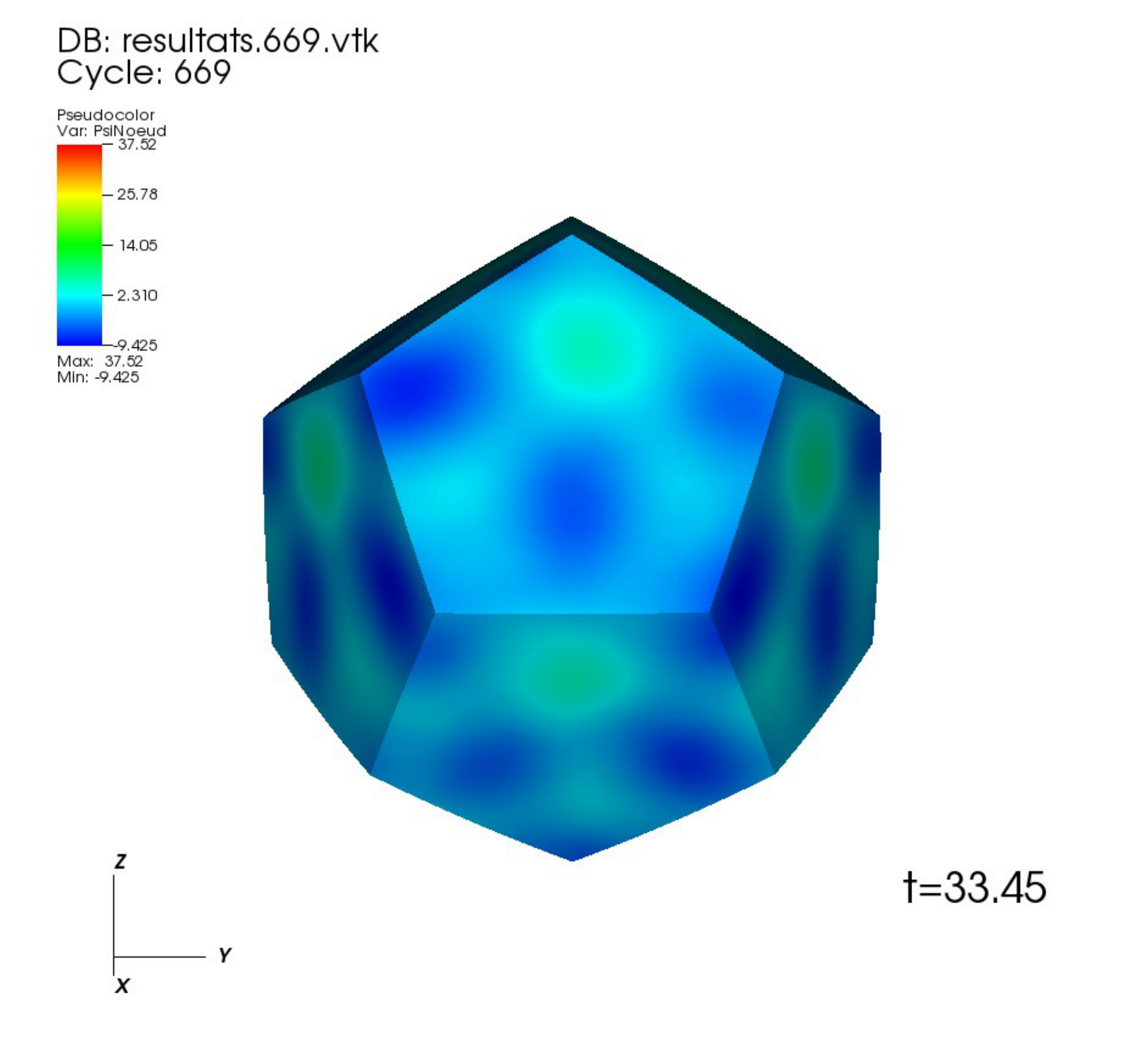}
\includegraphics[scale=0.085]{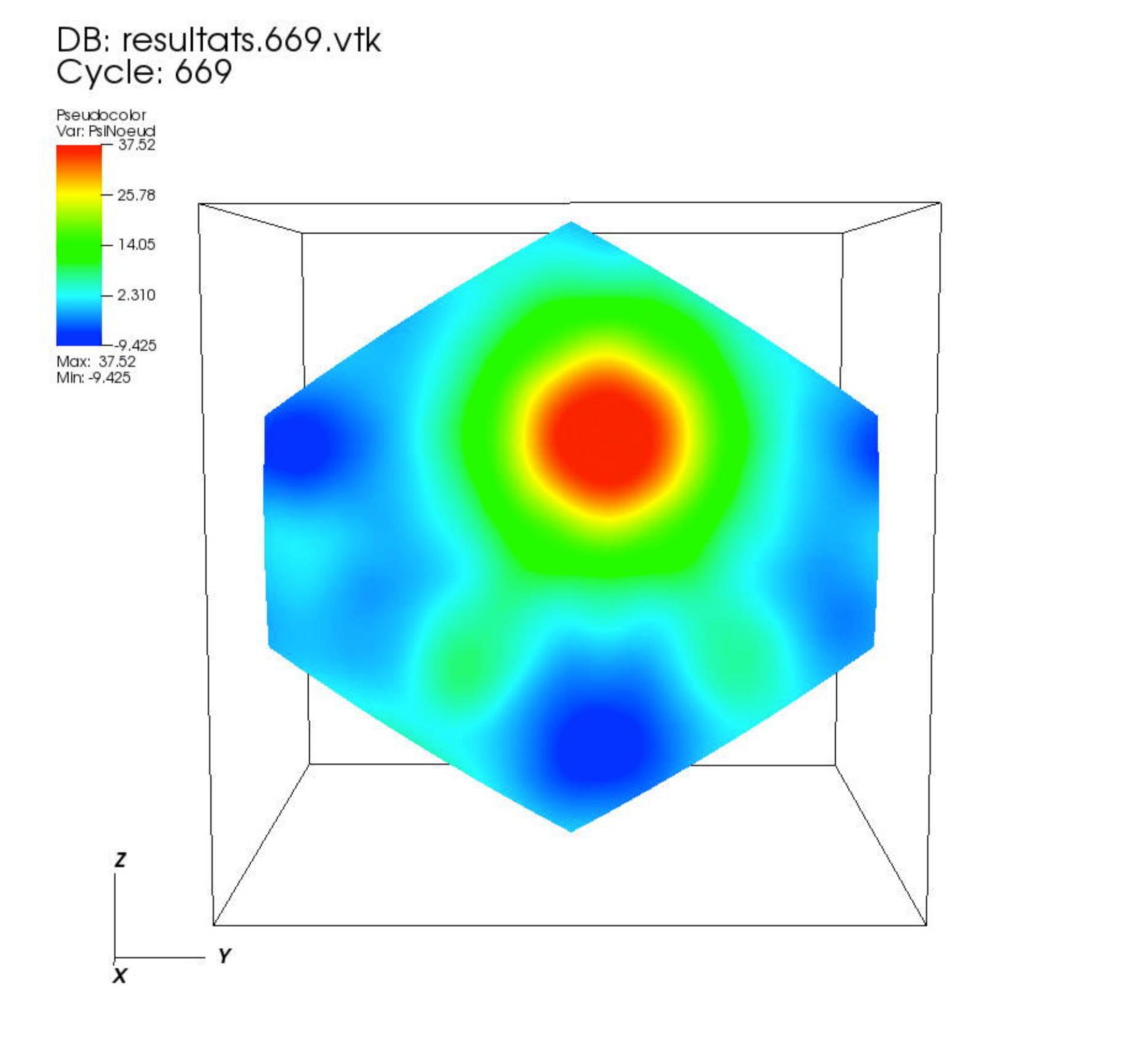}
\includegraphics[scale=0.085]{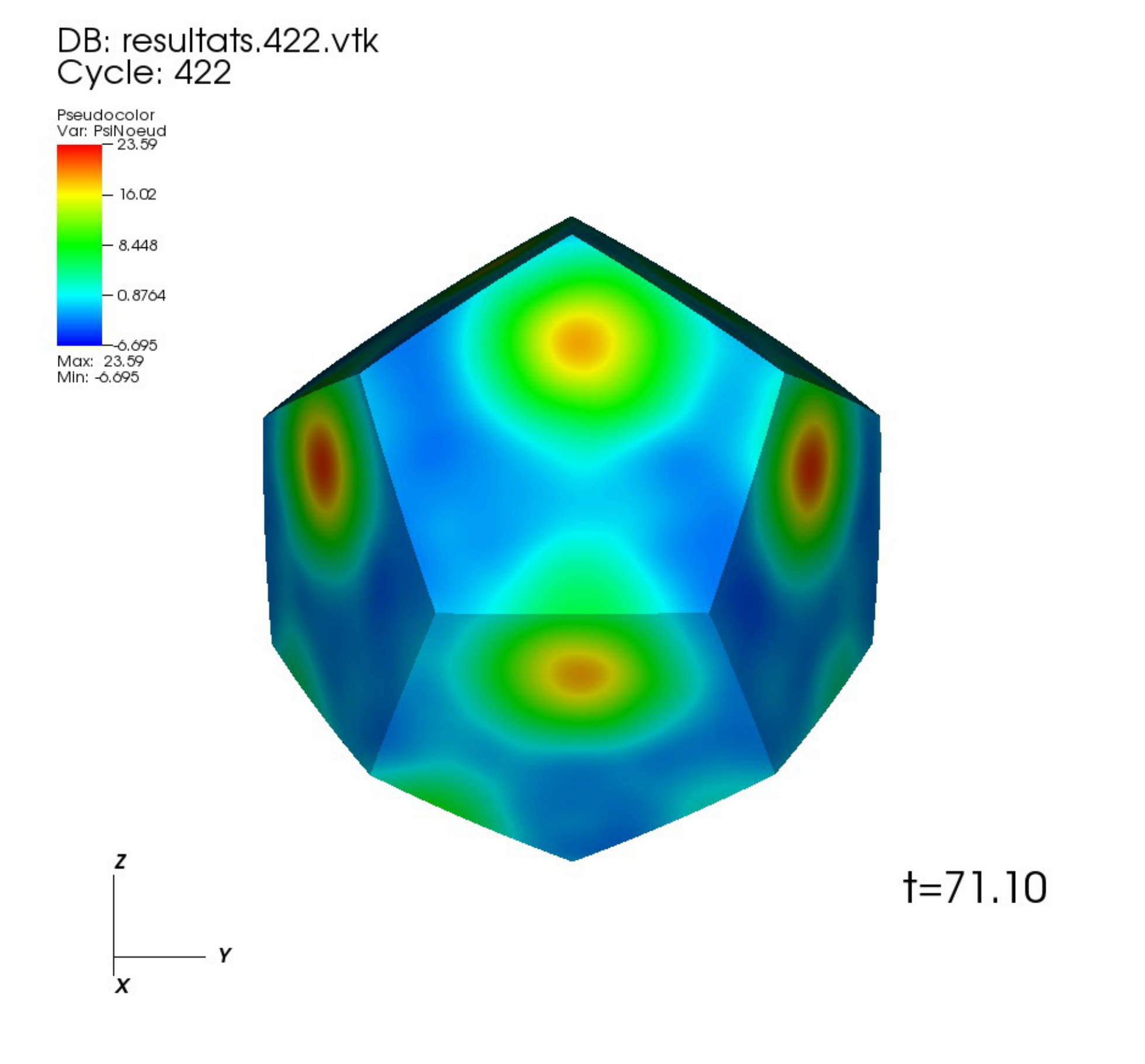}
\includegraphics[scale=0.085]{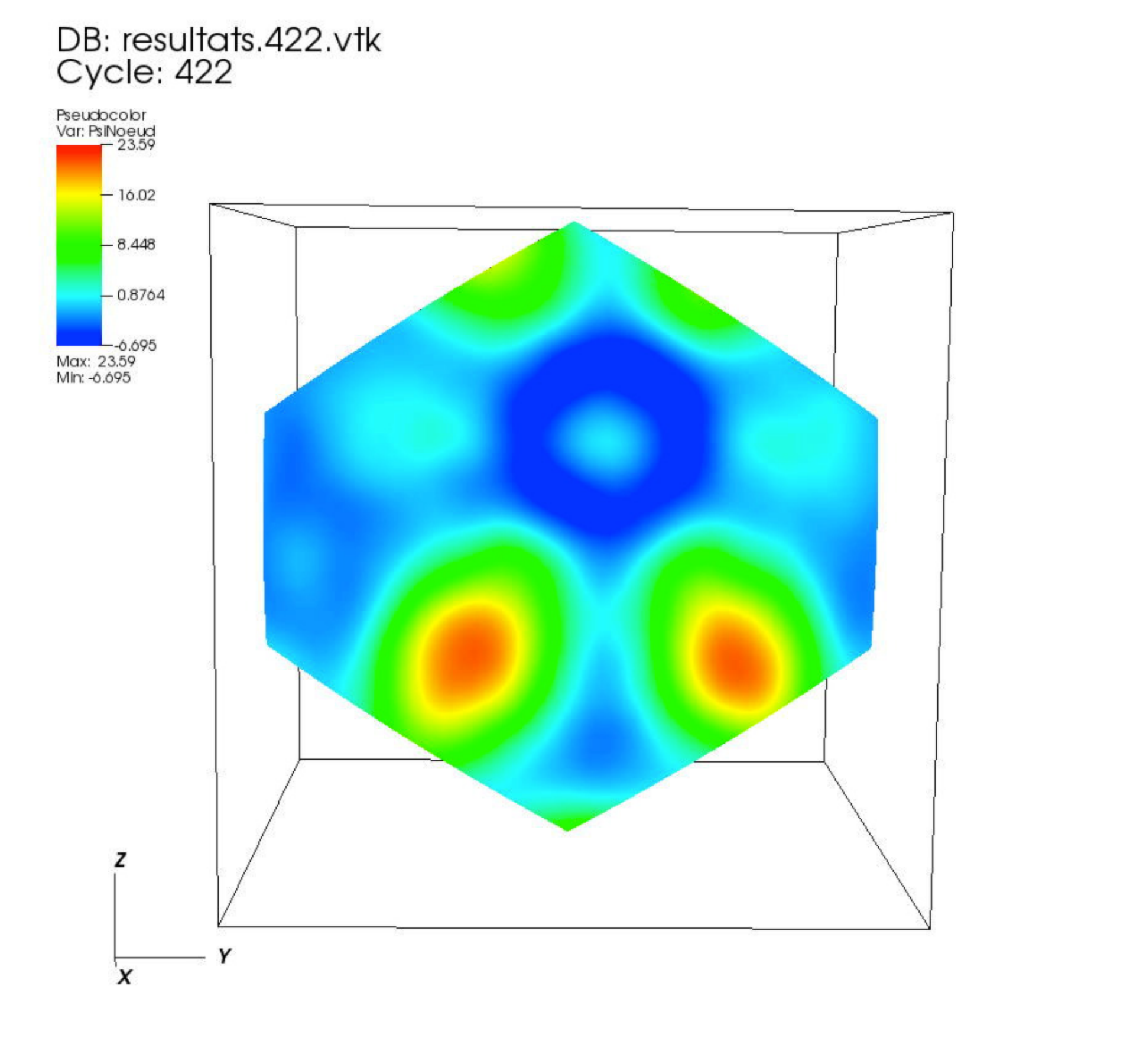}
\caption{\label{InitExc} The solution $\psi(t,X)$ with the initial data $Init_{exc}$ at $t=0$, $t=2.65$, $t=14.65$, $t=24.5$, $t=33.45$, $t=71.10$.}
\end{center}
\end{figure}
\begin{figure}[!h]
\begin{center}
\includegraphics[scale=0.085]{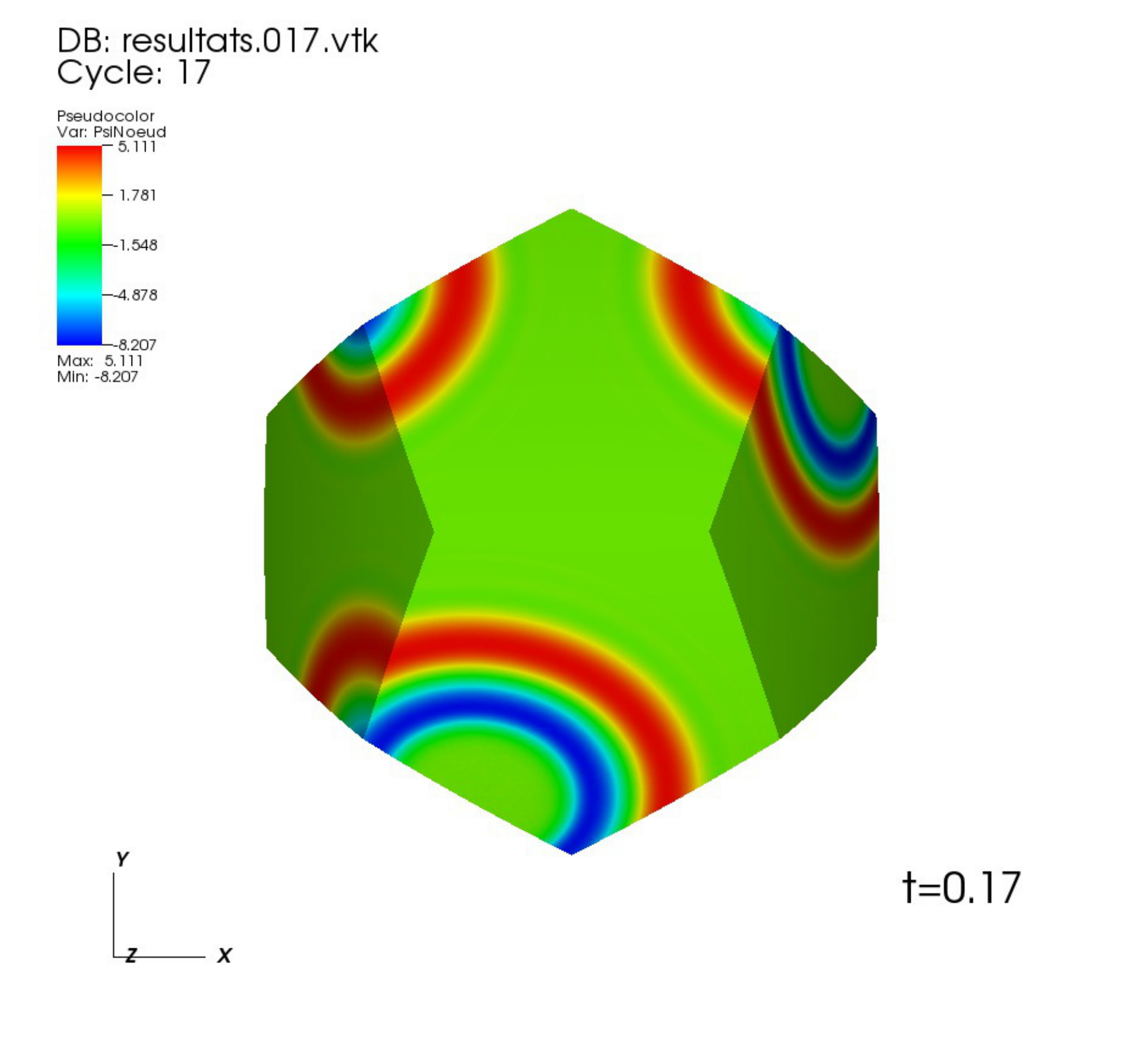}
\includegraphics[scale=0.085]{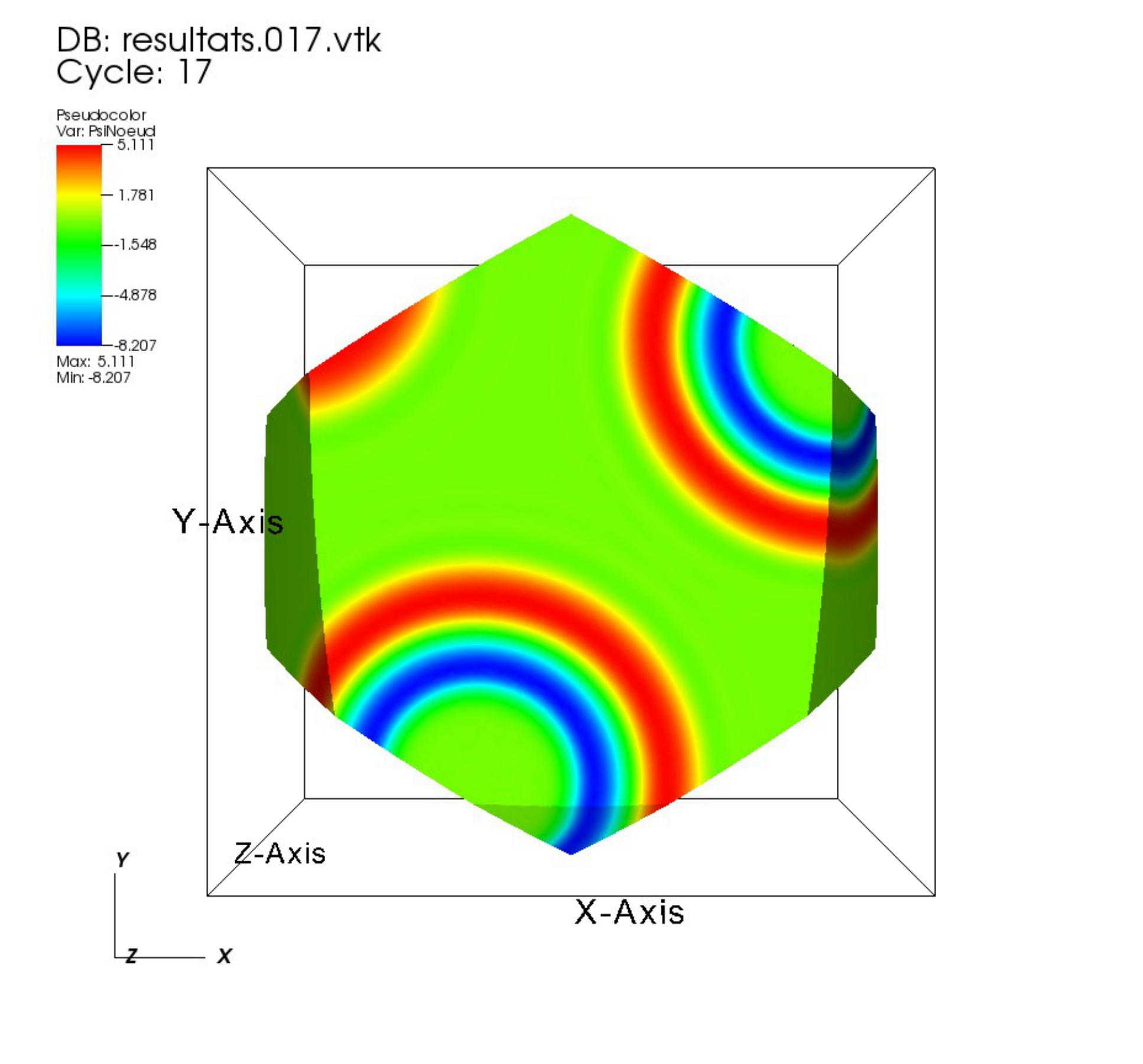}
\includegraphics[scale=0.085]{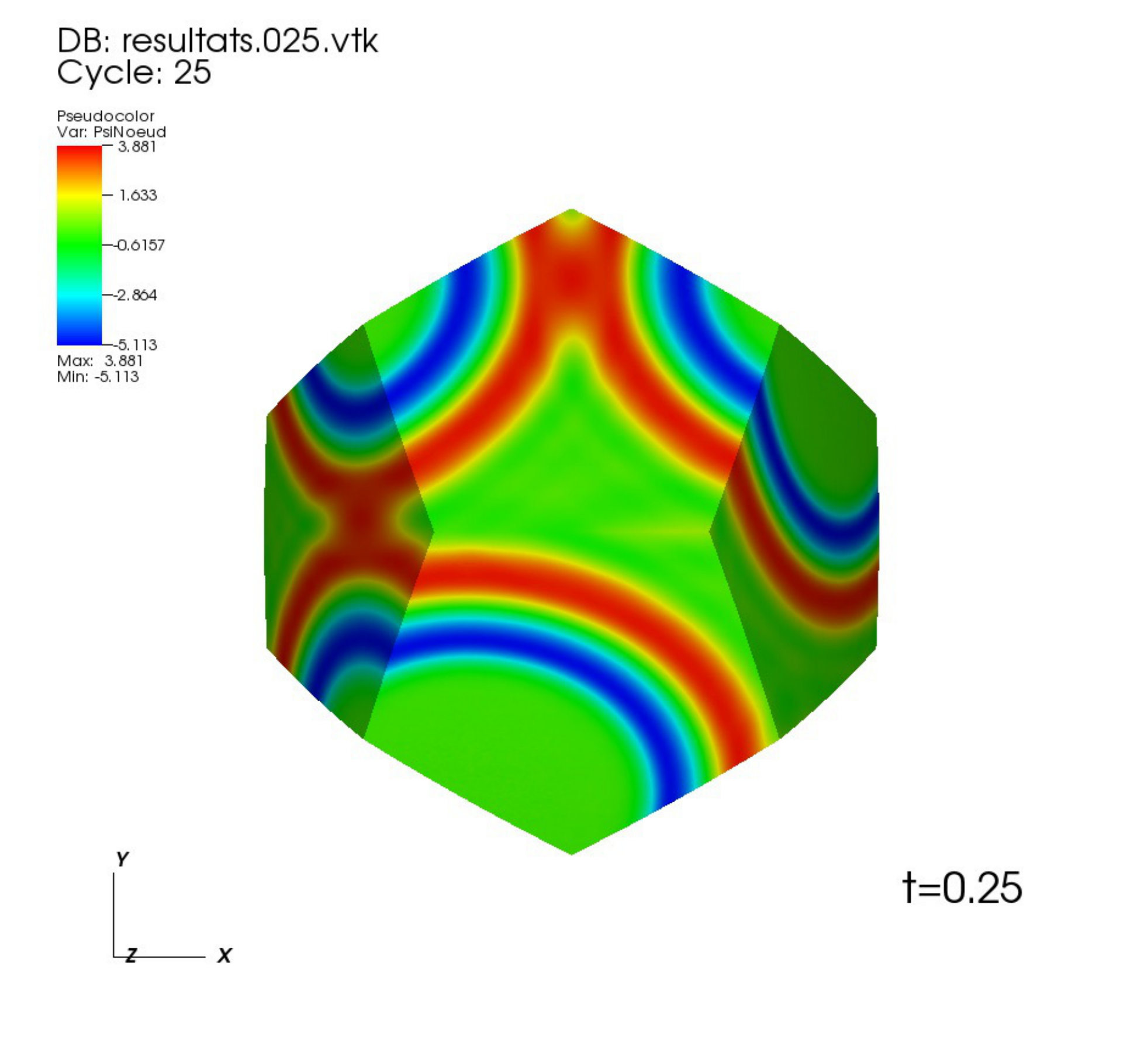}
\includegraphics[scale=0.085]{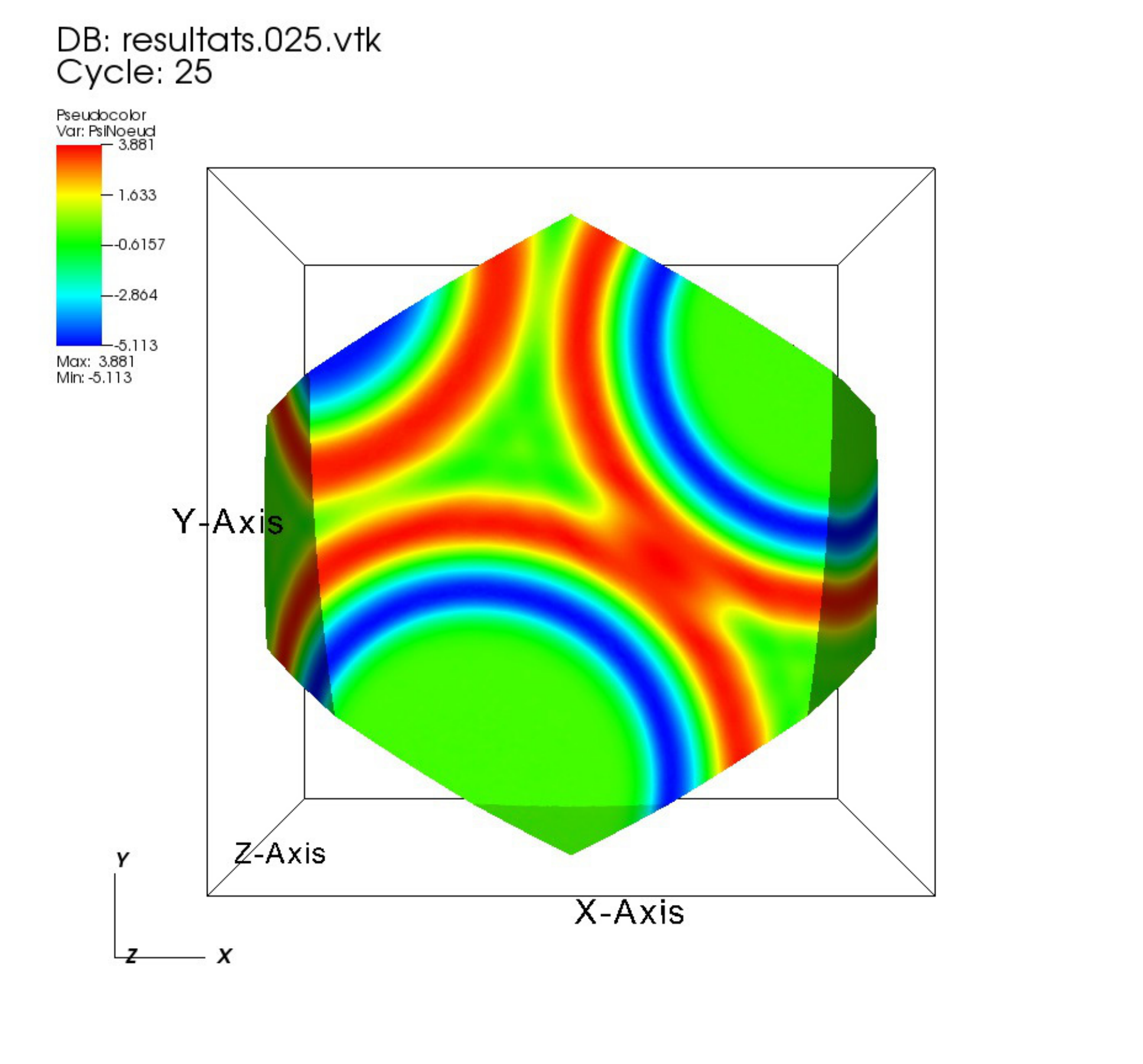}
\includegraphics[scale=0.085]{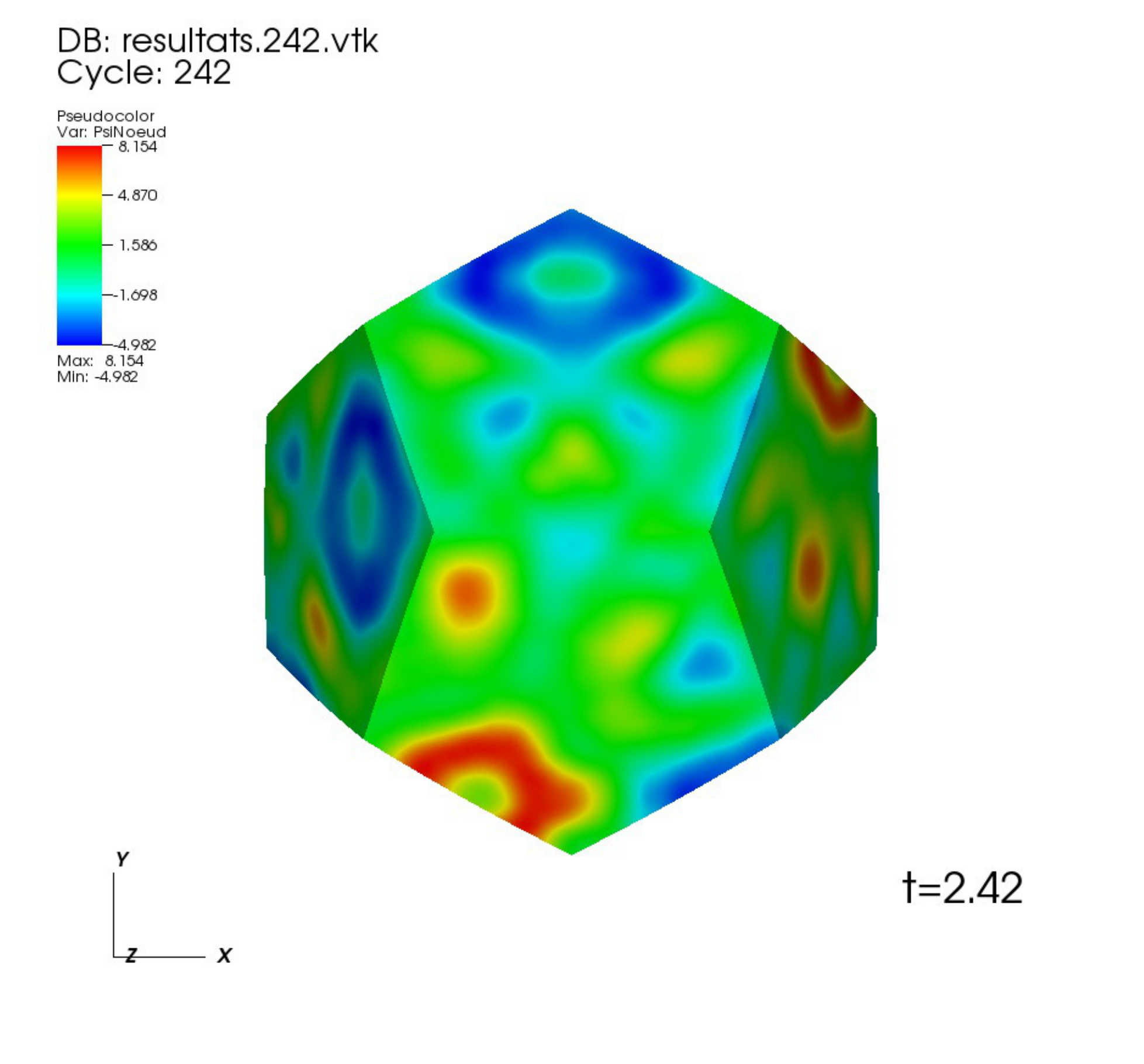}
\includegraphics[scale=0.085]{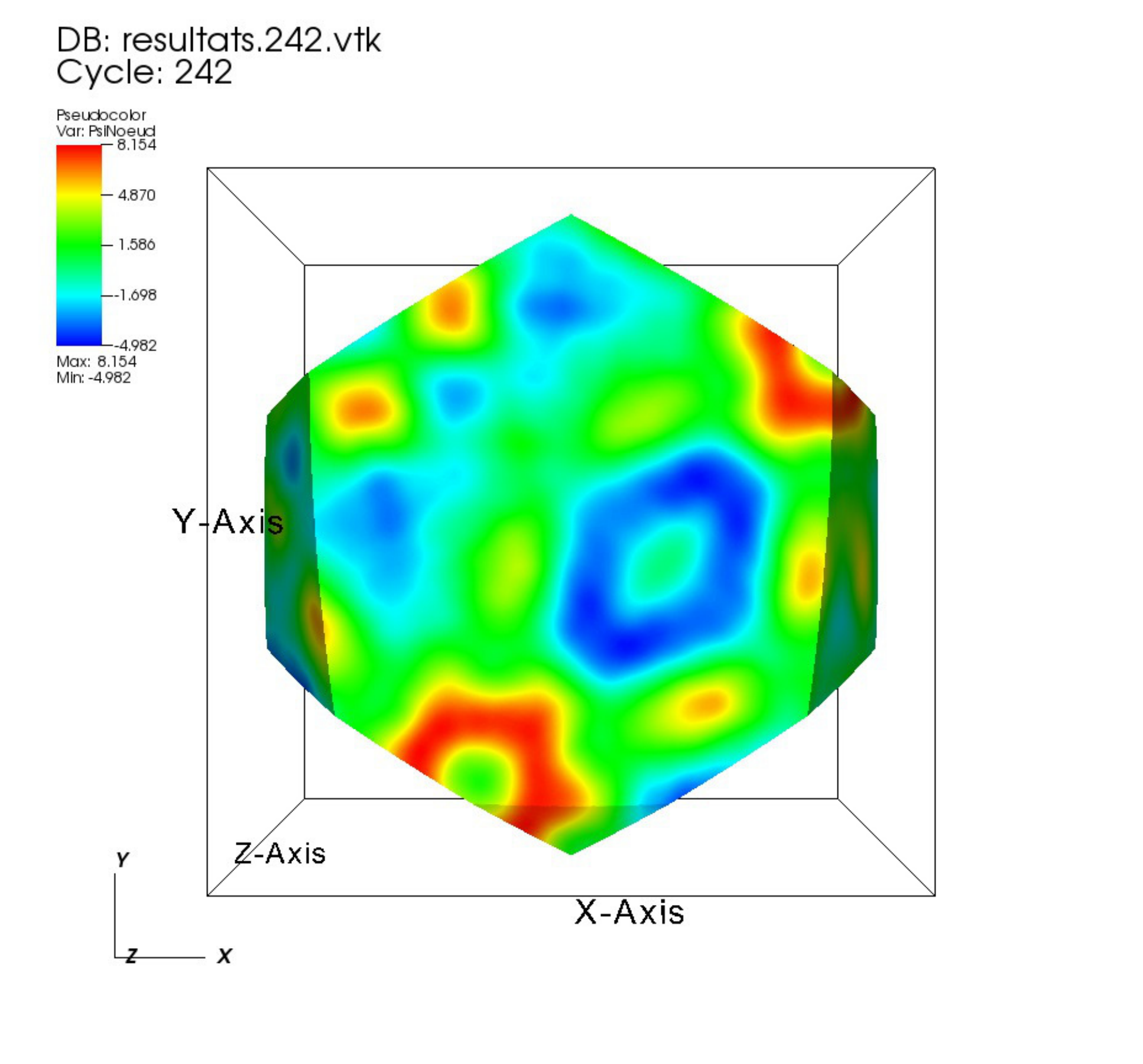}
\includegraphics[scale=0.085]{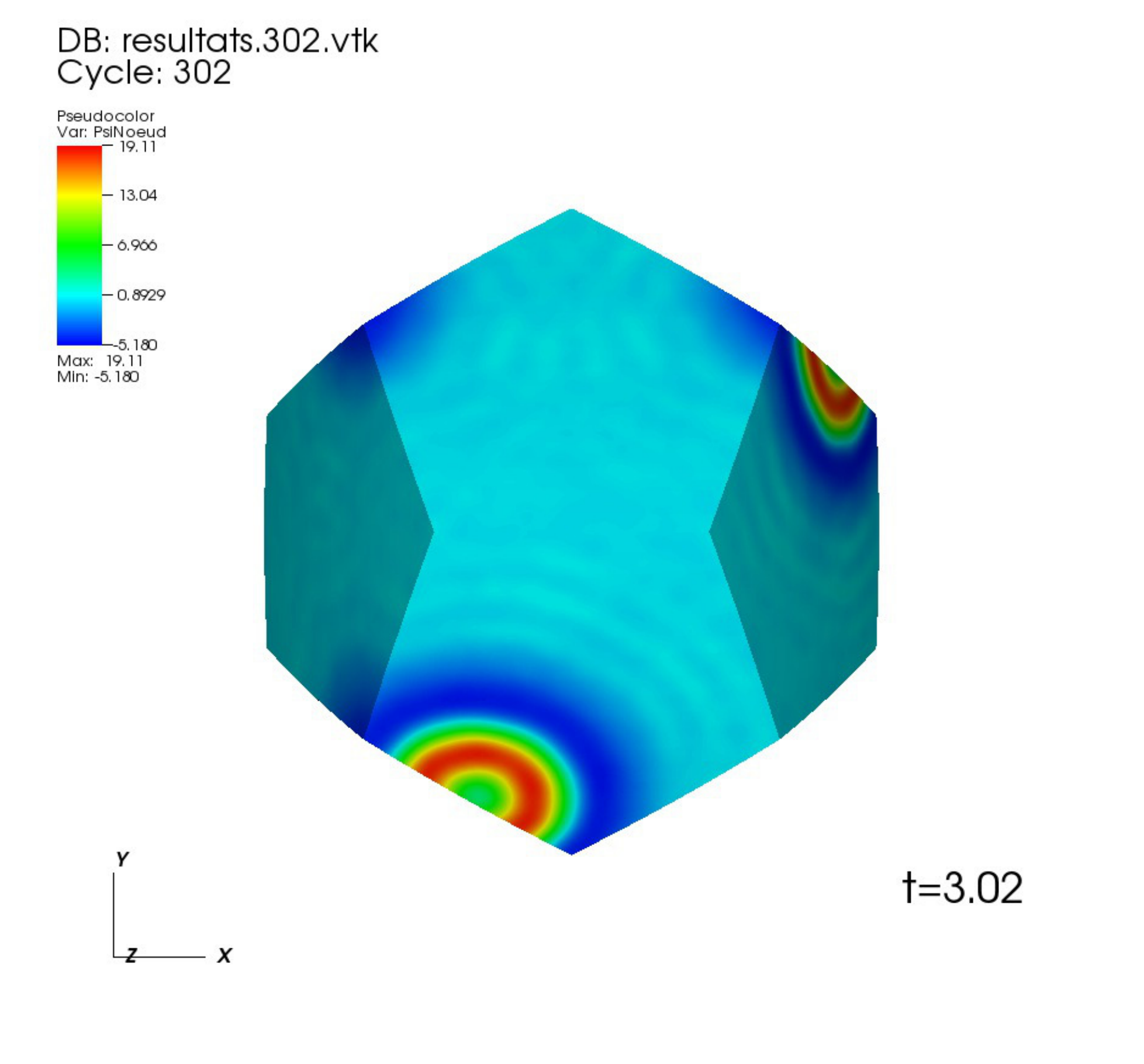}
\includegraphics[scale=0.085]{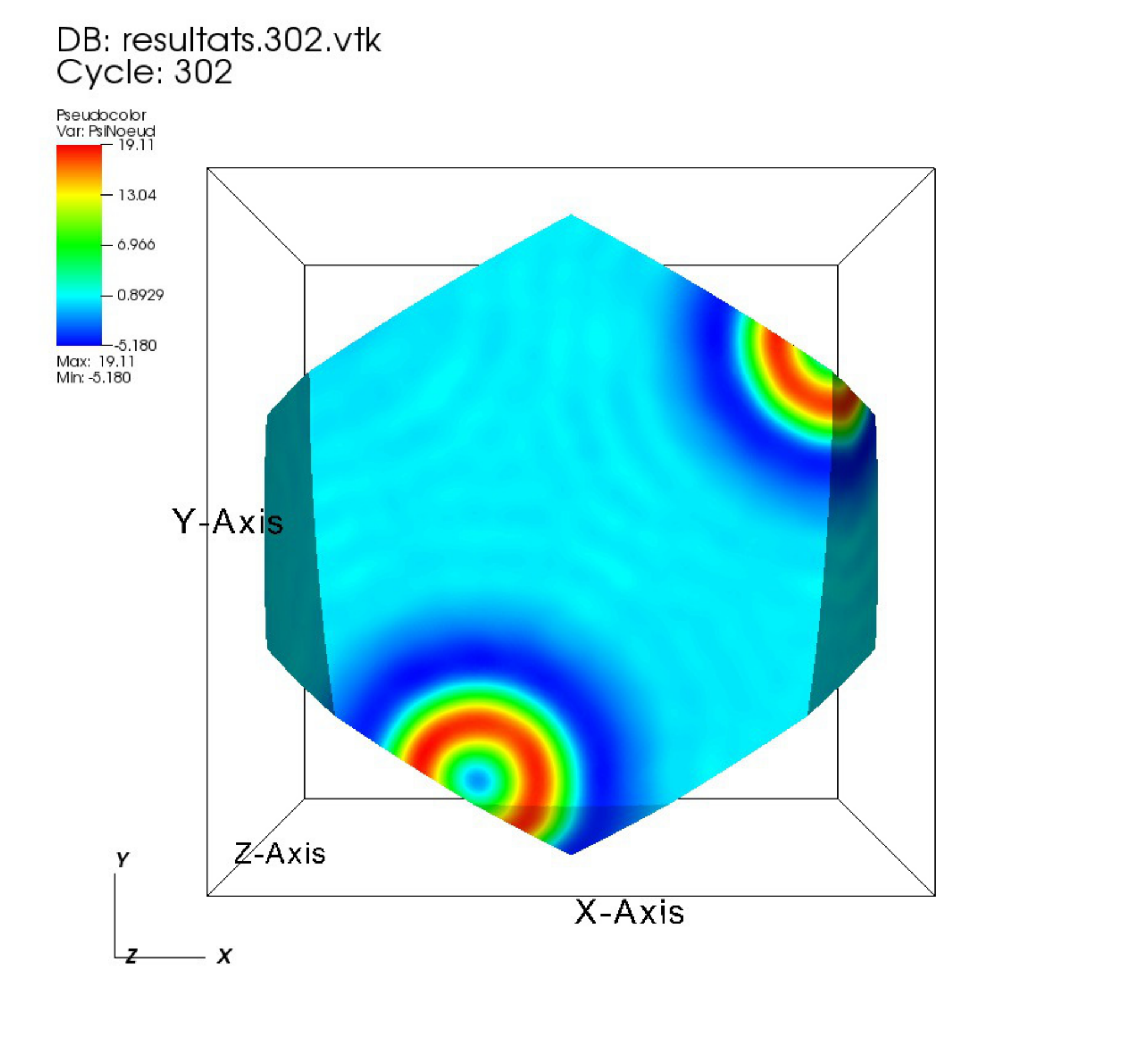}
\includegraphics[scale=0.085]{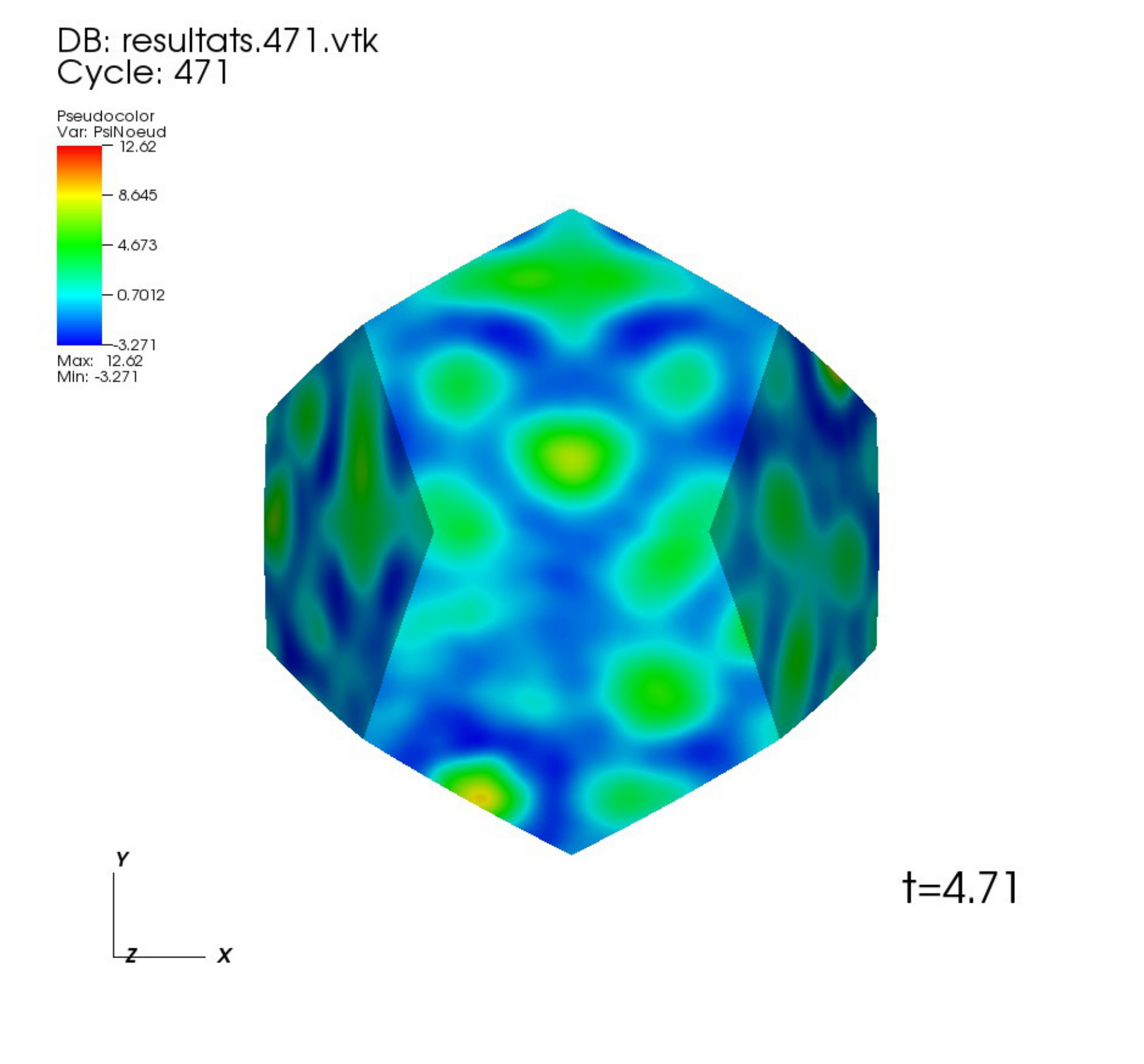}
\includegraphics[scale=0.085]{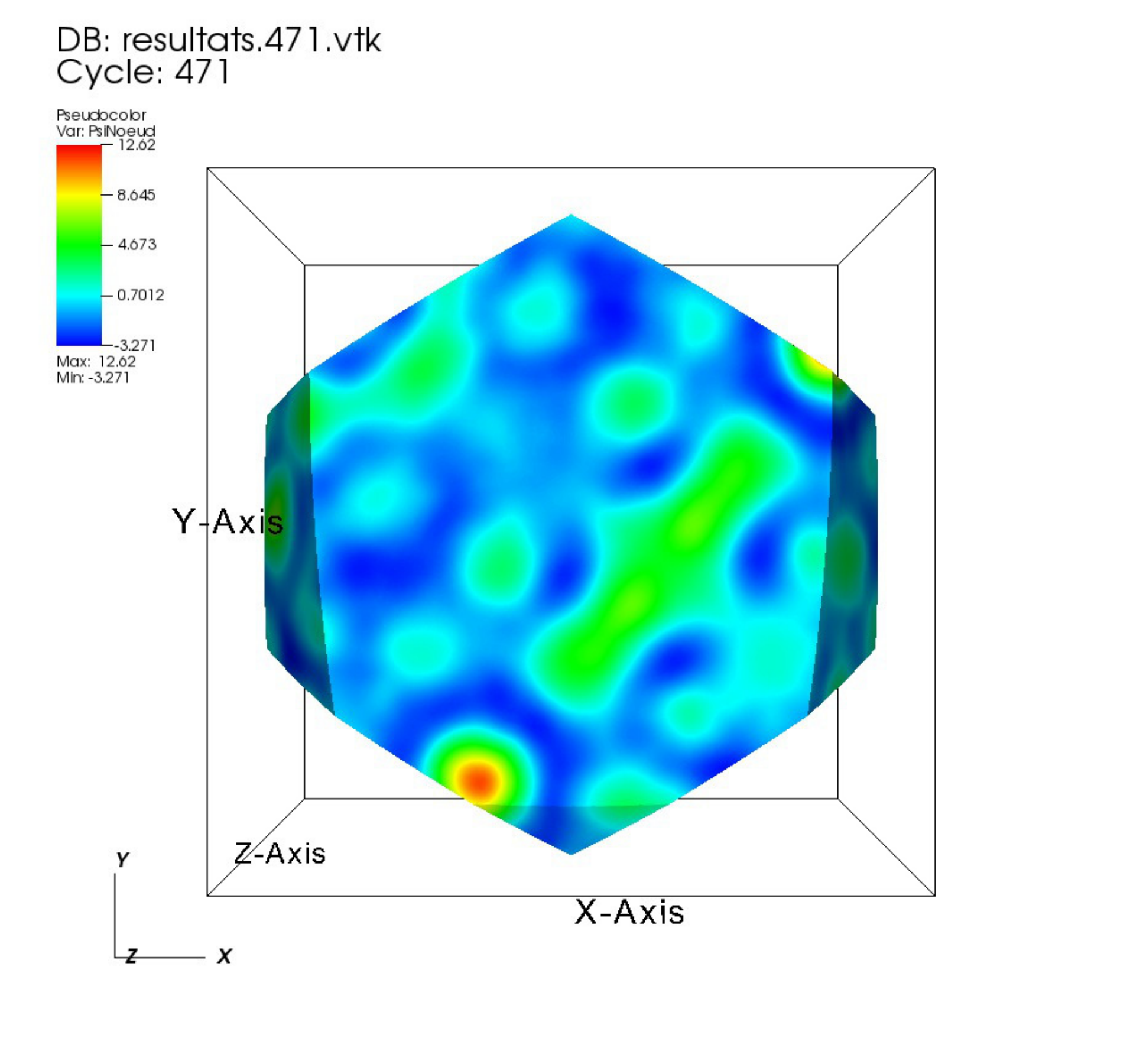}
\includegraphics[scale=0.085]{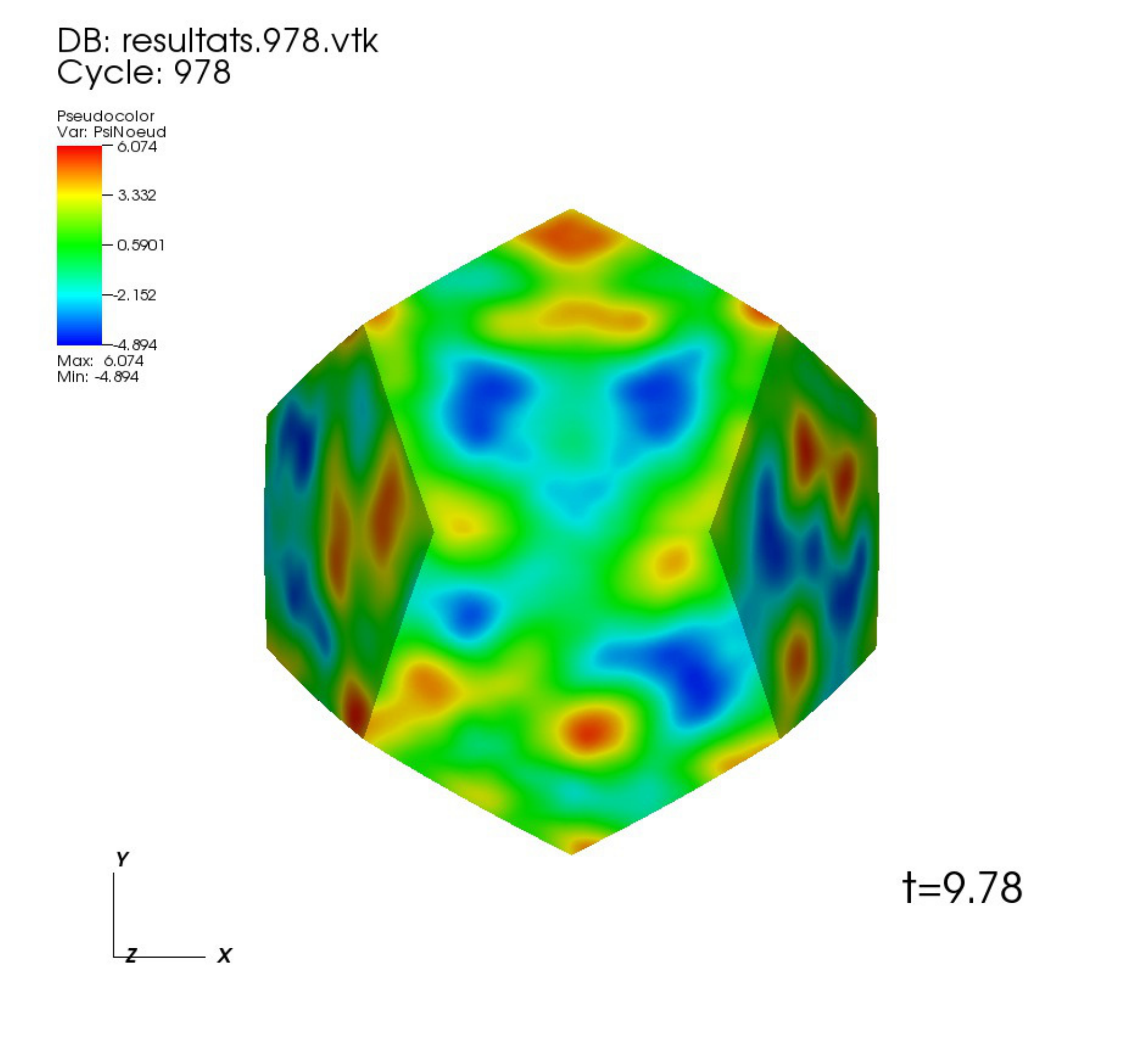}
\includegraphics[scale=0.085]{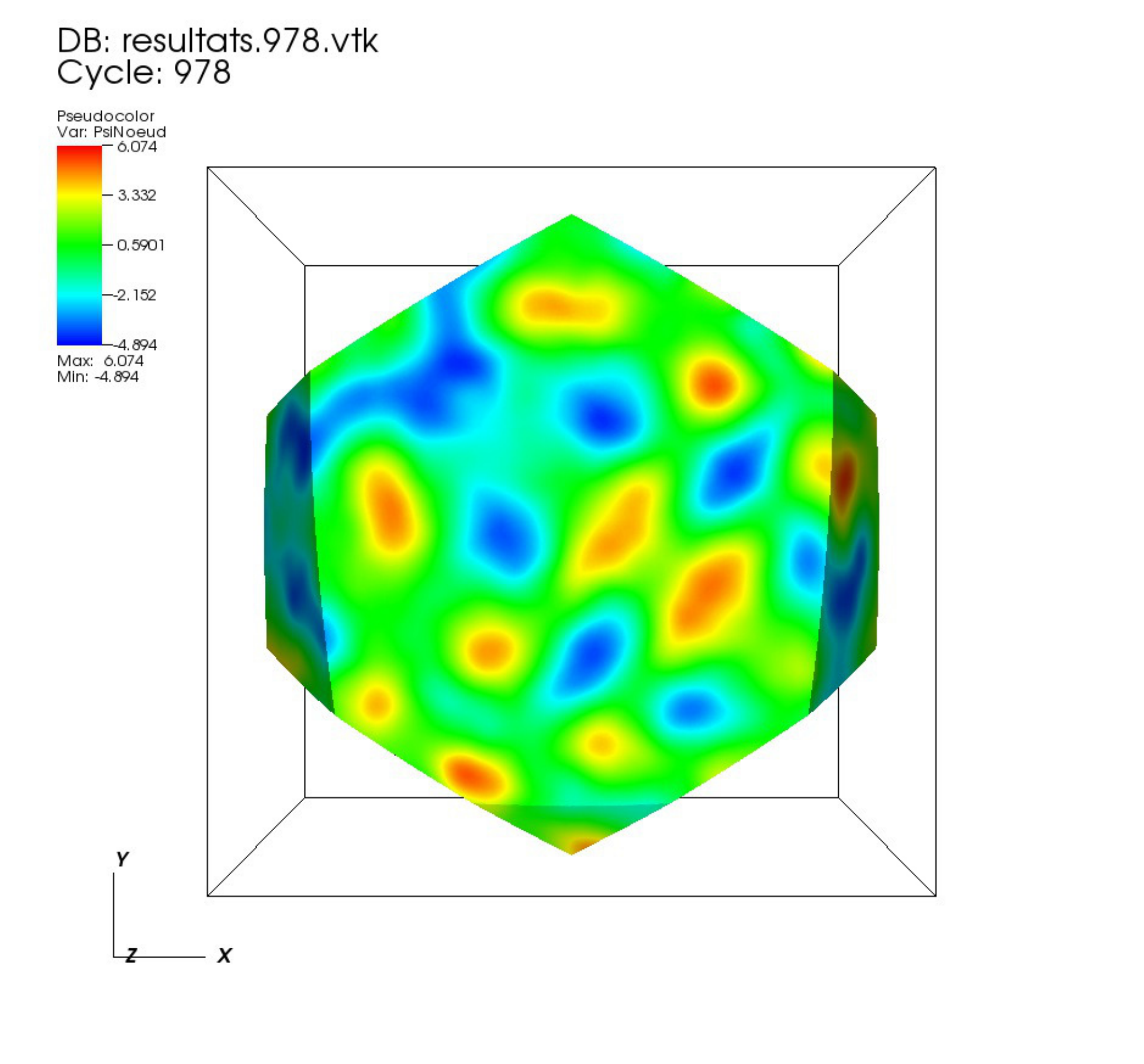}
\caption{\label{InitSum2} The solution $\psi(t,X)$ with the initial data $Init_{sum}$ at $t=0.17$, $t=0.25$, $t=2.42$, $t=3.02$, $t=4.71$, $t=9.78$.}
\end{center}
\end{figure}
\Fref{InitCentr}, \Fref{InitExc} and \Fref{InitSum2} contain some characteristic pictures for a given initial data. At each time there are a view of the solution on $\partial \mathcal{F}_v$, and, in a bounding box, a view on a cutting plane. On \Fref{InitExc} we can observe that the equivalence relation is satisfied. Progressively, equivalent faces are in the support of the solution and we have $\psi(t,x)=\psi(t,y)\neq 0$ for any equivalent points $x$ and $y$ of $\partial \mathcal{F}_v $. On \Fref{InitSum2} the cutting plane is the same as in \Fref{InitSum}. At $t=0.17$ we can observe that another non zero zone appears at the top left: it's around an equivalent vertex. The two initial non zero zones stay complementary of each other, as it can be seen on all the pictures.

Some movies are available on this personal home-page (http://www.math.u-bordeaux1.fr/$\sim$mbachelo/PDS.html).
\subsection {Eigenvalues}
We test our scheme in the time domain, by looking for the eigenvalues that are explicitly known (see \cite{Aur}, \cite{Ike}, \cite{Lach}, \cite{Leh-W-U-G-Lu}). Since the Laplace-Beltrami operator $\Delta_{\mathbf{K}}$ on PDS is a non positive, self-adjoint elliptic operator on a compact manifold, its spectrum is a discrete set of eigenvalues $-q^2\leq 0$,  and by the Hilbert-Schmidt theorem (see {\it e.g} Theorem 6.16 in \cite{Hilbert}), there exists an orthonormal basis in $L^2(\mathbf{K})$, formed of eigenfunctions $\left(\psi_q\right)_q\subset H^{\infty}(\mathbf{K})$  associated to $q^2$, i.e.
\begin{equation*}
-\Delta_{\mathcal{F}_v} \psi_q=q^2\psi_q,\;\;\psi_q\in
W^{\infty}(\mathcal{F}_v).
  \label{}
\end{equation*}
One has:
\[q^2=\beta ^2 -1,\]
with $\beta \in \{1,13,21,25,31,33,37,41,43,45,49,51,53,55,57\} \cup \{2n+1,\, n\geq 30\}$, and we take $\psi_0=\frac{1}{2\sqrt\pi}$. Therefore any finite energy solution $\psi(t,X)$ of  $\partial_t^2
\psi-\Delta_{\mathbf{K}}\psi=0$ has an expansion of the  form $at+b+\sum_q
\left(c_qe^{iqt}+c'_qe^{-iqt}\right)\psi_q(X)$. More precisely, if we denote $<,>$ the scalar product in
$L^2(\mathbf{K})$, we write
\begin{eqnarray}
\fl \psi(t,X)=\frac{1}{4\pi}(<\partial_t\psi(0,.),1>t+<\psi(0,.),1>)\nonumber\\
+\sum_{q\neq 0} <\partial_t\psi(0,.),\psi_q>\frac{\sin qt}{q}\;\psi_q(x,y)+\;<\psi(0,.),\psi_q>\cos qt\; \psi_q(X). \nonumber
  \label{}
\end{eqnarray}
To compute the eigenvalues $q^2$ we investigate the Fourier transform in time of
the signal $\psi(t,X_0)$ in the case where $\partial_t\psi(0,X)=0$, for different $X_0$. Practically, during the time resolution of the equation, we store the
values of the solution at some vertices $M(X)$ for the discrete time $k\Delta t$, $N_i\leq k\leq
N_f$. We choose the initial step $N_i$ in order to the transient wave
is stabilized, that to say $N_i\Delta t$ is at least greater than the diameter
of $\mathcal{F}_v$, i.e. $N_i \Delta t\geq 2*d(0,S_1)=d(S_1,S_{14})=\arccos \left(\frac{3\sigma -2}{8}\right)\simeq 0.776279$.
Then we compute a DFT of $(\psi_h(k\Delta t,X))_{N_i\leq k \leq N_f}$
with the free licensed (GNU GPL) FFT library fftw3. After we search the values $j_{\max}$ for which the previous result $(\Psi_j(X))_{0,N_f-N_i+1} $ has a local maximum. The eigenvalues found by the algorithm are expressed as:
\begin{equation*}
q=\frac{2\pi}{(N_f-N_i+1)\Delta t}j_{\max}.
\end{equation*}
We have made tests by varying parameters such as: mesh, initial data, $N_i$ and $N_f$. With F089 mesh, in all cases we have obtained the 36 first consecutive eigenvalues with an relative error on $q^2$ of about $10^{-2}$ for the first nine values, and $7\;10^{-2} $ for the following. The values are the same with all our initial data: $Init_c$, $Init_{exc}$, $Init_{exc,\infty}$ and $Init_{sum}$. We have also performed some calculations  with a mesh having $135$ mesh vertices on edges of $\mathcal{F}_v$. In this case the error is smaller. It was expected for two reasons: the approximation of $\mathcal{F}_v$ is best, and also, in order to respect the CFL condition \eref{CFL}, we must take a smaller time step. Therefore $\psi(t,X)$ is better approximated. In the next table we present the results for the nine first eigenvalues for this last mesh.
\[
\fl \begin{array}{|l|l|l|l||l|l|l|l|}
  \hline
  \beta& q^2 & \mbox{result}&\mbox{relative error}& \beta & q^2 & \mbox{result}&\mbox{relative error}\\
\hline
  13 &        168 &  167.6126  &   2.306\;10^{-3}&37 &       1368  & 1373.049 &     3.691\;10^{-3}\\
  21 &        440 &  439.107 &    2.029\;10^{-3}&41  &      1680  & 1687.761 &     4.619\;10^{-3}\\
  25 &        624  & 623.184 &     1.308\;10^{-3}&43  &      1848  & 1857.283  &    5.023\;10^{-3}\\
  31 &        960  & 959.561 &     4.571\;10^{-4}&45   &     2024  & 2034.916 &     5.393\;10^{-3}\\
  33  &      1088  & 1087.757 &     2.231\;10^{-4}&&&&\\
  \hline
\end{array}
\]

\section{Conclusion}
We note a  good agreement of the eigenvalues obtained by the spectral analysis of the transient waves that we have computed. We emphasize that this result is a strong evidence of the accuracy of our numerical approximation since the waves are rapidly oscillating due to the large value of the eigenvalues. Therefore we conclude that we have validated this computational method of the waves on the Poincar\'e dodecahedral space, and we plan to extend it, in the future, to the case of the non-linear waves on this manifold. The finite element method is much more appropriate to the nonlinear problems than the spectral method that has a very high complexity due to the nonlinearities. Among the nonlinear dynamics that could be treated, we can mention, in field theory on curved space-times, the Higgs field that has been studied for the Einstein universe $\RR_t\times{\mathcal S}^3$ by Y. Choquet-Bruhat and D. Christodoulou \cite{choquet}, and obeys the semilinear Klein-Gordon equation
$$
\boxvoid_g\Psi=F'(\Psi),\;\;F\in C^1(\RR),
$$
and also the more complicated Yang-Mills system (\cite{choquet}, \cite{chrusciel}) that has the form 
$$
\boxvoid_gF_{\mu\nu}=\mathcal{N}_{\mu\nu}(A,F,\partial A,\partial F),
$$
where the nonlinearities $\mathcal{N}_{\mu\nu}$ are cubic polynomials.
\ack
The author wishes to  thank J. Fresnel for his stimulating discussions on the Coxeter's book \cite{Cox}, R. Cools for having given me acces to the software CUBPACK.
\appendix 
\section*{Appendix A:} Description of $\mathcal{F}$ and $\mathcal{F}_v$.

\begin{enumerate}
\item Coordinates of the vertices $S_i$ of $\mathcal{F}$:
\[
 \begin{array}{ll}
S_1=\frac{1}{2\sqrt{2}}\left(\sigma^2,-\frac{1}{\sigma},\frac{1}{\sigma},-\frac{1}{\sigma}\right),&S_2=\frac{1}{2\sqrt{2}}\left(\sigma^2,1,\frac{1}{\sigma^2},0\right),\\
S_3=\frac{1}{2\sqrt{2}}\left(\sigma^2,-\frac{1}{\sigma},-\frac{1}{\sigma},\frac{1}{\sigma}\right),&S_4=\frac{1}{2\sqrt{2}}\left(\sigma^2,\frac{1}{\sigma},-\frac{1}{\sigma},-\frac{1}{\sigma}\right),\\
S_5=\frac{1}{2\sqrt{2}}\left(\sigma^2,0,-1,-\frac{1}{\sigma^2}\right),&S_6=\frac{1}{2\sqrt{2}}\left(\sigma^2,\frac{1}{\sigma},\frac{1}{\sigma},\frac{1}{\sigma}\right),\\
S_7=\frac{1}{2\sqrt{2}}\left(\sigma^2,-\frac{1}{\sigma^2},0,1\right),&S_8=\frac{1}{2\sqrt{2}}\left(\sigma^2,0,1,\frac{1}{\sigma^2}\right),\\
S_9=\frac{1}{2\sqrt{2}}\left(\sigma^2,-\frac{1}{\sigma},\frac{1}{\sigma},\frac{1}{\sigma}\right),&S_{10}=\frac{1}{2\sqrt{2}}\left(\sigma^2,\frac{1}{\sigma^2},0,1\right),\\
S_{11}=\frac{1}{2\sqrt{2}}\left(\sigma^2,0,1,-\frac{1}{\sigma^2}\right),&S_{12}=\frac{1}{2\sqrt{2}}\left(\sigma^2,-1,\frac{1}{\sigma^2},0\right),\\
S_{13}=\frac{1}{2\sqrt{2}}\left(\sigma^2,-\frac{1}{\sigma^2},0,-1\right),&S_{14}=\frac{1}{2\sqrt{2}}\left(\sigma^2,\frac{1}{\sigma},-\frac{1}{\sigma},\frac{1}{\sigma}\right),\\
S_{15}=\frac{1}{2\sqrt{2}}\left(\sigma^2,\frac{1}{\sigma^2},0,-1\right),&S_{16}=\frac{1}{2\sqrt{2}}\left(\sigma^2,-\frac{1}{\sigma},-\frac{1}{\sigma},-\frac{1}{\sigma}\right),\\
S_{17}=\frac{1}{2\sqrt{2}}\left(\sigma^2,\frac{1}{\sigma},\frac{1}{\sigma},-\frac{1}{\sigma}\right),&S_{18}=\frac{1}{2\sqrt{2}}\left(\sigma^2,-1,-\frac{1}{\sigma^2},0\right),\\
S_{19}=\frac{1}{2\sqrt{2}}\left(\sigma^2,1,-\frac{1}{\sigma^2},0\right),&S_{20}=\frac{1}{2\sqrt{2}}\left(\sigma^2,0,-1,\frac{1}{\sigma^2}\right).
\end{array}
\]

\item Images by the Clifford translation $g_i$ of the face $F_i$ of $\mathcal{F}$ and of its edges. \\
$g_1$ maps $F_1$ to $F_7$, and we have for the vertices and the edges:\\
$g_1(S_{3}) =S_{6},\quad g_1( S_{18})=S_{8},\quad g_1(S_{16})=S_{11},\quad g_1(S_{5} )=S_{17},\quad g_1(S_{20})=S_{2}$, 

\noindent
$g_1(\wideparen{S_{3}S_{18}})=\wideparen{S_{6}S_{8}},\; g_1(\wideparen{S_{18}S_{16}})=\wideparen{S_{8}S_{11}},\; g_1(\wideparen{S_{16}S_{5}})=\wideparen{S_{11}S_{17}},\; g_1(\wideparen{S_{5}S_{20}})=\wideparen{S_{17}S_{2}},$

\noindent
 $g_1(\wideparen{S_{20}S_{3}})=\wideparen{S_{2}S_{6}}$.\\

$g_2$ maps $F_2$ to $F_8$, and we have for the vertices and the edges:\\
$g_2(S_{18}) =S_{15},\quad g_2( S_{12})=S_{17},\quad g_2(S_{9})=S_{2},\quad g_2(S_{7} )=S_{19},\quad g_2(S_{3})=S_{4}$, 

\noindent
$g_2(\wideparen{S_{18}S_{12}})=\wideparen{S_{15}S_{17}},\; g_2(\wideparen{S_{17}S_{2}})=\wideparen{S_{8}S_{11}},\; g_2(\wideparen{S_{9}S_{7}})=\wideparen{S_{2}S_{19}},\; g_2(\wideparen{S_{7}S_{3}})=\wideparen{S_{19}S_{4}},$

\noindent
$ g_2(\wideparen{S_{3}S_{18}})=\wideparen{S_{4}S_{15}}$.\\

$g_3$ maps $F_3$ to $F_9$, and we have for the vertices and the edges:\\
$ g_3(S_{3}) =S_{1},\quad g_3( S_{7})=S_{11},\quad g_3(S_{10})=S_{17},\quad g_3(S_{14} )=S_{15},\quad g_3(S_{20})=S_{13}$, 

\noindent
$g_3(\wideparen{S_{3}S_{7}})=\wideparen{S_{1}S_{11}},\; g_3(\wideparen{S_{7}S_{10}})=\wideparen{S_{11}S_{17}},\; g_3(\wideparen{S_{10}S_{14}})=\wideparen{S_{17}S_{15}},$\\ 
$g_3(\wideparen{S_{14}S_{20}})=\wideparen{S_{15}S_{13}},\;g_3(\wideparen{S_{20}S_{3}})=\wideparen{S_{13}S_{1}}$.\\

$g_4$ maps $F_4$ to $F_{10}$, and we have for the vertices and the edges:\\
$g_4(S_{20}) =S_{9},\quad g_4( S_{14})=S_{8},\quad g_4(S_{19})=S_{11},\quad g_4(S_{4} )=S_{1},\quad g_4(S_{5})=S_{12}$, 

\noindent
$g_4(\wideparen{S_{20}S_{14}})=\wideparen{S_{9}S_{8}},\; g_4(\wideparen{S_{14}S_{19}})=\wideparen{S_{8}S_{11}},\; g_4(\wideparen{S_{19}S_{4}})=\wideparen{S_{11}S_{1}},\; g_4(\wideparen{S_{4}S_{5}})=\wideparen{S_{1}S_{12}},$

\noindent
$g_4(\wideparen{S_{5}S_{20}})=\wideparen{S_{12}S_{9}}$.\\

$g_5$ maps $F_5$ to $F_{11}$, and we have for the vertices and the edges:\\
$g_5(S_{5}) =S_{10},\quad g_5( S_{4})=S_{6},\quad g_5(S_{15})=S_{8},\quad g_5(S_{13} )=S_{9},\quad g_5(S_{16})=S_{7}$, 

\noindent
$g_5(\wideparen{S_{5}S_{4}})=\wideparen{S_{10}S_{6}},\; g_5(\wideparen{S_{4}S_{15}})=\wideparen{S_{6}S_{8}},\; g_5(\wideparen{S_{15}S_{13}})=\wideparen{S_{8}S_{9}},\; g_5(\wideparen{S_{13}S_{16}})=\wideparen{S_{9}S_{7}},$

\noindent
$g_5(\wideparen{S_{16}S_{5}})=\wideparen{S_{7}S_{10}}$\\

$g_6$ maps $F_6$ to $F_{12}$, and we have for the vertices and the edges:\\
$g_6(S_{16}) =S_{19},\quad g_6( S_{13})=S_{2},\quad g_6(S_{1})=S_{6},\quad g_6(S_{12} )=S_{10},\quad g_6(S_{18})=S_{14}$, 

\noindent
$g_6(\wideparen{S_{16}S_{13}})=\wideparen{S_{19}S_{2}},\; g_6(\wideparen{S_{13}S_{1}})=\wideparen{S_{2}S_{6}},\; g_6(\wideparen{S_{1}S_{12}})=\wideparen{S_{6}S_{10}},$\\
$g_6(\wideparen{S_{12}S_{18}})=\wideparen{S_{10}S_{14}},\;g_6(\wideparen{S_{18}S_{16}})=\wideparen{S_{14}S_{19}}$.\\

\item $F_i^b$, the set of all barycenters in $\RR^4$ of the vertices of $F_i$, is included in a $2$-plane of $\RR^4$. The equations of these $2$-planes are:
\[
\fl \begin{array}{ll}
F_1^b &\subset \left\{(x_0,x,y,z) \in \RR^4 ,\;  x_0=\frac{\sigma^2}{2\sqrt{2}}, \; -\frac{1}{\sigma} x-y=\frac{x_0}{\sigma^2}\right\},\\
F_7^b &\subset \left\{(x_0,x,y,z) \in \RR^4 ,\;  x_0=\frac{\sigma^2}{2\sqrt{2}}, \; +\frac{1}{\sigma} x+y=\frac{x_0}{\sigma^2}\right\},\\
F_2^b &\subset \left\{(x_0,x,y,z) \in \RR^4 ,\;  x_0=\frac{\sigma^2}{2\sqrt{2}}, \; -x+\frac{1}{\sigma} z=\frac{x_0}{\sigma^2}\right\},\\
F_8^b &\subset \left\{(x_0,x,y,z) \in \RR^4 ,\;  x_0=\frac{\sigma^2}{2\sqrt{2}}, \; +x-\frac{1}{\sigma} z=\frac{x_0}{\sigma^2}\right\},\\
F_3^b &\subset \left\{(x_0,x,y,z) \in \RR^4 ,\;  x_0=\frac{\sigma^2}{2\sqrt{2}}, \; -\frac{1}{\sigma} y+z=\frac{x_0}{\sigma^2}\right\},\\
F_9^b &\subset \left\{(x_0,x,y,z) \in \RR^4 ,\;  x_0=\frac{\sigma^2}{2\sqrt{2}}, \;+ \frac{1}{\sigma} y-z=\frac{x_0}{\sigma^2}\right\},\\
F_4^b &\subset \left\{(x_0,x,y,z) \in \RR^4 ,\;  x_0=\frac{\sigma^2}{2\sqrt{2}}, \; +\frac{1}{\sigma} x-y=\frac{x_0}{\sigma^2}\right\},\\
F_{10}^b &\subset \left\{(x_0,x,y,z) \in \RR^4 ,\;  x_0=\frac{\sigma^2}{2\sqrt{2}}, \; -\frac{1}{\sigma} x+y=\frac{x_0}{\sigma^2}\right\},\\
F_5^b &\subset \left\{(x_0,x,y,z) \in \RR^4 ,\;  x_0=\frac{\sigma^2}{2\sqrt{2}}, \; -\frac{1}{\sigma} y-z=\frac{x_0}{\sigma^2}\right\},\\
F_{11}^b &\subset \left\{(x_0,x,y,z) \in \RR^4 ,\;  x_0=\frac{\sigma^2}{2\sqrt{2}}, \; +\frac{1}{\sigma} y+z=\frac{x_0}{\sigma^2}\right\},\\
F_6^b &\subset \left\{(x_0,x,y,z) \in \RR^4 ,\;  x_0=\frac{\sigma^2}{2\sqrt{2}}, \; -x-\frac{1}{\sigma} z=\frac{x_0}{\sigma^2}\right\},\\
F_{12}^b &\subset \left\{(x_0,x,y,z) \in \RR^4 ,\;  x_0=\frac{\sigma^2}{2\sqrt{2}}, \; +x+\frac{1}{\sigma} z=\frac{x_0}{\sigma^2}\right\}.
\end{array}\]

\item So, after having normalized the points of $F_i^b$, we get that $F_i$ is included in an hyperplane of $\RR^4$. The equations of these $3$-planes are:
\[
\fl \begin{array}{ll}
F_1 &\subset \left\{(x_0,x,y,z) \in \mathcal{S}^3 ,\; -\frac{1}{\sigma} x-y=\frac{x_0}{\sigma^2}\right\},\\
F_7 &\subset \left\{(x_0,x,y,z) \in \mathcal{S}^3,\; +\frac{1}{\sigma} x+y=\frac{x_0}{\sigma^2}\right\},\\
F_2& \subset \left\{(x_0,x,y,z) \in \mathcal{S}^3 ,\; -x+\frac{1}{\sigma}z=\frac{x_0}{\sigma^2}\right\},\\
F_8 &\subset \left\{(x_0,x,y,z) \in \mathcal{S}^3 ,\;+ x-\frac{1}{\sigma}z=\frac{x_0}{\sigma^2}\right\},\\
F_3 &\subset \left\{(x_0,x,y,z) \in \mathcal{S}^3 ,\; -\frac{1}{\sigma}y+z=\frac{x_0}{\sigma^2}\right\},\\
F_9 &\subset \left\{(x_0,x,y,z) \in \mathcal{S}^3 ,\; +\frac{1}{\sigma}y-z=\frac{x_0}{\sigma^2}\right\},\\
F_4 &\subset \left\{(x_0,x,y,z) \in \mathcal{S}^3 ,\;+ \frac{1}{\sigma} x-y=\frac{x_0}{\sigma^2}\right\},\\
F_{10} &\subset \left\{(x_0,x,y,z) \in \mathcal{S}^3 ,\; -\frac{1}{\sigma} x+y=\frac{x_0}{\sigma^2}\right\},\\
F_5 &\subset \left\{(x_0,x,y,z) \in \mathcal{S}^3 ,\; -\frac{1}{\sigma}y-z=\frac{x_0}{\sigma^2}\right\},\\
F_{11} &\subset \left\{(x_0,x,y,z) \in \mathcal{S}^3 ,\; +\frac{1}{\sigma}y+z=\frac{x_0}{\sigma^2}\right\},\\
F_6& \subset \left\{(x_0,x,y,z) \in \mathcal{S}^3 ,\; -x-\frac{1}{\sigma}z=\frac{x_0}{\sigma^2}\right\},\\
F_{12} &\subset \left\{(x_0,x,y,z) \in \mathcal{S}^3 ,\; +x+\frac{1}{\sigma}z=\frac{x_0}{\sigma^2}\right\}.
\end{array}\]

\item We deduce from the previous items that $F_{i,v}^b$, the set of all barycenters in $\RR^3$ of the vertices of $F_{i,v}$, is included in a plane of $\RR^3$. The equations of these planes are:
\[
\fl \begin{array}{ll}
F_{1,v}^b &\subset \left\{(x,y,z) \in \RR^3 ,\;  -\frac{1}{\sigma} x-y=\frac{1}{2\sqrt{2}}\right\},\\
F_{7,v}^b &\subset \left\{(x,y,z) \in \RR^3 ,\;  +\frac{1}{\sigma} x+y=\frac{1}{2\sqrt{2}}\right\},\\
F_{2,v}^b &\subset \left\{(x,y,z) \in \RR^3 ,\; -x+\frac{1}{\sigma} z=\frac{1}{2\sqrt{2}}\right\},\\
F_{8,v}^b &\subset \left\{(x,y,z) \in \RR^3 ,\; +x-\frac{1}{\sigma} z=\frac{1}{2\sqrt{2}}\right\},\\
F_{3,v}^b &\subset \left\{(x,y,z) \in \RR^3 ,\; -\frac{1}{\sigma} y+z=\frac{1}{2\sqrt{2}}\right\},\\
F_{9,v}^b &\subset \left\{(x,y,z) \in \RR^3 ,\; +\frac{1}{\sigma} y-z=\frac{1}{2\sqrt{2}}\right\},\\
F_{4,v}^b &\subset \left\{(x,y,z) \in \RR^3 ,\; +\frac{1}{\sigma} x-y=\frac{1}{2\sqrt{2}}\right\},\\
F_{10,v}^b &\subset \left\{(x,y,z) \in \RR^3 ,\; -\frac{1}{\sigma} x+y=\frac{1}{2\sqrt{2}}\right\},\\
F_{5,v}^b &\subset \left\{(x,y,z) \in \RR^3 ,\; -\frac{1}{\sigma} y-z=\frac{1}{2\sqrt{2}}\right\},\\
F_{11,v}^b &\subset \left\{(x,y,z) \in \RR^3 ,\; +\frac{1}{\sigma} y+z=\frac{1}{2\sqrt{2}}\right\},\\
F_{6,v}^b &\subset \left\{(x,y,z) \in \RR^3 ,\;  -x-\frac{1}{\sigma} z=\frac{1}{2\sqrt{2}}\right\},\\
F_{12,v}^b &\subset \left\{(x,y,z) \in \RR^3 ,\;  +x+\frac{1}{\sigma} z=\frac{1}{2\sqrt{2}}\right\}.
\end{array}\]

\item As $x_0^2=1-x^2-y^2-z^2$, it follows from item (iv) that the faces $F_{i,v}$ of $\mathcal{F}_v $ are included in an ellipsoid.\\
$F_{1,v}$ and $F_{7,v}$ are included in the same ellipsoid:\[\left\{(x,y,z) \in \RR^3 ,\; (\sigma+2)x^2+3\sigma^2 y^2+z^2+2\sigma^3 xy=1\right\}.\]
$F_{2,v}$ and $F_{8,v}$ are included in the same ellipsoid:\[\left\{(x,y,z) \in \RR^3 ,\; 3\sigma^2 x^2+y^2+(\sigma+2)z^2-2\sigma^3 xz=1\right\}.\]
$F_{3,v}$ and $F_{9,v}$ are included in the same ellipsoid:\[\left\{(x,y,z) \in \RR^3 ,\; x^2+(\sigma+2)y^2+3\sigma^2 z^2-2\sigma^3yz=1\right\}.\]
$F_{4,v}$ and $F_{10,v}$ are included in the same ellipsoid:\[\left\{(x,y,z) \in \RR^3 ,\;  (\sigma+2)x^2+3\sigma^2 y^2+z^2-2\sigma^3 xy=1\right\}.\]
$F_{5,v}$ and $F_{11,v}$ are included in the same ellipsoid:\[\left\{(x,y,z) \in \RR^3 ,\; x^2+(\sigma+2)y^2+3\sigma^2 z^2+2\sigma^3yz=1\right\}.\]
$F_{6,v}$ and $F_{12,v}$ are included in the same ellipsoid:\[\left\{(x,y,z) \in \RR^3 ,\; 3\sigma^2 x^2+y^2+(\sigma+2)z^2+2\sigma^3 xz=1\right\}.\]
\end{enumerate}

\section*{Appendix B} We give the expression of the metric matrix on $\RR^2$ endowed by the metric of $\mathcal{S}^3$ which is written in \ref{metricR2}.
We denote by $f$ the application that transfoms the 2-D mesh to a mesh of $F_1$, that is $f:$
\[ 
\fl \begin{array}{cclclcl}
\RR^2&\rightarrow &\RR^3&\rightarrow  &F_{1,v}^b\subset \RR^3 &\rightarrow &F_1^b\subset \RR^4\\
 (x,y) & \mapsto &(x,y,0)&\mapsto&(x',y',z') :=(r\circ t)^{-1}(x,y,0)&\mapsto &\left(\frac{\sigma^2}{2\sqrt{2}},x',y',z'\right)
\end{array}
\]
followed by
\[
\begin{array}{lcl}
F_1^b\subset \RR^4&\rightarrow &F_1 \subset \mathcal{S}^3\\
\left(\frac{\sigma^2}{2\sqrt{2}},x',y',z'\right)& \mapsto & \frac{1}{\left\|\left(\frac{\sigma^2}{2\sqrt{2}},x',y',z'\right)\right\|}\left(\frac{\sigma^2}{2\sqrt{2}},x',y',z'\right)
\end{array}
\]

We simplify: \[\fl \left\|\left(\frac{\sigma^2}{2\sqrt{2}},x',y',z'\right)\right\|^2=\frac{5}{20^2}\left[80x^2+80y^2 +8\sqrt{2}\sqrt{5}x+\sqrt{2}(20+4\sqrt{5})y+45+17\sqrt{5}\right],\]
and we define $g(x,y)$ by 
\[\frac{g(x,y)}{20^2}:=\left\|\left(\frac{\sigma^2}{2\sqrt{2}},x',y',z'\right)\right\|^2.\]
So:
\[
\fl \begin{array}{rl}
 f_1\, := \,( {x,y} )\mapsto &\,\frac52\, \frac {\sqrt {2} \left( 3+\sqrt {5} \right) }{\sqrt {g ( x,y) }},\\
f_2\, := \,( {x,y} )\mapsto &\frac { \left( 10+2\,\sqrt{5} \right) x-4\,\sqrt{5}y-\sqrt{10}}{\sqrt {g(x,y)}},\\
f_3\, := \,( {x,y} )\mapsto &\frac12\,\frac {-8\,\sqrt {5}x+ \left( 20-4\,\sqrt {5} \right) y- \left( 5+\sqrt {5} \right) \sqrt {2}}{\sqrt {g ( x,y)}}, \\
f_4\, := \,( {x,y} )\mapsto &\frac {\sqrt {5+\sqrt {5}} \left(  \left( 5-\sqrt {5} \right) \sqrt {2}x+2\,\sqrt {10}y+\sqrt {5} \right) }{\sqrt {g( x,y)}}.
\end{array}
\]
Thanks to Maple we obtain:
\[
\fl \begin{array}{rl}
m_{11} := &320\left[ 80\,{x}^{2}+80\,{y}^{2}+8\, \sqrt{2} \sqrt{5} x+20\, \sqrt{2} y+4\, \sqrt{2} \sqrt{5} y+45+17\, \sqrt{5} \right] ^{-3}\\
&\\
\times &\left[ 1600\,{x}^{2}{y}^{2}+1600\,{y}^{4}+80\, \sqrt{2} \sqrt{5}{x}^{2}y+400\, \sqrt{2}{x}^{2}y+160\, \sqrt{2} \sqrt{5}x{y}^{2}\right.\\
&\left. +800\, \sqrt{2}{y}^{3} +160\, \sqrt{2} \sqrt{5}{y}^{3}+860\,{x}^{2}+340\, \sqrt{5}{x}^{2}+80\,xy+80\, \sqrt{5}xy\right.\\
&\left. +760\, \sqrt{5}{y}^{2}+2000\,{y}^{2}+86\, \sqrt{2} \sqrt{5}x+170\, \sqrt{2}x+258\, \sqrt{2} \sqrt{5}y+610\, \sqrt{2}y\right.\\
&\left.+845+374\, \sqrt{5}\right],
\end{array}\]
\[
\fl \begin{array}{rl}
m_{22} := &160 \left[ 80\,{x}^{2}+80\,{y}^{2}+8\, \sqrt{2} \sqrt{5}x+20\, \sqrt{2}y+4\, \sqrt{2} \sqrt{5}y+45+17\, \sqrt{5} \right] ^{-3}\\
& \\
\times &\left[3200x^4+3200x^2y^2+640\sqrt{2}\sqrt{5}x^3+800\sqrt{2}x^2y+160 \sqrt{2} \sqrt{5} x^2 y\right.\\
&\left. +320 \sqrt{5} \sqrt{2} x y^2 +3800 x^2+1320 \sqrt{5} x^2+160 \sqrt{5} x y+160 x y+1680 y^2\right.\\
&\left.+640 \sqrt{5} y^2+660 \sqrt{2} x +348 \sqrt{2} \sqrt{5} x +244 \sqrt{2} \sqrt{5} y+580 \sqrt{2} y\right.\\
&\left.+717 \sqrt{5}+1625\right],
\end{array}\]
\[
\fl \begin{array}{rl}
m_{12}:= &32000\left[80x^2+80y^2+8\sqrt{2}\sqrt{5}x+20\sqrt{2}y+4\sqrt{2}\sqrt{5}y+45+17\sqrt{5}\right]^{-6} \\
\times &\left[8192000x^9y +32768000x^7y^3+49152000x^5y^5+32768000x^3y^7\right.\\
&\left. +8192000xy^9+204800\sqrt{2}\sqrt{5}x^9 +1024000\sqrt{2}x^9 +3686400\sqrt{2}\sqrt{5}x^8y \right.\\
&\left. +2457600\sqrt{5}\sqrt{2}x^7y^2 +12288000\sqrt{2}x^7y^2 +11468800\sqrt{2}\sqrt{5}x^6y^3\right.\\
&\left.+6144000\sqrt{5}\sqrt{2}x^5y^4 +30720000\sqrt{2}x^5y^4+12288000\sqrt{2}\sqrt{5}x^4y^5\right.\\
&\left. +5734400\sqrt{5}\sqrt{2}x^3y^6 +28672000\sqrt{2}x^3y^6+4915200 \sqrt{5} \sqrt{2} x^2 y^7\right.\\
&\left.+9216000 \sqrt{2} x y^8+1843200 \sqrt{5} \sqrt{2} x y^8 +409600 \sqrt{5} \sqrt{2} y^9 +921600 x^8\right.\\
&\left. +921600 \sqrt{5} x^8+7782400 \sqrt{5} x^7 y+27443200 x^7 y +8601600 x^6 y^2\right.\\
&\left.+8601600 \sqrt{5} x^6 y^2 +84787200 x^5 y^3+25804800 \sqrt{5} x^5 y^3 +15360000 \sqrt{5} x^4 y^4 \right.\\
&\left. +15360000 x^4 y^4+87244800 x^3 y^5+28262400 \sqrt{5} x^3 y^5 +8601600 x^2 y^6 \right.\\
&\left.+8601600 \sqrt{5} x^2 y^6+29900800 x y^7+10240000 \sqrt{5} x y^7 +921600 \sqrt{5} y^8 \right.\\
&\left. +921600 y^8 +1495040 \sqrt{2} \sqrt{5} x^7+3993600 \sqrt{2} x^7 \right.\\
&\left.+7884800 \sqrt{2} \sqrt{5} x^6 y+13619200 \sqrt{2} x^6 y +13578240 \sqrt{2} \sqrt{5} x^5 y^2 \right.\\
&\left.+35328000 \sqrt{2} x^5 y^2+18329600 \sqrt{2} \sqrt{5} x^4 y^3 +32256000 \sqrt{2} x^4 y^3 \right.\\
&\left.+22835200 \sqrt{2} \sqrt{5} x^3 y^4+57856000 \sqrt{2} x^3 y^4+11857920 \sqrt{2} \sqrt{5} x^2 y^5 \right.\\
&\left.+21196800 \sqrt{2} x^2 y^5 +26521600 \sqrt{2} x y^6+10752000 \sqrt{2} \sqrt{5} x y^6 \right.\\
&\left. +1413120 \sqrt{5} \sqrt{2} y^7 +2560000 \sqrt{2} y^7+4802560 x^6+2365440 \sqrt{5} x^6 \right.\\
&\left.+43054080 x^5 y +18201600 \sqrt{5} x^5 y+31795200 x^4 y^2+15513600 \sqrt{5} x^4 y^2\right.\\
&\left.+95078400 x^3 y^3 +40704000 \sqrt{5} x^3 y^3+15820800 \sqrt{5} x^2 y^4+32716800 x^2 y^4\right.\\
&\left.+51655680 x y^5+22379520 \sqrt{5} x y^5+5232640 y^6+2508800 \sqrt{5} y^6\right.\\
&\left.+2740224 \sqrt{2} \sqrt{5} x^5 +6259200 \sqrt{2} x^5+19347200 \sqrt{2} x^4 y+8878336 \sqrt{2} \sqrt{5} x^4 y\right.\\
&\left.+16392192 \sqrt{2} \sqrt{5} x^3 y^2 +37248000 \sqrt{2} x^3 y^2+27302400 \sqrt{2} x^2 y^3\right.\\
&\left.+12486144 \sqrt{2} \sqrt{5} x^2 y^3 +13570048 \sqrt{2} \sqrt{5} x y^4 +30681600 \sqrt{2} x y^4\right.\\
&\left.+5241600 \sqrt{2} y^5+2389248 \sqrt{2} \sqrt{5} y^5+5671424 x^4+2559744 \sqrt{5} x^4\right.\\
&\left.+32037888 x^3 y+14255616 \sqrt{5} x^3 y +9666048 \sqrt{5} x^2 y^2+21451776 x^2 y^2\right.\\
&\left.+36815872 x y^3+16409088 \sqrt{5} x y^3+6244864 y^4+2809600 \sqrt{5} y^4\right.\\
&\left.+1914304 \sqrt{2} \sqrt{5} x^3+4288192 \sqrt{2} x^3 +8825088 \sqrt{2} x^2 y+3954816 \sqrt{2} \sqrt{5} x^2 y \right.\\
&\left.+5696448 \sqrt{5} \sqrt{2} x y^2+12752064 \sqrt{2} x y^2+3657984 \sqrt{2} y^3+1638528 \sqrt{2} \sqrt{5} y^3 \right.\\
&\left. +2327712 x^2 +1041696 \sqrt{5} x^2+3759032 \sqrt{5} x y\right.\\
&\left.+8408152 x y+1119072 \sqrt{5} y^2+2501088 y^2+458214 \sqrt{2} \sqrt{5} x+1024706 \sqrt{2} x\right.\\
&\left.+357278 \sqrt{2} \sqrt{5} y+798798 \sqrt{2} y+192091+85909 \sqrt{5}\right].
\end{array}
\]
\section*{Appendix C} 
We present the rotations that map $F_{1,v}^b$ to each $F_{i,v}^b$ for $i$ belonging to $\{2,6\}$. Using them, we can built a mesh of the five adjacent faces $F_{2,v}^b$, ....., $F_{6,v}^b$ of $F_{1,v}^b$.

\textbullet $ F_{1,v}^b$ is sent to $ F_{6,v}^b$ and $ F_{3,v}^b$ by rotations in $\RR^3$ with an angle $\pm\frac{2\pi}{5}$ and an axis $\vec{u}:=\frac{1}{\|\overrightarrow{OG_2}\|}\overrightarrow{OG_2}$, where $G_2$ denotes the center of $F_{2,v}^b$. We have: $\;\vec{u}=\frac{1}{\sqrt{5}\sqrt{2+\sigma}} \left(-(\sigma+2) \mathbf{i}+\sqrt{5}\,\mathbf{k}\right)$. It is easier to calculate a rotation with an angle equal to $\frac{4\pi}{5}$ than with an angle equal to $\frac{2\pi}{5}$. Hence we begin to calculate $\rho$ a rotation in $\RR^3$ with an angle $\frac{4\pi}{5}$ and an axis $\vec{u}$:
\[
\fl \begin{array}{ll}
\rho&\left(x \mathbf{i}+y \mathbf{j}+z \mathbf{k}\right)\\
&= \left[\cos\left(\frac{2\pi}{5}\right) \mathbf{1}+\sin\left(\frac{2\pi}{5}\right) \vec{u}\right]\left[x \mathbf{i}+y \mathbf{j}+z \mathbf{k}\right]\left[\cos\left(\frac{2\pi}{5}\right) \mathbf{1}-\sin\left(\frac{2\pi}{5}\right) \vec{u}\right]\\
&\\
&= \left[\frac{1}{2\sigma}\mathbf{1}+\frac{\sqrt{2+\sigma}}{2}\frac{1}{\sqrt{5}\sqrt{2+\sigma}} \left(-(2+\sigma)\mathbf{i}+ \sqrt{5} \,\mathbf{k}\right)\right]\left[x \mathbf{i}+y \mathbf{j}+z \mathbf{k}\right]\\
&\qquad \qquad \qquad \qquad \qquad \qquad  \left[\frac{1}{2\sigma}\mathbf{1}-\frac{\sqrt{2+\sigma}}{2}\frac{1}{\sqrt{5}\sqrt{2+\sigma}} \left(-(2+\sigma)\mathbf{i}+ \sqrt{5} \,\mathbf{k}\right)\right].
\end{array}
\]
So 
\[
\rho(x,y,z)=\frac12\left(\begin{array}{rrr}
1&-\frac{1}{\sigma}&-\sigma\\
\frac{1}{\sigma}&-\sigma &1\\
-\sigma&-1&-\frac{1}{\sigma}
\end{array} \right) 
\left(\begin{array}{l}
x\\
y\\
z
\end{array} \right). 
\]
Hence $\rho^2:\, F_{1,v}^b \rightarrow F_{6,v}^b$, and $\rho^3:\, F_{1,v}^b \rightarrow F_{3,v}^b$. We have:
\[
\fl \rho^2(x,y,z)=\frac12\left(\begin{array}{rrr}
\sigma&1&-\frac{1}{\sigma}\\
-1&\frac{1}{\sigma}&-\sigma \\
-\frac{1}{\sigma}&\sigma&1
\end{array} \right) 
\left(\begin{array}{l}
x\\
y\\
z
\end{array} \right),\,
\rho^3(x,y,z)=\frac12\left(\begin{array}{rrr}
\sigma&-1&-\frac{1}{\sigma}\\
1&\frac{1}{\sigma}&\sigma \\
-\frac{1}{\sigma}&-\sigma&1
\end{array} \right) 
\left(\begin{array}{l}
x\\
y\\
z
\end{array} \right). 
\]

\textbullet $ F_{1,v}^b$ is sent to $ F_{2,v}^b$ and $ F_{4,v}^b$ by rotations in $\RR^3$ with an angle $\pm\frac{2\pi}{5}$ and an axis $\vec{u}:=\frac{1}{\|\overrightarrow{OG_3}\|}\overrightarrow{OG_3}$, where $G_3$ denotes the center of $F_{3,v}^b$. We have: $\;\vec{u}=\frac{1}{\sqrt{5}\sqrt{2+\sigma}} \left(-\sqrt{5}\,\mathbf{j }+(\sigma+2) \mathbf{k}\right)$. Once more we begin to calculate $\rho$ a rotation in $\RR^3$ with an angle $\frac{4\pi}{5}$ and an axis $\vec{u}$:
\[
\fl \begin{array}{ll}
\rho&\left(x \mathbf{i}+y \mathbf{j}+z \mathbf{k}\right)=\\
& \left[\frac{1}{2\sigma}\mathbf{1}+\frac{\sqrt{2+\sigma}}{2}\frac{1}{\sqrt{5}\sqrt{2+\sigma}} \left(- \sqrt{5} \,\mathbf{j}+(2+\sigma)\mathbf{k}\right)\right]\left[x \mathbf{i}+y \mathbf{j}+z \mathbf{k}\right]\\
&\qquad \qquad \qquad \qquad \qquad \qquad \left[\frac{1}{2\sigma}\mathbf{1}-\frac{\sqrt{2+\sigma}}{2}\frac{1}{\sqrt{5}\sqrt{2+\sigma}} \left(- \sqrt{5} \,\mathbf{j}+(2+\sigma)\mathbf{k}\right)\right].
\end{array}
\]
So 
\[
\rho(x,y,z)=\frac12\left(\begin{array}{rrr}
-\sigma&-1&-\frac{1}{\sigma}\\
1&-\frac{1}{\sigma}&-\sigma \\
\frac{1}{\sigma}&-\sigma&1
\end{array} \right) 
\left(\begin{array}{l}
x\\
y\\
z
\end{array} \right) .
\]
Hence $\rho^2:\, F_{1,v}^b \rightarrow F_{2,v}^b$, and $\rho^3:\, F_{1,v}^b \rightarrow F_{4,v}^b$. We have:
\[
\fl \rho^2(x,y,z)=\frac12\left(\begin{array}{rrr}
\frac{1}{\sigma}&\sigma&1\\
-\sigma&1&-\frac{1}{\sigma}\\
-1&-\frac{1}{\sigma}&\sigma
\end{array} \right) 
\left(\begin{array}{l}
x\\
y\\
z
\end{array} \right),\,
\rho^3(x,y,z)=\frac12\left(\begin{array}{rrr}
\frac{1}{\sigma}&-\sigma&-1\\
\sigma&1&-\frac{1}{\sigma}\\
1&-\frac{1}{\sigma}&\sigma
\end{array} \right) 
\left(\begin{array}{l}
x\\
y\\
z
\end{array} \right) .
\]
\textbullet $ F_{1,v}^b$ is sent to $ F_{5,v}^b$ by a rotation in $\RR^3$ with an angle $\pm\frac{2\pi}{5}$ and an axis $\vec{u}:=\frac{1}{\|\overrightarrow{OG_6}\|}\overrightarrow{OG_6}$, where $G_6$ denotes the center of $F_{6,v}^b$. We have: $\;\vec{u}=\frac{1}{\sqrt{5}\sqrt{2+\sigma}} \left(-(\sigma+2) \mathbf{i}-\sqrt{5}\,\mathbf{k }\right)$. Once more we begin to calculate $\rho$ a rotation in $\RR^3$ with an angle $\frac{4\pi}{5}$ and an axis $\vec{u}$:
\[
\fl \begin{array}{ll}
\rho&\left(x \mathbf{i}+y \mathbf{j}+z \mathbf{k}\right)=\\
& \left[\frac{1}{2\sigma}\mathbf{1}+\frac{\sqrt{2+\sigma}}{2}\frac{1}{\sqrt{5}\sqrt{2+\sigma}} \left(-(2+\sigma)\mathbf{i}- \sqrt{5} \,\mathbf{k}\right)\right]\left[x \mathbf{i}+y \mathbf{j}+z \mathbf{k}\right]\\
&\qquad \qquad \qquad \qquad \qquad \qquad  \left[\frac{1}{2\sigma}\mathbf{1}-\frac{\sqrt{2+\sigma}}{2}\frac{1}{\sqrt{5}\sqrt{2+\sigma}} \left(-(2+\sigma)\mathbf{i}- \sqrt{5} \,\mathbf{k}\right)\right].
\end{array}
\]
So 
\[
\rho(x,y,z)=\frac12\left(\begin{array}{rrr}
1&\frac{1}{\sigma}&\sigma\\
-\frac{1}{\sigma}&-\sigma&1 \\
\sigma&-1&-\frac{1}{\sigma}
\end{array} \right) 
\left(\begin{array}{l}
x\\
y\\
z
\end{array} \right) .
\]
Hence $\rho^2:\, F_{1,v}^b \rightarrow F_{5,v}^b$. We have:
\[
\fl \rho^2(x,y,z)=\frac12\left(\begin{array}{rrr}
\sigma&-1&\frac{1}{\sigma}\\
1&\frac{1}{\sigma}&-\sigma\\
\frac{1}{\sigma}&\sigma&1
\end{array} \right) 
\left(\begin{array}{l}
x\\
y\\
z
\end{array} \right).
\]


\end{document}